\documentclass[twocolumn]{aastex62}

\newcommand{\todo}{\ifmmode \text{\color{red}\Huge{\(\bullet\)}} \else {\color{red}{\Huge$\bullet$}}\fi}
\newcommand{\tido}{\ifmmode {{\color{red}\bullet}} \else {\color{red}$\bullet$}\fi}

\newcommand{\E        }[1]{\ifmmode 10^{#1} \else $10^{#1}$\fi}
\newcommand{\tE        }[1]{\ifmmode \times10^{#1} \else $\times10^{#1}$\fi}
\newcommand{\til}{\ifmmode \sim \else $\sim$\fi}
\renewcommand{\~} {\ifmmode \sim \else $\sim$\fi}

\newcommand{\pc}	{\ifmmode {\rm pc} \else pc\fi}
\newcommand{\kpc}	{\ifmmode {\rm kpc} \else kpc\fi}
\newcommand{\ld}	{\ifmmode {\rm l.d.} \else l.d.\fi}
\newcommand{\kms}	{\ifmmode {\rm km\,s}^{-1} \else km\,s$^{-1}$\fi}
\newcommand{\cc}	{\ifmmode {\rm cm}^{-3}    \else cm$^{-3}$\fi}
\newcommand{\cmii}	{\ifmmode {\rm cm}^{-2}    \else cm$^{-2}$\fi}
\newcommand{\ergs}	{\ifmmode {\rm erg\,s}^{-1} \else erg s$^{-1}$\fi}
\newcommand{\ergcms}	{\ifmmode {\rm erg\,cm}^{-2}\,{\rm s}^{-1} \else erg\,cm$^{-2}$\,s$^{-1}$\fi}
\newcommand{\ergcmsA}	{\ifmmode {\rm erg\,cm}^{-2}\,{\rm s}^{-1}\,{\rm\AA}^{-1}
\else erg\,cm$^{-2}$\,s$^{-1}$\,\AA$^{-1}$\fi}
\newcommand{  \ergcmsHz  }{\ifmmode{\rm erg\,cm}^{-2}\,{\rm s}^{-1}\,{\rm Hz}^{-1}
                       \else ergs\,cm$^{-2}$\,s$^{-1}$\,Hz$^{-1}$\fi}
\newcommand{\kev}	{\ifmmode {\rm keV} \else keV\fi}

\newcommand{\mic}	{\ifmmode {\rm \mu m} \else $\mu$m\fi}
\newcommand{\vFWHM}	{\ifmmode v_{\mbox{\tiny FWHM}} \else $v_{\mbox{\tiny FWHM}}$\fi}
\newcommand{\vBLR}	{\ifmmode v_{\mbox{\tiny BLR}} \else $v_{\mbox{\tiny BLR}}$\fi}
\newcommand{\sigBLR}	{\ifmmode \sigma_{\mbox{\tiny BLR}} \else $\sigma_{\mbox{\tiny BLR}}$\fi}
\newcommand{\vNLR}	{\ifmmode v_{\mbox{\tiny NLR}} \else $v_{\mbox{\tiny NLR}}$\fi}
\newcommand{\tauBLR}	{\ifmmode \tau_{\mbox{\tiny BLR}} \else $\tau_{\mbox{\tiny BLR}}$\fi}

\newcommand{\Hubble}	{\ifmmode {\rm km\,s}^{-1}\,{\rm Mpc}^{-1} \else km\,s$^{-1}$\,Mpc$^{-1}$\fi}
\newcommand{\NDunit}	{\ifmmode {\rm Mpc}^{-3} \else Mpc$^{-3}$\fi}
\newcommand{\LFunit}	{\ifmmode {\rm Mpc}^{-3}\,{\rm mag}^{-1} \else Mpc$^{-3}$\,mag$^{-1}$\fi}
\newcommand{\MFunit}	{\ifmmode {\rm Mpc}^{-3}\,{\rm dex}^{-1} \else Mpc$^{-3}$\,dex$^{-1}$\fi}

\newcommand{\Msun}{\ifmmode M_{\odot} \else $M_{\odot}$\fi}
\newcommand{\Lsun}{\ifmmode L_{\odot} \else $L_{\odot}$\fi}
\newcommand{\Zsun}{\ifmmode Z_{\odot} \else $Z_{\odot}$\fi}
\newcommand{\mpyr}{\ifmmode \Msun\,{\rm yr}^{-1} \else $\Msun\,{\rm yr}^{-1}$\fi}

\newcommand{\Msol}{\Msun}
\newcommand{\Lsol}{\Lsun}

\newcommand{\qnote}{\ifmmode q_{0} \else $q_{0}$\fi}
\newcommand{\Hnote}{\ifmmode H_{0} \else $H_{0}$\fi}
\newcommand{\hnote}{\ifmmode h_{0} \else $h_{0}$\fi}
\newcommand{\anote}{\ifmmode a_{0} \else $a_{0}$\fi}
\newcommand{\tnote}{\ifmmode t_{0} \else $t_{0}$\fi}

\newcommand{\ltsim}{\raisebox{-.5ex}{$\;\stackrel{<}{\sim}\;$}}

\def\gsim{\;\rlap{\lower 2.5pt \hbox{$\sim$}}\raise 1.5pt\hbox{$>$}\;}
\def\lsim{\;\rlap{\lower 2.5pt \hbox{$\sim$}}\raise 1.5pt\hbox{$<$}\;}

\newcommand{  \Halpha   }{\ifmmode {\rm H}\alpha \else H$\alpha$\fi}

\newcommand{  \ha       }{\Halpha}
\newcommand{  \Hbeta    }{\ifmmode {\rm H}\beta \else H$\beta$\fi}

\newcommand{  \hb       }{\Hbeta}
\newcommand{  \Hgamma   }{\ifmmode {\rm H}\gamma \else H$\gamma$\fi}
\newcommand{  \Hdelta   }{\ifmmode {\rm H}\delta \else H$\delta$\fi}
\newcommand{  \Lya      }{\ifmmode {\rm Ly}\alpha \else Ly$\alpha$\fi}
\newcommand{  \Lyb      }{\ifmmode {\rm Ly}\beta \else Ly$\beta$\fi}
\newcommand{  \Pa       }{\ifmmode {\rm P}\alpha \else P$\alpha$\fi}
\newcommand{  \Pb       }{\ifmmode {\rm P}\beta \else P$\beta$\fi}
\newcommand{  \Bra      }{\ifmmode {\rm Br}\alpha \else Br$\alpha$\fi}
\newcommand{  \Brg      }{\ifmmode {\rm Br}\gamma \else Br$\gamma$\fi}
\newcommand{  \hii      }{\ifmmode {\rm H}\,\textsc{ii} \else H\,\textsc{ii}\fi}
\newcommand{  \hei      }{\ifmmode {\rm He}\,\textsc{i} \else He\,\textsc{i}\fi}
\newcommand{  \heii     }{\ifmmode {\rm He}\,\textsc{ii} \else He\,\textsc{ii}\fi}
\newcommand{  \HeIIuv   }{\ifmmode {\rm He}\,\textsc{ii}\,\lambda1640 \else He\,\textsc{ii}\,$\lambda1640$\fi}
\newcommand{  \HeIIop   }{\ifmmode {\rm He}\,\textsc{ii}\,\lambda4686 \else He\,\textsc{ii}\,$\lambda4686$\fi}
\newcommand{  \CII	}{\ifmmode \left[{\rm C}\,\textsc{ii}\right]\,\lambda157.74\,\mu{\rm m} \else [C\,{\sc ii}]\ $\lambda157.74\,\mu{\rm m}$\fi}
\newcommand{  \cii	}{\ifmmode \left[{\rm C}\,\textsc{ii}\right] \else [C\,{\sc ii}]\fi}

\newcommand{  \ciii     }{\ifmmode {\rm C}\,\textsc{iii}\right] \else C\,\textsc{iii}]\fi}
\newcommand{  \CIII     }{\ifmmode {\rm C}\,\textsc{iii}\right]\,\lambda1909 \else C\,\textsc{iii}]\,$\lambda1909$\fi}
\newcommand{  \civ      }{\ifmmode {\rm C}\,\textsc{iv}  \else C\,\textsc{iv}\fi}
\newcommand{  \CIV      }{\ifmmode {\rm C}\,\textsc{iv}\,\lambda1549 \else C\,\textsc{iv}\,$\lambda1549$\fi}
\newcommand{  \NIIopt   }{\ifmmode \left[{\rm N}\,\textsc{ii}\right]\,\lambda6584 \else [N\,\textsc{ii}]\,$\lambda6584$\fi}
\newcommand{  \nii      }{\ifmmode \left[{\rm N}\,\textsc{ii}\right]  \else [N\,\textsc{ii}]\fi}
\newcommand{  \niii     }{\ifmmode {\rm N}\,\textsc{iii} \else N\,\textsc{iii}\fi}
\newcommand{  \NIII     }{\ifmmode {\rm N}\,\textsc{iii}\,\lambda4640 \else N\,\textsc{iii}\,$\lambda4640$\fi}
\newcommand{  \niv      }{\ifmmode {\rm N}\,\textsc{iv}  \else N\,\textsc{iv}\fi}
\newcommand{  \NIVuv    }{\ifmmode {\rm N}\,\textsc{iv}\,\lambda1486 \else N\,\textsc{iv}\,$\lambda1486$\fi}
\newcommand{  \nv       }{\ifmmode {\rm N}\,\textsc{v}   \else N\,\textsc{v}\fi}
\newcommand{\oi}{\ifmmode \left[{\rm O}\,\textsc{i}\right] \else [O\,{\sc i}]\fi}
\newcommand{\OI}{\ifmmode \left[{\rm O}\,\textsc{i}\right]\,\lambda6300 \else [O\,{\sc i}]$\,\lambda6300$\fi}
\newcommand{\oii}{\ifmmode \left[{\rm O}\,\textsc{ii}\right] \else [O\,{\sc ii}]\fi}
\newcommand{\OII}{\ifmmode \left[{\rm O}\,\textsc{ii}\right]\,\lambda3727 \else [O\,{\sc ii}]\,$\lambda3727$\fi}
\newcommand{\oiii}{\ifmmode \left[{\rm O}\,\textsc{iii}\right] \else [O\,{\sc iii}]\fi}
\newcommand{\OIII}{\ifmmode \left[{\rm O}\,\textsc{iii}\right]\,\lambda5007 \else [O\,{\sc iii}]\,$\lambda5007$\fi}
\newcommand{  \OIIIbf   }{\ifmmode {\rm O}\,\textsc{iii}\,\lambda3133 \else O\,\textsc{iii}\,$\lambda3133$\fi}
\newcommand{  \OIIIuv   }{\ifmmode {\rm O}\,\textsc{iii}\,\lambda1663 \else O\,\textsc{iii}\,$\lambda1663$\fi}
\newcommand{  \oiv      }{\ifmmode {\rm O}\,\textsc{iv}  \else O\,\textsc{iv}\fi}
\newcommand{  \OIVuv    }{\ifmmode {\rm O}\,\textsc{iv}\,\lambda1402  \else O\,\textsc{iv}\,$\lambda1402$\fi}
\newcommand{  \OIVIR    }{\ifmmode {\rm O}\,\textsc{iv}\,25.9\,\mu {\rm m} \else O\,\textsc{iv}\,$25.9\,\mu$m\fi}
\newcommand{  \ovi      }{\ifmmode {\rm O}\,\textsc{vi}   \else O\,\textsc{vi}\fi}
\newcommand{  \Ovi      }{\ifmmode {\rm O}\,\textsc{vi}\,\lambda1035 \else O\,\textsc{vi}\,$\lambda1035$\fi}
\newcommand{  \nei      }{\ifmmode {\rm Ne}\,\textsc{i}   \else Ne\,\textsc{i}\fi}
\newcommand{  \neii     }{\ifmmode {\rm Ne}\,\textsc{ii}  \else Ne\,\textsc{ii}\fi}
\newcommand{  \NeiiIR   }{\ifmmode {\rm Ne}\,\textsc{ii}\,12.8\,\mu {\rm m} \else Ne\,\textsc{ii}\,$12.8\,\mu$m\fi}
\newcommand{  \neiii    }{\ifmmode {\rm Ne}\,\textsc{iii} \else Ne\,\textsc{iii}\fi}
\newcommand{  \neiv     }{\ifmmode {\rm Ne}\,\textsc{iv}  \else Ne\,\textsc{iv}\fi}
\newcommand{  \nev      }{\ifmmode {\rm Ne}\,\textsc{v}   \else Ne\,\textsc{v}\fi}
\newcommand{  \NevIR    }{\ifmmode {\rm Ne}\,\textsc{v}\,24.3\,\mu {\rm m} \else Ne\,\textsc{v}\,$24.3\,\mu$m\fi}
\newcommand{  \nevi     }{\ifmmode {\rm Ne}\,\textsc{vi}  \else Ne\,\textsc{vi}\fi}
\newcommand{  \mgi      }{\ifmmode {\rm Mg}\,\textsc{i} \else Mg\,\textsc{i}\fi}
\newcommand{  \mgii     }{\ifmmode {\rm Mg}\,\textsc{ii} \else Mg\,\textsc{ii}\fi}
\newcommand{  \MgII     }{\ifmmode {\rm Mg}\,\textsc{ii}\,\lambda2798 \else Mg\,\textsc{ii}\,$\lambda2798$\fi}
\newcommand{  \sii      }{\ifmmode {\rm S}\,\textsc{ii} \else S\,\textsc{ii}\fi}
\newcommand{  \siii     }{\ifmmode {\rm S}\,\textsc{iii} \else S\,\textsc{iii}\fi}
\newcommand{  \siv      }{\ifmmode {\rm S}\,\textsc{iv} \else S\,\textsc{iv}\fi}
\newcommand{  \sili     }{\ifmmode {\rm Si}\,\textsc{i}   \else Si\,\textsc{i}\fi}
\newcommand{  \silii    }{\ifmmode {\rm Si}\,\textsc{ii}  \else Si\,\textsc{ii}\fi}
\newcommand{  \Siliv    }{\ifmmode {\rm Si}\,\textsc{iv}  \else Si\,\textsc{iv}\fi}
\newcommand{  \SilIVuv  }{\ifmmode {\rm Si}\,\textsc{iv}\,\lambda1400  \else Si\,\textsc{iv}\,$\lambda1400$\fi}
\newcommand{  \AlIII   }{\ifmmode {\rm Al}\,\textsc{iii}\,\lambda1857 \else Al\,\textsc{iii}\,$\lambda1857$\fi}
\newcommand{  \Aliii   }{\ifmmode {\rm Al}\,\textsc{iii} \else Al\,\textsc{iii}\fi}
\newcommand{  \caii     }{\ifmmode {\rm Ca}\,\textsc{ii} \else Ca\,\textsc{ii}\fi}
\newcommand{  \feii     }{\ifmmode {\rm Fe}\,\textsc{ii} \else Fe\,\textsc{ii}\fi}
\newcommand{  \feiii    }{\ifmmode {\rm Fe}\,\textsc{iii} \else Fe\,\textsc{iii}\fi}
\newcommand{  \Kalpha   }{\ifmmode {\rm K}\alpha \else K$\alpha$\fi}

\newcommand{ \Lhb   }{\ifmmode L_{\hb} \else $L_{\hb}$\fi}
\newcommand{ \Lha   }{\ifmmode L_{\ha} \else $L_{\ha}$\fi}
\newcommand{ \fwhb  }{\ifmmode {\rm FWHM}\left(\hb\right) \else FWHM(\hb)\fi}
\newcommand{\sighb  }{\ifmmode \sigma\left(\hb\right) \else $\sigma\left(\hb\right)$\fi}
\newcommand{ \ewhb  }{\ifmmode {\rm EW}\left(\hb\right) \else EW(\hb)\fi}
\newcommand{ \fwha  }{\ifmmode {\rm FWHM}\left(\ha\right) \else FWHM(\ha)\fi}
\newcommand{ \ewha  }{\ifmmode {\rm EW}\left(\ha\right) \else EW(\ha)\fi}
\newcommand{ \Lmg   }{\ifmmode L\left(\mgii\right) \else $L\left(\mgii\right)$\fi}
\newcommand{ \fwmg  }{\ifmmode {\rm FWHM}\left(\mgii\right) \else FWHM(\mgii)\fi}
\newcommand{ \Lciv  }{\ifmmode L\left(\civ\right) \else $L\left(\civ\right)$\fi}
\newcommand{ \fwciv }{\ifmmode {\rm FWHM}\left(\civ\right) \else FWHM(\civ)\fi}
\newcommand{ \fwhm  }{\ifmmode {\rm FWHM} \else FWHM\fi} 
\newcommand{ \voff  }{\ifmmode v_{\rm off} \else $v_{\rm off}$\fi} 
\newcommand{ \vmax  }{\ifmmode v_{\rm max} \else $v_{\rm max}$\fi} 

\newcommand{ \mumg  }{\ifmmode \mu\left(\mgii\right) \else $\mu\left(\mgii\right)$\fi}
\newcommand{ \fmg   }{\ifmmode f\left(\mgii\right) \else $f\left(\mgii\right)$\fi}
\newcommand{ \muciv }{\ifmmode \mu\left(\civ\right) \else $\mu\left(\civ\right)$\fi}
\newcommand{ \fciv  }{\ifmmode f\left(\civ\right) \else $f\left(\civ\right)$\fi}
\newcommand{  \auvo     }{\ifmmode \alpha_{\nu,{\rm UVO}} \else $\alpha_{\nu,{\rm UVO}}$\fi}
\newcommand{  \Ledd     }{\ifmmode L_{\rm Edd} \else $L_{\rm Edd}$\fi}
\newcommand{  \lamLlam  }{\ifmmode \lambda L_{\lambda} \else $\lambda L_{\lambda}$\fi}
\newcommand{  \lLl      }{\ifmmode \lambda L_{\lambda} \else $\lambda L_{\lambda}$\fi}
\newcommand{  \nuLnu    }{\ifmmode \nu L_{\nu} \else $\nu L_{\nu}$\fi}
\newcommand{  \nLn      }{\ifmmode \nu L_{\nu} \else $\nu L_{\nu}$\fi}
\newcommand{  \Luv      }{\ifmmode L_{1450} \else $L_{1450}$\fi}
\newcommand{  \Lop      }{\ifmmode L_{5100} \else $L_{5100}$\fi}
\newcommand{  \lLop     }{\ifmmode \log\left(\Lop/\ergs\right) \else $\log\left(\Lop/\ergs\right)$\fi}
\newcommand{  \Lthree   }{\ifmmode L_{3000} \else $L_{3000}$\fi}
\newcommand{  \lLthree  }{\ifmmode \log\left(\Lthree/\ergs\right) \else $\log\left(\Lthree/\ergs\right)$\fi}
\newcommand{  \Lsix      }{\ifmmode L_{6200} \else $L_{6200}$\fi}
\newcommand{  \lLisx     }{\ifmmode \log\left(\Lop/\ergs\right) \else $\log\left(\Lop/\ergs\right)$\fi}
\newcommand{  \Lxray    }{\ifmmode L_{\rm X} \else $L_{\rm X}$\fi}

\newcommand{  \Lhard    }{\ifmmode L_{\rm 2-10} \else $L_{\rm 2-10}$\fi}
\newcommand{  \Lsoft    }{\ifmmode L_{\rm 0.5-2} \else $L_{\rm 0.5-2}$\fi}

\newcommand{\Fthree}{\ifmmode F_{3000} \else $F_{3000}$\fi}
\newcommand{\fuv}{\ifmmode f_{\lambda}\left(1450{\rm \AA}\right) \else $f_{\lambda}\left(1450 {\rm \AA}\right)$\fi}
\newcommand{\fthree}{\ifmmode f_{\lambda}\left(3000{\rm \AA}\right) \else $f_{\lambda}\left(3000{\rm \AA}\right)$\fi}
\newcommand{\fH}{\ifmmode f_{\lambda}\left(1.65\micron\right) \else
$f_{\lambda}\left(1.65\micron\right)$\fi}

\newcommand{\fbol}{\ifmmode f_{\rm bol} \else $f_{\rm bol}$\fi}
\newcommand{\fbolwv}{\ifmmode f_{\rm bol}\left(\lambda\right) \else $f_{\rm bol}\left(\lambda\right)$\fi}
\newcommand{\fbolopt}{\ifmmode f_{\rm bol}\left(5100{\rm \AA}\right) \else $f_{\rm bol}\left(5100{\rm \AA}\right)$\fi}
\newcommand{\fbolthree}{\ifmmode f_{\rm bol}\left(3000{\rm \AA}\right) \else $f_{\rm bol}\left(3000{\rm \AA}\right)$\fi}
\newcommand{\fboluv}{\ifmmode f_{\rm bol}\left(1450{\rm \AA}\right) \else $f_{\rm bol}\left(1450{\rm \AA}\right)$\fi}

\newcommand{\fbolbat}{\ifmmode f_{\rm bol}\left(14-150\,\kev\right) \else $f_{\rm bol}\left(14-150\,\kev\right)$\fi}
\newcommand{\fbolhard}{\ifmmode f_{\rm bol}\left(2-10\,\kev\right) \else $f_{\rm bol}\left(2-10\,\kev\right)$\fi}

\newcommand{\fobs}{\ifmmode f_{\rm obs} \else $f_{\rm obs}$\fi}

\newcommand{  \mbh      }{\ifmmode M_{\rm BH} \else $M_{\rm BH}$\fi}
\newcommand{  \lmbh     }{\ifmmode \log\left(\mbh/\Msun\right) \else $\log\left(\mbh/\Msun\right)$\fi} 
\newcommand{  \lledd    }{\ifmmode L/L_{\rm Edd} \else $L/L_{\rm Edd}$\fi}
\newcommand{  \mmedd    }{\ifmmode \dot{m}/\dot{m}_{\rm \,Edd} \else $\dot{m}/\dot{m}_{\rm \,Edd}$\fi}
\newcommand{  \Lbol     }{\ifmmode L_{\rm bol} \else $L_{\rm bol}$\fi}
\newcommand{  \lbol     }{\ifmmode L_{\rm bol} \else $L_{\rm bol}$\fi}
\newcommand{  \lLbol    }{\ifmmode \log\left(\Lbol/\ergs\right) \else $\log\left(\Lbol/\ergs\right)$\fi} 
\newcommand{  \Lagn     }{\ifmmode L_{\rm AGN} \else $L_{\rm AGN}$\fi}
\newcommand{  \lagn     }{\ifmmode L_{\rm AGN} \else $L_{\rm AGN}$\fi}

\newcommand{  \tgrow     }{\ifmmode t_{\rm growth} \else $t_{\rm growth}$\fi}
\newcommand{  \tAD     }{\ifmmode t_{\rm acc} \else $t_{\rm acc}$\fi}
\newcommand{  \tacc    }{\ifmmode t_{\rm acc} \else $t_{\rm acc}$\fi}
\newcommand{  \tUni      }{\ifmmode t_{\rm Universe} \else $t_{\rm Universe}$\fi}

\newcommand{  \Mdotin	}{\ifmmode \dot{M}_{\rm infall} \else $\dot{M}_{\rm infall}$\fi}
\newcommand{  \Mdotbh	}{\ifmmode \dot{M}_{\rm BH} \else $\dot{M}_{\rm BH}$\fi}
\newcommand{  \Mdotad	}{\ifmmode \dot{M}_{\rm AD} \else $\dot{M}_{\rm AD}$\fi}
\newcommand{  \Mdotacc	}{\ifmmode \dot{M}_{\rm acc} \else $\dot{M}_{\rm acc}$\fi}
\newcommand{  \Mdotthin	}{\ifmmode \dot{M}_{\rm thin} \else $\dot{M}_{\rm thin}$\fi}
\newcommand{  \Mdotdisk	}{\ifmmode \dot{M}_{\rm disk} \else $\dot{M}_{\rm disk}$\fi}

\newcommand{  \Mindot	}{\ifmmode \dot{M}_{\rm infall} \else $\dot{M}_{\rm infall}$\fi}
\newcommand{  \Mbhdot	}{\ifmmode \dot{M}_{\rm BH} \else $\dot{M}_{\rm BH}$\fi}
\newcommand{  \Maddot	}{\ifmmode \dot{M}_{\rm AD} \else $\dot{M}_{\rm AD}$\fi}
\newcommand{  \Maccdot	}{\ifmmode \dot{M}_{\rm acc} \else $\dot{M}_{\rm acc}$\fi}
\newcommand{  \Mthdot	}{\ifmmode \dot{M}_{\rm thin} \else $\dot{M}_{\rm thin}$\fi}
\newcommand{  \Mdsdot	}{\ifmmode \dot{M}_{\rm disk} \else $\dot{M}_{\rm disk}$\fi}

\newcommand{  \as	}{\ifmmode a_{\rm *} \else $a_{\rm *}$\fi}
\newcommand{  \avec	}{\ifmmode \vec{a}_{\rm *} \else $\vec{a}_{\rm *}$\fi}
\newcommand{  \re	}{\ifmmode \eta      	 \else $\eta$\fi}
\newcommand{  \RISCO	}{\ifmmode R_{\rm ISCO}  \else $R_{\rm ISCO}$\fi}

\newcommand{  \mseed    }{\ifmmode M_{\rm seed} \else $M_{\rm seed}$\fi}
\newcommand{  \mbul     }{\ifmmode M_{\rm bulge} \else $M_{\rm bulge}$\fi} 
\newcommand{  \mstar    }{\ifmmode M_{*} \else $M_{*}$\fi} 
\newcommand{  \mgal     }{\ifmmode M_{*} \else $M_{*}$\fi} 
\newcommand{  \mhost    }{\ifmmode M_{\rm host} \else $M_{\rm host}$\fi}
\newcommand{  \mmsmall  }{\ifmmode M_{\rm BH}/M_{*} \else $M_{\rm BH}/M_{*}$\fi}
\newcommand{  \mmlarge  }{\ifmmode M_{*}/M_{\rm BH} \else $M_{*}/M_{\rm BH}$\fi}

\newcommand{  \mmdotlarge}{\ifmmode \dot{M}_*/\Mbhdot \else $\dot{M}_*/\Mbhdot$\fi}
\newcommand{  \mmdotsmall}{\ifmmode \Mbhdot/\dot{M}_* \else $\Mbhdot/\dot{M}_*$\fi}

\newcommand{  \mmwp     }{\ifmmode \left(M_{*}/M_{\rm BH}\right) \else $\left(M_{*}/M_{\rm BH}\right)$\fi}
\newcommand{  \ml       }{\ifmmode M_{*}/L_{*} \else $M_{*}/L_{*}$\fi}
\newcommand{  \mlwp     }{\ifmmode \left(M_{*}/L\right) \else $\left(M_{*}/L\right)$\fi}
\newcommand{  \mlk      }{\ifmmode \left(M_{*}/L_{K}\right) \else $\left(M_{*}/L_{K}\right)$\fi}
\newcommand{  \sigs     }{\ifmmode \sigma_{*} \else $\sigma_{*}$\fi}
\newcommand{  \Reff     }{\ifmmode R_{\rm e} \else $R_{\rm e}$\fi}
\newcommand{  \Rvir     }{\ifmmode R_{\rm vir} \else $R_{\rm vir}$\fi}
\newcommand{  \Rtwo     }{\ifmmode R_{200} \else $R_{200}$\fi}
\newcommand{  \Rfive    }{\ifmmode R_{500} \else $R_{500}$\fi}
\newcommand{  \Rgrp     }{\ifmmode R_{\rm grp} \else $R_{\rm grp}$\fi}
\newcommand{  \nser     }{\ifmmode n_{\rm s} \else $n_{\rm s}$\fi}
\newcommand{  \LSF      }{\ifmmode L_{\rm SF}  \else $L_{\rm SF}$\fi}
\newcommand{  \LFIR     }{\ifmmode L_{\rm FIR} \else $L_{\rm FIR}$\fi}
\newcommand{  \Lfir     }{\ifmmode L_{\rm FIR} \else $L_{\rm FIR}$\fi}
\newcommand{  \LTIR     }{\ifmmode L_{\rm TIR} \else $L_{\rm TIR}$\fi}
\newcommand{  \Ltir     }{\ifmmode L_{\rm TIR} \else $L_{\rm TIR}$\fi}

\newcommand{  \mdyn     }{\ifmmode M_{\rm dyn} \else $M_{\rm dyn}$\fi} 
\newcommand{  \mgas     }{\ifmmode M_{\rm gas} \else $M_{\rm gas}$\fi} 
\newcommand{  \mh       }{\ifmmode M_{\rm h} \else $M_{\rm h}$\fi}
\newcommand{  \mhalo    }{\ifmmode M_{\rm halo} \else $M_{\rm halo}$\fi}
\newcommand{  \sfr      }{\ifmmode {\rm SFR} \else SFR\fi}

\newcommand{ \Lcii     }{\ifmmode L_{\cii} \else $L_{\cii}$\fi}
\newcommand{ \fwcii  }{\ifmmode {\rm FWHM}\cii \else FWHM\cii\fi}

\newcommand{  \herschel} {{\it Herschel}}

\newcommand{\bj}{\ifmmode b_{\rm J} \else $b_{\rm J}$\fi}

\newcommand{\iab}{\ifmmode i_{\rm AB} \else $i_{\rm AB}$\fi}

\newcommand{\jab}{\ifmmode J_{\rm AB} \else $J_{\rm AB}$\fi}
\newcommand{\hab}{\ifmmode H_{\rm AB} \else $H_{\rm AB}$\fi}
\newcommand{\kab}{\ifmmode K_{\rm AB} \else $K_{\rm AB}$\fi}

\newcommand{\jveg}{\ifmmode J_{\rm Vega} \else $J_{\rm Vega}$\fi}
\newcommand{\hveg}{\ifmmode H_{\rm Vega} \else $H_{\rm Vega}$\fi}
\newcommand{\kveg}{\ifmmode K_{\rm Vega} \else $K_{\rm Vega}$\fi}

\def\arcsec{\hbox{$^{\prime\prime}$}}

\newcommand{  \Chisq    }{\ifmmode \chi^{2} \else $\chi^{2}$}
\newcommand{  \nelec    }{\ifmmode n_{e} \else $n_{e}$\fi}     % electron density
\newcommand{  \nh       }{\ifmmode n_{\rm H} \else $n_{\rm H}$\fi}     % hydrogen density
\newcommand{  \Ncol     }{\ifmmode N_{\rm col} \else $N_{\rm col}$\fi} % column density
\newcommand{  \NH       }{\ifmmode N_{\rm H} \else $N_{\rm H}$\fi}     % column density

\newcommand{\zfpe}{$z \simeq 4.8$}

\def\sun{\hbox{$\odot$}}

\def\arcsec{\hbox{$^{\prime\prime}$}}
\def\ion#1#2{#1$\;${\small\rm\@Roman{#2}}\relax}

\received{}
\revised{}
\accepted{}
\submitjournal{ApJ}

\shorttitle{Quasar hosts at $z \sim 4.8$}
\shortauthors{Nguyen et al.}

\usepackage{default}
\usepackage{amsmath}
\usepackage{multirow}
\usepackage{graphicx}
\usepackage{hyperref}
\usepackage{hhline}
\usepackage[all]{hypcap}
\usepackage{bold-extra}

\newcommand{\zrange}{\ifmmode z\simeq4.6-4.9 \else $z\simeq4.6-4.9$\fi}
\newcommand{\zpaper}{\ifmmode z\simeq4.8 \else $z\simeq4.8$\fi}
\newcommand{\tln}{$^{\rm T17}$}

\begin{document}

\title{ALMA Observations of Quasar Host Galaxies at $z\simeq4.8$}

\correspondingauthor{Nathen Nguyen}

\author[0000-0003-4134-6596]{Nathen H. Nguyen}
\affil{Departamento de Astronom\'{i}a, Universidad de Chile, Camino el Observatorio 1515, Las Condes, Santiago, Casilla 36-D, Chile.}
\email{nnguyen@das.uchile.cl}

\author{Paulina Lira}
\affil{Departamento de Astronom\'{i}a, Universidad de Chile, Camino el Observatorio 1515, Las Condes, Santiago, Casilla 36-D, Chile.}

\author[0000-0002-6766-0260]{Benny Trakhtenbrot, Hagai Netzer}
\affil{School of Physics and Astronomy, Tel Aviv University,
Tel Aviv 69978, Israel.}

\author{Claudia Cicone}
\affil{INAF - Osservatorio Astronomico di Brera, Via Brera 28, 20121 Milano, Italy}
\affil{Institute of Theoretical Astrophysics, University of Oslo, P.O. Box 1029 Blindern, 0315 Oslo, Norway}
  
\author{Roberto Maiolino}
\affil{Cavendish Laboratory, University of Cambridge, 19 J. J. Thomson Ave., Cambridge CB3 0HE, UK.}
\affil{Kavli Institute of Cosmology Cambridge, Madingley Road, Cambridge CB3 0HA, UK.}

\author[0000-0003-4327-1460]{Ohad Shemmer}
\affil{Department of Physics, University of North Texas, Denton, TX 76203,
USA.}

\begin{abstract}

We present ALMA band-7 data of the \CII\ emission line and underlying
far-infrared (FIR) continuum for twelve luminous quasars at \zfpe\,
powered by fast-growing supermassive black holes (SMBHs). Our total
sample consists of eighteen quasars, twelve of which are presented here
for the first time. The new sources consists of six \herschel/SPIRE
detected systems, which we define as "FIR-bright" sources, and six
\herschel/SPIRE undetected systems, which we define as "FIR-faint"
sources.  We determine dust masses for the quasars hosts of $M_{dust}
\le 0.2-25.0\times 10^8 \Msol$, implying ISM gas masses comparable to
the dynamical masses derived from the \cii\ kinematics. It is found
that on average the \mgii\ line is blueshifted by $\sim 500\,\kms$
with respect to the \cii\ emission line, which is also observed when
complementing our observations with data from the literature. We find
that all of our "FIR-bright" subsample and most of the "FIR-faint"
objects lie above the main sequence of star forming galaxies at $z
\sim 5$. We detect companion sub-millimeter galaxies (SMGs) for two
sources, both FIR-faint, with a range of projected distances of
$\sim20-60$ \kpc\ and with typical velocity shifts of $\left|\Delta
v\right|\ltsim200\,\kms$ from the quasar hosts. Of our total sample of
eighteen quasars, 5/18 are found to have dust obscured starforming
companions.
\end{abstract}


\section{Introduction}
\label{sec:intro}

Most galaxies are believed to host a Super Massive Black Hole (SMBH)
at their center \citep{Kormendy2013}. Both Active Galactic Nuclei
(AGN) and star formation (SF) luminosity functions are found to peak
at $z \sim 2$, declining towards lower redshifts
\citep{Aird2015,Fiore2017}. Hence, a coordinated growth of SMBHs and
the stellar mass of their hosts has been proposed. These SMBHs can
grow through accretion during an AGN phase \citep{Salpeter1964}, while
the growth of their stellar mass can be measured through their SF.  It
is commonly believed that accretion onto SMBHs and intense starburst
activity occur nearly simultaneously, with both processes pulling from
a shared reservoir of cold gas. These reservoirs of cold gas are
commonly proposed to be fed by major mergers
\citep{Dimatteo2005,Hopkins2006,Somerville2008}.

Testing these scenarios observationally has proven to be extremely
challenging since it requires to characterize accreting SMBHs
and their hosts for well defined samples. The AGN-related emission
dominates over most of the optical-NIR spectral regime, significantly
limiting the prospects of determining the host properties. The best
strategy is to observe these systems in the far-IR (FIR), where dust
heated by the starformation dominates the continuum emission and
interstellar emission lines allow us to determine the host
kinematics. For high-$z$ sources, this can be readily achieved through
Atacama Large Millimeter Array (ALMA) sub-mm observations.

Following the work in our pilot sample of \citet[][T17
  hereafter]{Trakhtenbrot2017} we continue to probe the connection
between SMBHs and their host galaxies using an optically selected,
flux-limited sample of the most luminous quasars at $z \sim
4.8$. These fast-growing SMBHs should also be experiencing fast
stellar growth, as seen in high-\lagn\ systems studied at $z\sim 1-3$
\citep{Netzer2007,Rosario2012,Lutz2014}.

Throughout this work we assume a cosmological model with
$\Omega_{\Lambda}=0.7$, $\Omega_{\rm M}=0.3$, and
$H_{0}=70\,\kms\,{\rm Mpc}^{-1}$, which provides an angular scale of
about $6.47\,\kpc/\arcsec$ at $z=4.8$, the typical redshift of our
sources. In Section \ref{sec:data} we describe our data sample,
observations, and methods of data reduction and analysis. In Section
\ref{sec:results} we present results on the host galaxy properties of
our sample, and compare the occurrence of companions to other ALMA
samples. Finally, in Section \ref{sec:summary} we summarize the
results and findings of our work. We further assume the stellar
initial mass function (IMF) of \cite{Chabrier2003}.

\section{Sample, ALMA Observations, and Data Analysis} 
\label{sec:data}

\subsection{Previous Observations and Sample Properties}

Our original sample is a selection of the 38 brightest (\lbol
$\,\~3-23 \times 10^{46}$ \ergs ) unobscured quasars from the sixth
data release of the Sloan Digital Sky Survey
\cite[SDSS/DR6;][]{York2000_SDSS,Adelman2008} at redshifts z $\~4.65 -
4.92$. This redshift range, which we will often refer to as \zfpe, was
selected to allow follow up observations of the \MgII\ emission line
and nearby 3000 \AA\ continuum luminosity. Observations of \mgii\ were
carried out using VLT/SINFONI and Gemini-North/NIRI and presented in
T11, which provided estimates of the SMBH masses ($M_{BH}$) and
accretion rates of the quasars (\lledd). These results indicated that
the sample, on average, has higher accretion rates ($\lledd \~ 0.6$)
and lower masses ($\~8.4 \times 10^8 \Msol $) than AGN observed at
lower redshifts.

Further observations were carried out with the \herschel\ Spectral and
Photometric Imaging Receiver (SPIRE) \citep[M12 and N14
  henceforth]{Mor_all2012,Netzer2014}, and relied on data from the
\textit{Spitzer} Infrared Array Camera (IRAC) (also from N14) 3.6 and
4.5 \micron\ bands for positional priors for the
\herschel\ photometry. While the majority of sources were detected
using Spitzer, only nine source were detected in all three SPIRE
bands. We define these \herschel/SPIRE detections as "FIR-bright"
sources, having on average $L_{FIR} \~ 8.5 \times 10^{46}\, \ergs$ ($2.2
\times 10^{13} \Lsol$). By using the standard conversion factor based
on the IMF of \cite{Chabrier2003} we calculated star formation rates
as $SFR / \mpyr = L_{FIR}\ / 10^{10} \Lsol$, giving SFRs $\~ 1000 -
4000$ \mpyr\ for our nine FIR-bright sources. To determine the SFRs of
the \herschel\ non-detected sources, which we refer to as "FIR-faint"
sources, stacking analysis was carried out in \cite{Netzer2014} and
gave a median SFR of $\~ 400$ \mpyr. The work of N14 and M12 indicate
that there is a wide variation of SFRs in our sample, while we see in
T11 that the variation of SMBH and AGN properties are more uniform
across the sample.

The goal of the \herschel/SPIRE campaign was to determine the peak of
the SF heated dust continuum (M12, N14), and if possible, to observe
evidence for merger activity. However, the size of the field of view
and the spatial resolution of the data ($\sim18\arcsec$, or
$\gtrsim100\,\kpc$ at \zfpe) was insufficient to determine the
presence of close nearby systems.

\subsection{ALMA Observations}
\label{sec:alma_obs}

\capstartfalse
\begin{deluxetable*}{lccccccl}
\tablecolumns{8}
\tablewidth{0pt}
\tablecaption{Observations Log \label{tab:obs_log}}
\tablehead{
\colhead{sub-sample} &
\colhead{Target ID}  &
\colhead{ $N_{\rm Ant}\dag$} &
\colhead{$T_{\rm exp}$} &
\colhead{$F_\nu$ rms y} &
\colhead{Beam Size} &
\colhead{Pixel Size} & 
\colhead{ALMA Companions} \\
       &         &       & [sec]  & [mJy/beam] & [\arcsec]       & [\arcsec] & 
}
\startdata
Bright & SDSS~J080715.11$+$132805.1 & 43 & 2054 & 5.1 $\times 10^{-2}$ & 0.37 $\times$ 0.21 & 0.06& ...\\
       & SDSS~J140404.63$+$031403.9 & 42 & 1184 & 6.2 $\times 10^{-2}$ & 0.36 $\times$ 0.29 & 0.06&   ...\\
       & SDSS~J143352.21$+$022713.9 & 40 & 1001 & 5.1 $\times 10^{-2}$ & 0.37 $\times$ 0.32 & 0.06&  ...\\ 
       & SDSS~J161622.10$+$050127.7 & 43 & 1690 & 3.6 $\times 10^{-2}$ & 0.23 $\times$ 0.19 & 0.06&  ...\\
       & SDSS~J165436.85$+$222733.7 & 42 & 1305 & 5.5 $\times 10^{-2}$ & 0.27 $\times$ 0.21 & 0.06&  ...\\
       & SDSS~J222509.19$-$001406.9 & 40 & 1486 & 5.4 $\times 10^{-2}$ & 0.29 $\times$ 0.23 & 0.06&   ...\\
\hline \\ [-1.75ex]
Faint  & SDSS~J101759.63$+$032739.9 & 41 & 2064 & 2.8 $\times 10^{-2}$ & 0.36 $\times$ 0.24 & 0.06&  ...\\
       & SDSS~J115158.25$+$030341.7 & 42 & 1851 & 5.1 $\times 10^{-2}$ & 0.33 $\times$ 0.28 & 0.06&  ...\\
       & SDSS~J132110.81$+$003821.7 & 40 & 2276 & 2.8 $\times 10^{-2}$ & 0.33 $\times$ 0.30 & 0.06&  ...\\
       & SDSS~J144734.09$+$102513.1 & 39 & 1871 & 5.1 $\times 10^{-2}$ & 0.54 $\times$ 0.31 & 0.06 & SMG (w/ \cii)\\
       & SDSS~J205724.14$-$003018.7 & 39 & 1550 & 4.4 $\times 10^{-2}$ & 0.28 $\times$ 0.21 & 0.06 & SMG (w/ \cii), ``B'' (w/o \cii)\\
       & SDSS~J224453.06$+$134631.6 & 40 & 1881 & 3.4 $\times 10^{-2}$ & 0.32 $\times$ 0.29 & 0.06&  ...\\
\enddata
\tablenotetext{\dag}{Number of antennae used, averaging after antennae flagging.}
\end{deluxetable*}
\capstarttrue

Previously, in T17, we observed three FIR-bright, and three FIR-faint
quasar from our sample using ALMA. In this paper we present an
additional six FIR-bright and six FIR-faint quasars observed with
ALMA. Thus, all of our original FIR-bright objects and 9/29 of our
FIR-faint objects have been observed.

The twelve new targets were observed using ALMA band-7 during the
Cycle-4 period of 2016 November 9 to 2017 May 6. Our main goal is to
detect and resolve the [C II] emission line, which is expected to have
a width of several hundred \kms\ (T17), as well as line-free dust
emission continuum.

For consistency we aimed to have the same spectral and spatial
resolutions as T17. The observations were done with the C40-5
configuration, and the exposure time ranged from 1001 - 2276 seconds,
with an observed angular resolution variation of 0.19-0.33\arcsec\ and
a central frequency range of 317 - 349 GHZ. The observed angular
resolution corresponds to $\sim2 \,\ \kpc$ at $z \simeq 4.8$. We chose
the TDM correlator mode which provides four spectral windows, each
covering an effective bandwidth of 1875 MHz, which corresponds to
$\sim1650\,\kms$ at the observed frequencies. This spectral range is
sampled by 128 channels with a frequency of 15.625 MHz or $\sim
15\,\kms$ per channel. The default spectral resolution of ALMA is
given as roughly twice the size of the channels, i.e. $\sim 30\,\kms$.
Two such spectral windows were centered on the frequency corresponding
to the expected peak of the \cii\ line, estimated from the \mgii-based
redshifts of our targets (as determined in T11). Because of the
specific redshifts of the sources, the spectral windows were found to
be more affected by poor atmosphere transmission than those used
during the observations of the six objects presented in T17, resulting
in noisier \cii\ data. The other two adjacent windows were placed at
higher frequencies and separated from the first pair by about 12
GHz. Each of these pairs of spectral windows overlapped by roughly 50
MHz. However, the rejection of a few channels at the edge of the
windows due to divergent flux values (a common flagging procedure in
ALMA data reduction), leads to a small spectral gap between pairs of
windows. This presents some issues for certain targets (Section
\ref{sec:reduction}). Given this spectral setup of four bands, the
ALMA observations could in principle probe \cii\ line emission over a
spectral region corresponding to roughly $\sim 3000\, \kms$ ($\Delta z
\simeq 0.06$). \autoref{tab:obs_log} is an observation log with
additional details of the ALMA observations. We will use abbreviated
object names (i.e., ``JHHMM'') in the rest of this paper.

\begin{figure*}
\center
\includegraphics[scale=0.35]{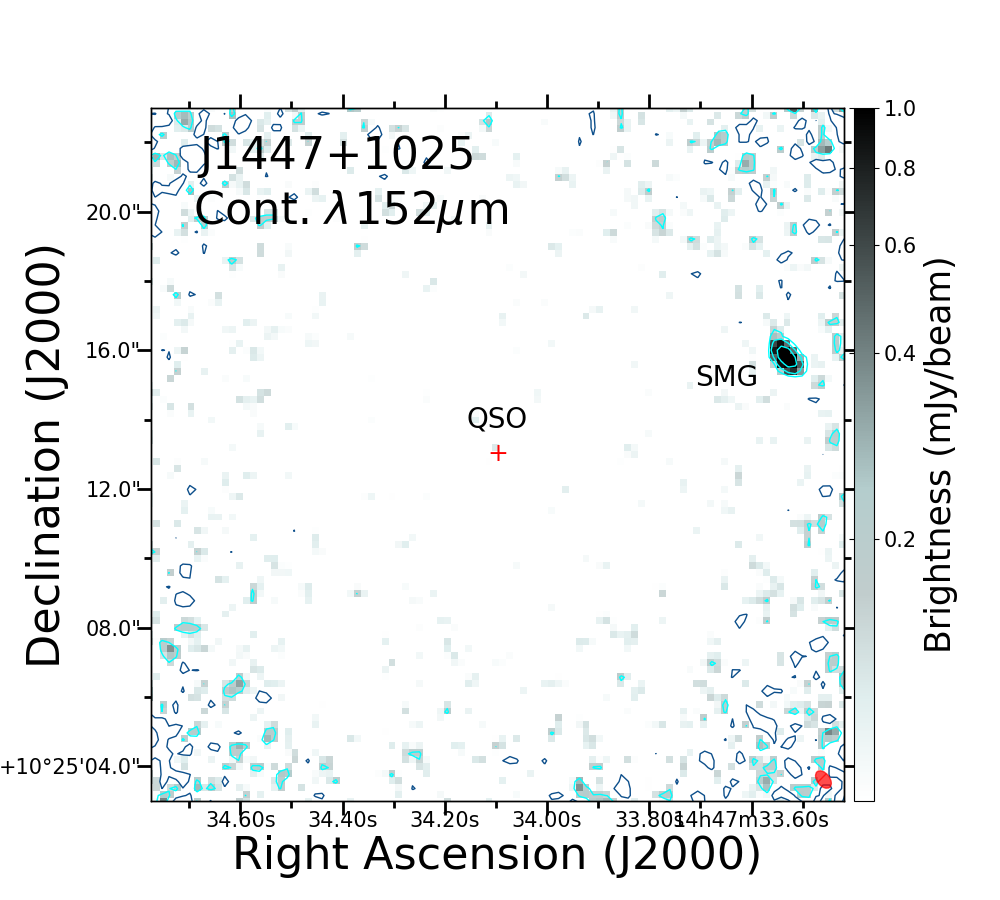} %
\includegraphics[scale=0.35]{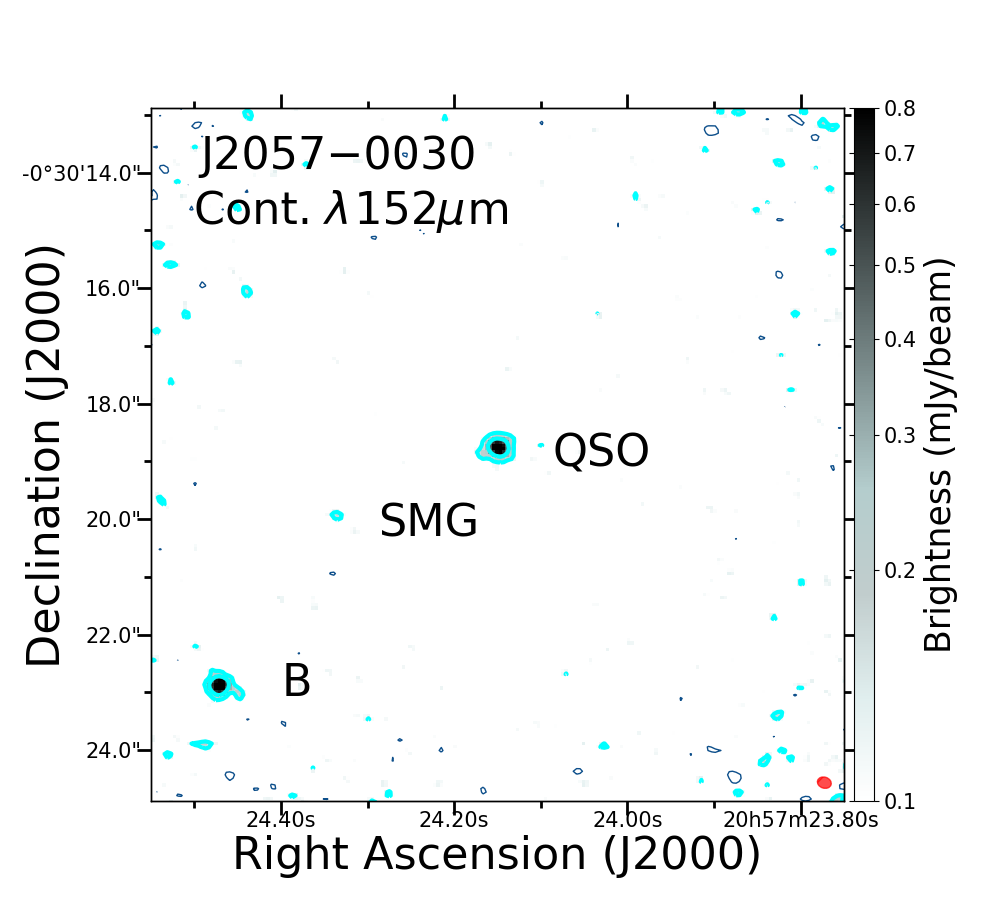} \\
\includegraphics[scale=0.35]{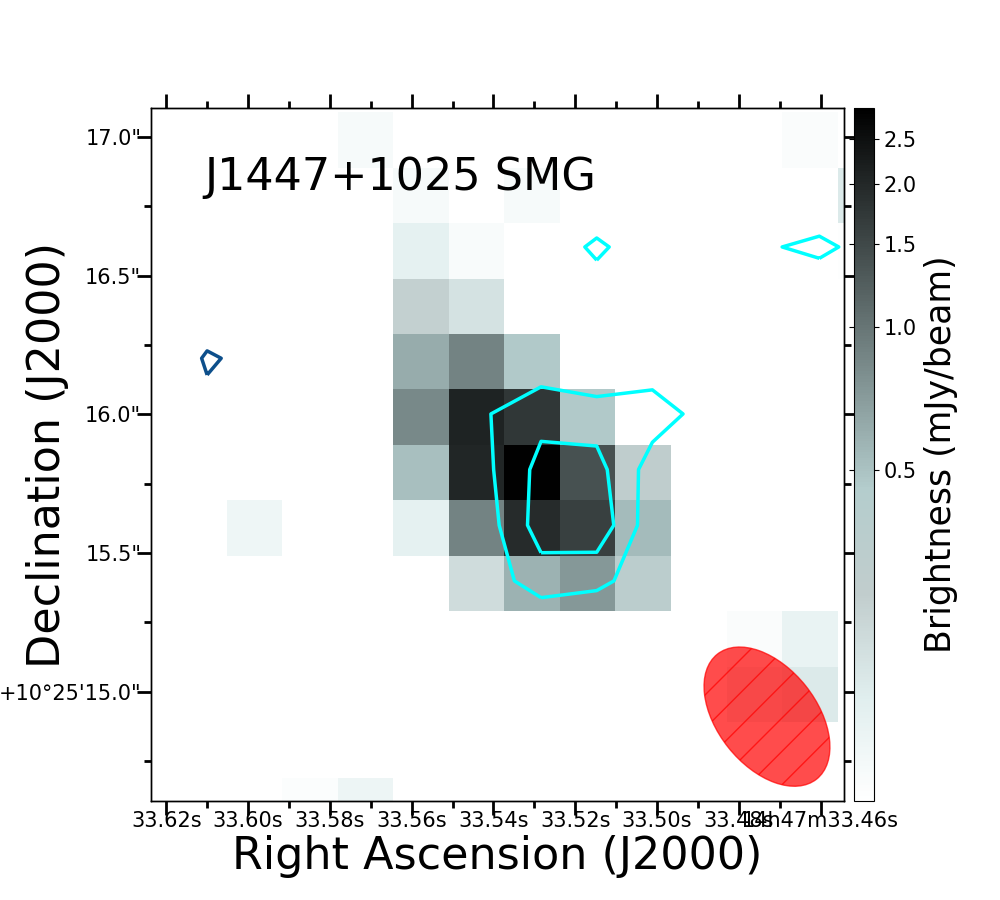}%
\includegraphics[scale=0.35]{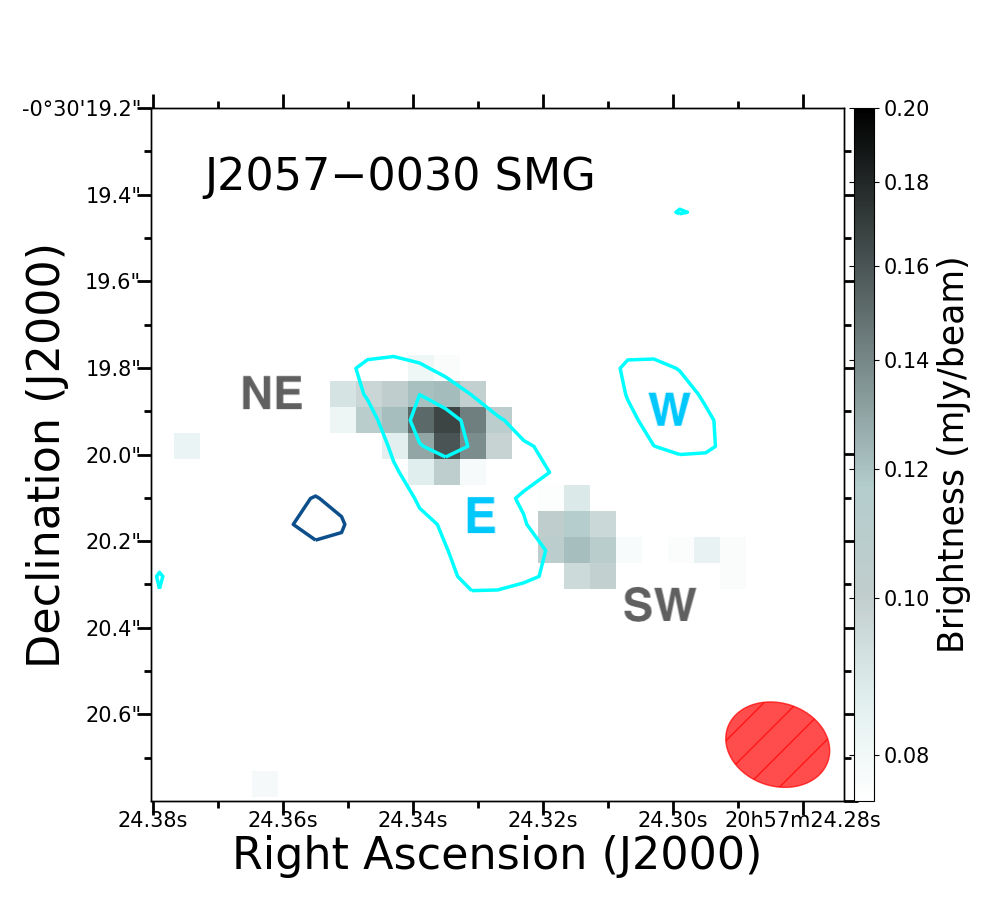}\\
\caption{Top: Large-scale continuum images for the two FIR-faint quasars in
  our sample where companions have been found: J2057 and J1447. Note
  that J1447 is {\em not\/} detected in dust continuum. The gray-scale
  maps show the continuum emission determined from the line-free ALMA
  spectral windows. Cyan and blue contours trace emission levels at
  different positive and negative significance levels, respectively,
  with the first contour tracing the region where the continuum
  emission exceeds 2$\sigma$, and consecutive contours plotted in
  steps of $2\sigma$. The ALMA beams are shown as red ellipses near
  the bottom-right of each panel. Physical companions, i.e., sources
  that have clear \cii\ detections with redshifts consistent with
  those of the quasars, are marked as ``SMG''. The continuum source
  accompanying J2057 that lacks significant \cii\ emission is marked
  as ``B''. 
   Bottom: Small-scale continuum and \cii\ line emission maps derived
  for the SMGs accompanying J1447 and J2057. For each source, the
  gray-scale map traces the continuum emission, while the contours
  trace the \cii\ line emission (i.e., surface brightness) at
  significance levels of 3 and 6$-\sigma$. For each
  source, the line fluxes used for the contours were extracted from a
  spectral window spanning $\pm500\,\kms$ around the \cii\ line
  peak. The ALMA beams are shown as red ellipses near the bottom-right
  of each panel. The two J2057 components observed in \cii\ emission
  are labeled E and W, while the two components seen in continuum are
  labeled NE and SW.}
\label{fig:smgs}
\label{fig:withcompanion}

\end{figure*}

\subsection{Data Reduction}
\label{sec:reduction}

Data reduction was performed using the CASA package version 4.7.2
\citep{McMullin2007}. \texttt{CLEAN} algorithms were ran with "briggs"
weighting and a robustness parameter of 0.5 in order to create
continuum and emission line images. Continuum emission images were
constructed using the line-free spectral window pair, while the
\texttt{UVCONTSUB} command was used to subtract continuum emission
from the \cii\ window pair, resulting in continuum-subtracted cubes.
Observed flux densities and beam deconvolved continuum source sizes
are presented in Table \ref{tab:spec_measurements}.

Sizes of the continuum emitting regions were determined from the
respective images by fitting spatial 2D Gaussians to the sources,
which are characterized by a peak flux, semi-major and semi-minor
axes, and a position angle. The fluxes were measured by integrating
over these spatial 2D Gaussians. The SMG companion to J2057 (see
Section \ref{sec:sources}), however, seems to be composed of two
separate sources which were not properly fitted by the CASA 2D
Gaussian routine. Instead, sizes were obtained directly from the
continuum images using an azimuthally averaged Gaussian fit. Since
these values are not beam-corrected, they are quoted as upper limits
in Table \ref{tab:spec_measurements}.

Various \texttt{IMMOMENTS} commands gave the velocity fields and
velocity dispersion maps (first and second moment, respectively) from
the \cii\ continuum subtracted cubes. To measure the properties of the
$\cii$ emission lines, we used both a "spatial" and "spectral" method.
In the "spatial" approach, we created zero-moment images (i.e.,
integrated over the spectral axis) for all sources and fitted the
spatial distribution of line emission with 2D Gaussian profiles. Line
fluxes were obtained as described before for the continuum flux
determinations.

\begin{longrotatetable}
\begin{table}[h]
\caption{Spectral Measurements}
\label{tab:spec_measurements}
\begin{tabular}{lllccccccccccc}
\hline \hline
Subsample     & \multicolumn{2}{c}{Target}  & Cont. Flux & $\nu$ & Cont. Size & $F_{\cii}$ & FWHM$_{\cii}$ & $\nu_{0,\cii}$ & \cii\ size & $L_{\cii}$ & $\Delta d$ & $\Delta v$ \\
\cline{2-3}
& ID  &  comp. & [mJy] &  [GHz]    &    [\arcsec]   & [Jy \kms]  & [\kms] &  [GHz]    &   "   & [$10^{9}\,\Lsun$]     & [kpc] & [\kms]  \\
\hline
Bright & J0807 & QSO                      &  6.80 $\pm$ 0.20 & 334.87 & 0.23 $\times$ 0.19 & 5.8  $\pm$ 1.40 & 398.6 $\pm$ 19.2 & 323.27 $\pm$ 0.010 & 0.52 $\times$ 0.14 & 4.01 & . . . & . . .\\
       & J1404 & QSO                      & 11.31 $\pm$ 0.27 & 333.86 & 0.28 $\times$ 0.25 & 5.81 $\pm$ 0.71 & 483.3 $\pm$ 21.3 & 320.86 $\pm$ 0.009 & 0.52 $\times$ 0.43 & 4.08 & . . . & . . .\\
       & J1433 & QSO                      &  7.61 $\pm$ 0.33 & 334.05 & 0.32 $\times$ 0.26 & 4.79 $\pm$ 0.38 & 397.0 $\pm$ 13.7 & 331.78 $\pm$ 0.006 & 0.43 $\times$ 0.37 & 3.17 & . . . & . . .\\
       & J1616 & QSO                      &  6.29 $\pm$ 0.28 & 335.74 & 0.23 $\times$ 0.16 & 10.1 $\pm$ 1.50 & 469.5 $\pm$ 24.1 & 322.99 $\pm$ 0.011 & 0.60 $\times$ 0.36 & 7.00 & . . . & . . .\\
       & J1654 & QSO                      &  4.73 $\pm$ 0.10 & 344.53 & 0.10 $\times$ 0.08 & 2.07 $\pm$ 0.46 & 543.0 $\pm$ 34.9 & 331.81 $\pm$ 0.016 & 0.31 $\times$ 0.08 & 1.36 & . . . & . . .\\
       & J2225 & QSO                      & 13.13 $\pm$ 0.21 & 334.61 & 0.22 $\times$ 0.17 & 8.05 $\pm$ 0.73 & 445.5 $\pm$ 22.4 & 322.50 $\pm$ 0.010 & 0.44 $\times$ 0.29 & 5.60 & . . . & . . .\\ 
\hline                                    
Faint  & J1017 & QSO                      &  1.36 $\pm$ 0.10 & 331.76 & 0.23 $\times$ 0.20 & 1.93 $\pm$ 0.27 & 223.8 $\pm$  8.3 & 319.49 $\pm$ 0.004 & 0.32 $\times$ 0.30 & 1.37 & . . . & . . .\\ 
       & J1151\tablenotemark{a} & QSO     &  $<0.81$         & 346.13 & . . .              & $<0.31$         & . . .            & . . .              & . . .              & . . .& . . . & . . .\\ 
       & J1321 & QSO                      &  1.56 $\pm$ 0.07 & 343.73 & 0.29 $\times$ 0.22 & 1.72 $\pm$ 0.21 & 480.7 $\pm$ 26.4 & 322.12 $\pm$ 0.012 & 0.46 $\times$ 0.27 & 1.13 & . . . & . . .\\
       & J1447 & QSO\tablenotemark{b}     & $<0.12$          & 346.61 & . . .              & 0.14 $\pm$ 0.09 & 293.2 $\pm$ 113.6& 334.50 $\pm$ 0.027 & 0.30 $\times$ 0.28 & 0.09 & . . . & . . .\\
       & J1447 & SMG\tablenotemark{c}     &  3.86 $\pm$ 0.17 & 346.61 & 0.40 $\times$ 0.15 & 0.88 $\pm$ 0.27 & 215 $\pm$ 22 & 334.37 $\pm$ 0.008 &  $< 0.3$           & 0.57 & 59    &  206.5\\
       & J1447 & SMG\tablenotemark{c}     &                  &        &                    & 0.54 $\pm$ 0.16 & 199 $\pm$ 33 & 333.72 $\pm$ 0.008 &  $< 0.3$           & 0.35 & 59    &  701.2\\
       & J2057 & QSO                      &  2.03 $\pm$ 0.14 & 346.52 & 0.24 $\times$ 0.21 & 2.51 $\pm$ 0.31 & 331.4 $\pm$ 20.5 & 334.44 $\pm$ 0.009 & 0.40 $\times$ 0.18 & 1.63 & . . . & . . .\\
       & J2057 &SMG\tablenotemark{d}\ NE,E  &  0.28 $\pm$ 0.04 & 346.52 & $< 0.3$            & 0.63 $\pm$ 0.11 & 475.4 $\pm$ 84 & 334.62 $\pm$ 0.009 &  $< 0.3$           & 0.41 & 20    & -161.4\\
       & J2057 &SMG\tablenotemark{d}\ SW,W  &  0.17 $\pm$ 0.06 & 346.52 & $< 0.3$            & 0.37 $\pm$ 0.07 & 336.3 $\pm$ 68 & 334.32 $\pm$ 0.009 & 0.57 $\times$ 0.14 & 0.24 & 20    &  107.7\\
       & J2244 & QSO                      &  3.34 $\pm$ 0.09 & 346.95 & 0.20 $\times$ 0.19 & 3.86 $\pm$ 0.29 & 283.1 $\pm$  7.4 & 335.71 $\pm$ 0.003 & 0.40 $\times$ 0.30 & 2.49 & . . . & . . .\\
\hline
\end{tabular}
\tablenotetext{a}{3-$\sigma$ upper limit of the calculated RMS at the expected position of the source.}
\tablenotetext{b}{Line fluxes were determined by aperture photometry at the position of the source.}
\tablenotetext{c}{Two Gaussian profiles were fitted to the \cii\ line spectra. Source sizes have upper limits only.}
\tablenotetext{d}{Two components are seen in continuum (NE and SW) and \cii\ (E and W). Most source sizes have upper limits only.}
\end{table}
\end{longrotatetable} 
In the "spectral" approach, we extracted 1D spectra
from the $\cii\ $ continuum subtracted cubes. A Gaussian profile was
fitted to the emission line profiles, from which we obtained the
integrated line flux.

We found that the two different methods described above to be in good
agreement, with a median difference of 0.05 dex. As stated in T17 the
"spatial" approach is less sensitive to the low Signal-to-Noise (S/N)
outer regions of the sources and the low S/N "wings" of the line
profiles, thus we adopt this method for our own analysis. J2057,
however, has a spectral gap (as described in \ref{sec:alma_obs}) lying
in the center of the \cii\ line. This proved difficult for the
"spatial" method as no interpolation of the missing line flux was
possible. Hence, the line flux reported in Table
\ref{tab:spec_measurements} was obtained with the "spectral" approach.
Also, both SMG companions to J1447 and J2057 (see next Section) show
separate dynamical components. In the case of J1447 the "spectral"
approach was used to determine their properties. The J2057 SMG also
breaks into two components in continuum emission, which are not
clearly related to the \cii\ emission. Both components are
characterized in Table \ref{tab:spec_measurements}.

\subsection{Source Detections}
\label{sec:sources}

Ten of our twelve new quasars are clearly detected in both continuum
and $\cii$ emission with 6-12$\sigma$ significance. While J1447 is
only detected at a 3$\sigma$ level in $\cii$ line, and J1151 is not
detected at all. J1447 and J1151 are both FIR-faint sources. Because
J1447 has a very weak signal, it was not possible to fit a Gaussian to
the spatial distribution of its line and continuum emission. Instead,
`aperture' photometry was carried out with an area corresponding to
roughly the beam size. The \cii\ emission of J1447 was found to have a
S/N \~ 3.6, while there was a non-significant signal in the
continuum. The continuum values listed in Table
\ref{tab:spec_measurements} for J1447 and J1151 correspond to 3 times
the average RMS noise about the expected quasar positions.

Two FIR-faint quasars show the presence of companions detected in both
continuum and \cii\ emission with a significance of 6-9
$\sigma$. Continuum maps for these two sources are presented in Figure
\ref{fig:withcompanion}. A continuum-only source is found separated
from J2057 by $6\farcs3$ in the SE direction, which corresponds 41.8
\kpc\ at the redshift of the quasar and is marked with a 'B' in Figure
\ref{fig:withcompanion}. We can put a lower limit of $\sim \pm 1500$
\kms\ to the velocity shift of any \cii\ emission from this source and
the \cii\ emission from the quasar host. Because of the separation and
lack of a $\cii$ line detection, we conclude that this continuum
source is most likely a source only seen in projection. A similar
continuum-only source is found in T17, which is concluded to be a
background/foreground projection. Information about companions can be
found in Tables \ref{tab:obs_log} and \ref{tab:spec_measurements}.

For all our quasars detected in both, continuum and \cii, the two
emissions follow each other well. The exceptions are the two detected
SMGs. Their detailed continuum and \cii\ maps are presented in Figure
\ref{fig:smgs}. In the case of J1447, the continuum emission seems
more extended towards the north than the \cii\ emission, although
weaker, redshifted \cii\ emission appears towards the north in
dynamical maps (see next Section). The SMG to J2057 has secondary
peaks in \cii\ and continuum emission. These are labeled as E, W and
NE, SW in Figure \ref{fig:smgs}, respectively. We will see in the next
Section that there is strong indication of gravitational perturbations
in these two SMG sources.

\begin{figure*}
\center
\renewcommand{\tabcolsep}{1pt}
\begin{tabular}{c c c}
&FIR-Bright Objs.& \\ 
\includegraphics[width=6cm,height=4cm]{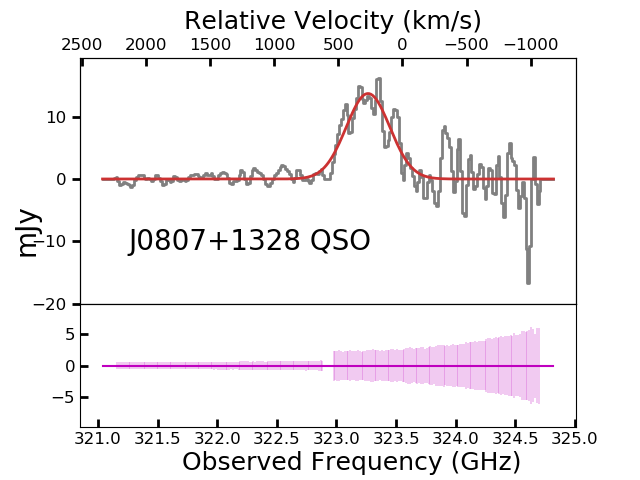} & \includegraphics[width=6cm,height=4cm]{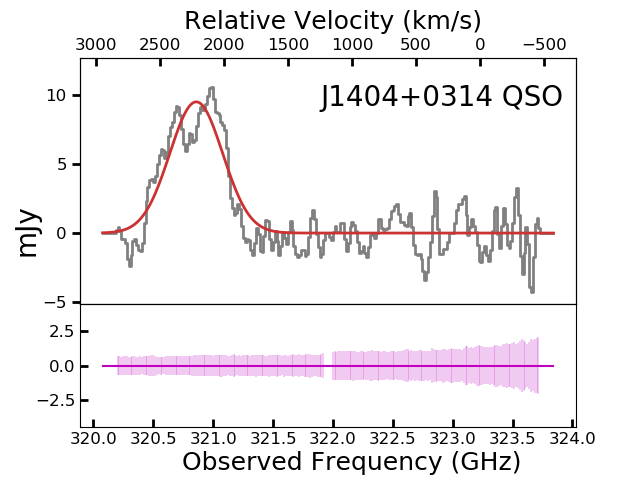} & \includegraphics[width=6cm,height=4cm]{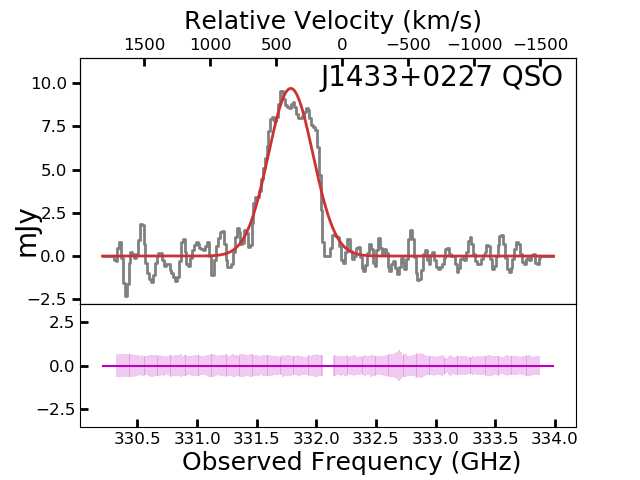} \\
\includegraphics[width=6cm,height=4cm]{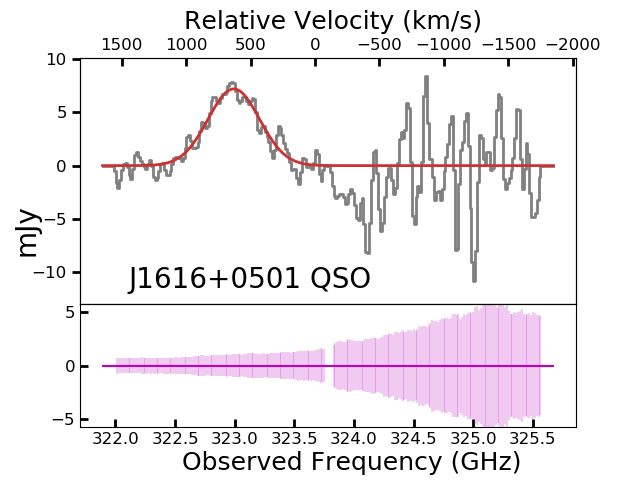} & \includegraphics[width=6cm,height=4cm]{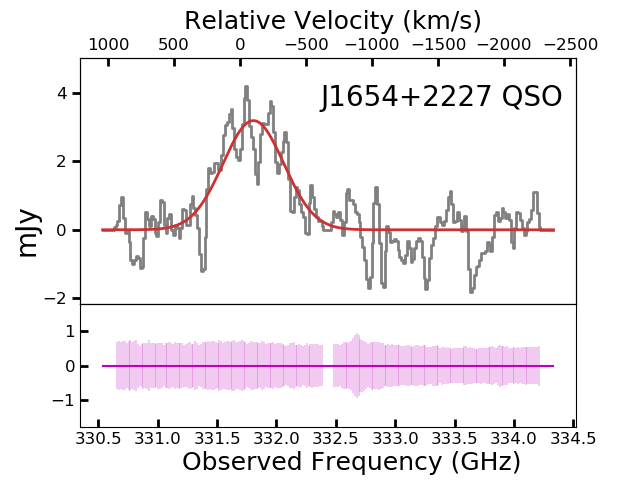} & \includegraphics[width=6cm,height=4cm]{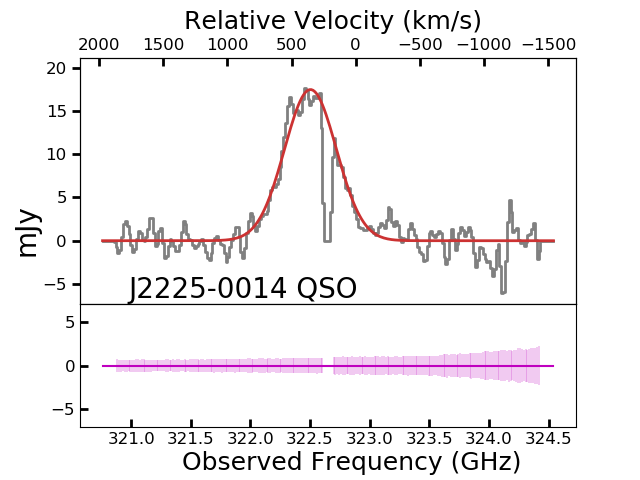} \\
&FIR-Faint Objs.& \\ 
\includegraphics[width=6cm,height=4cm]{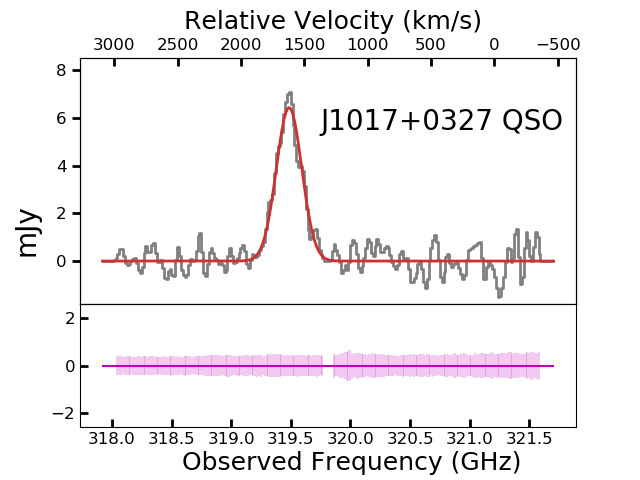} & \includegraphics[width=6cm,height=4cm]{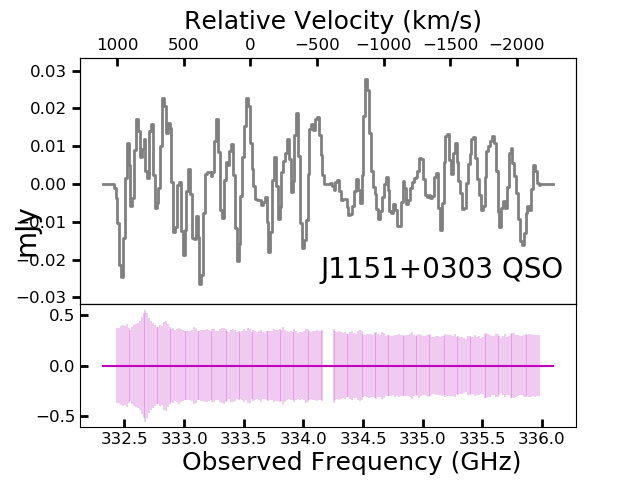} & \includegraphics[width=6cm,height=4cm]{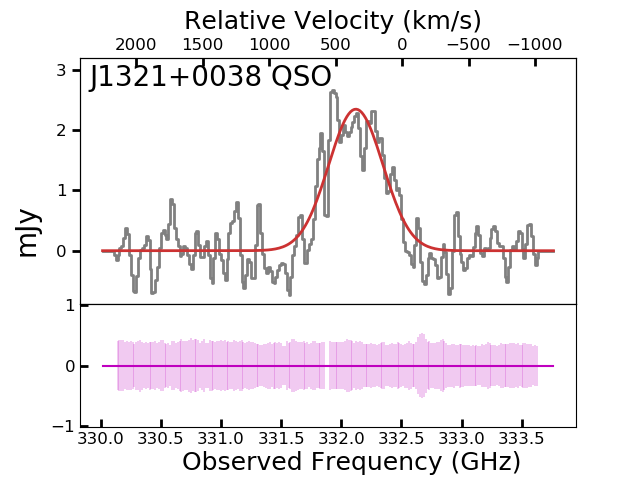} \\
\includegraphics[width=6cm,height=4cm]{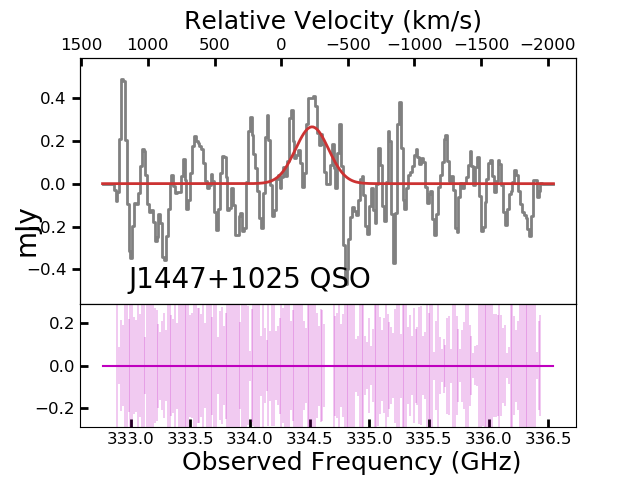} & \includegraphics[width=6cm,height=4cm]{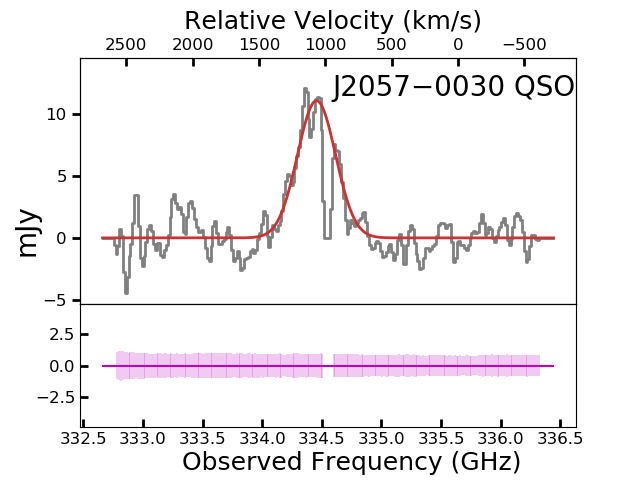} & \includegraphics[width=6cm,height=4cm]{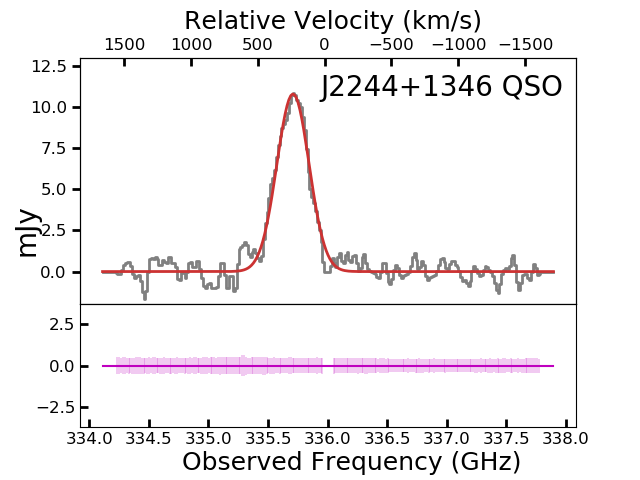}  \\
&SMGs& \\
\multicolumn{3}{c}{\includegraphics[width=5cm,height=4cm]{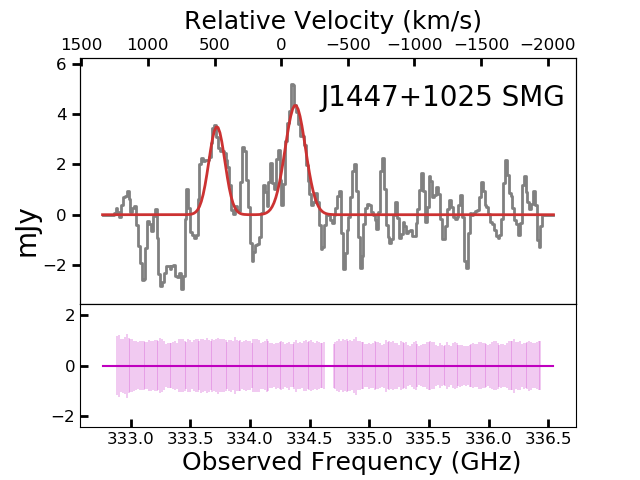} \includegraphics[width=5cm,height=4cm]{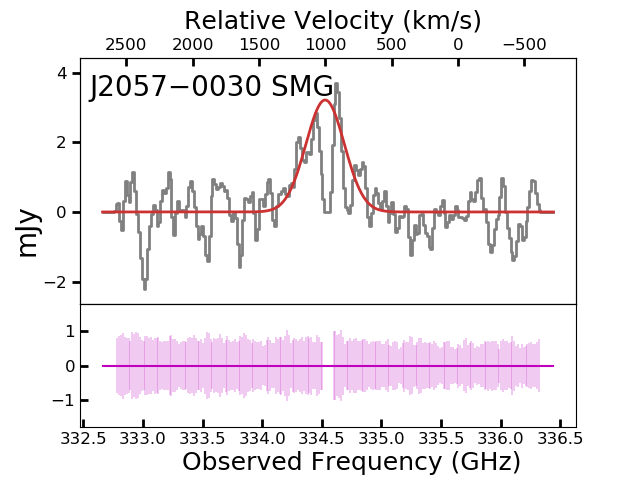} } \\
\end{tabular}
\caption{Spectra of the \CII\ emission line for all the new ALMA
  observations reported in this work. FIR-bright sources are presented in the top two rows, FIR-faint sources in
  the middle two rows, and accompanying SMGs are presented in the bottom
  row. For each spectrum the upper x-axis denote the velocity
  offsets with respect to the redshift derived from the \mgii\ broad
  emission lines (T11). Red lines show the Gaussian fits to the line
  profiles. RMS spectra are also included in the same scale as the
  flux spectrum except for J1151 where there is no \cii\ detection.}
\label{fig:spectra}
\end{figure*}

\begin{figure*}
\center
FIR-Bright Objs.\\ 
\includegraphics[scale=0.2]{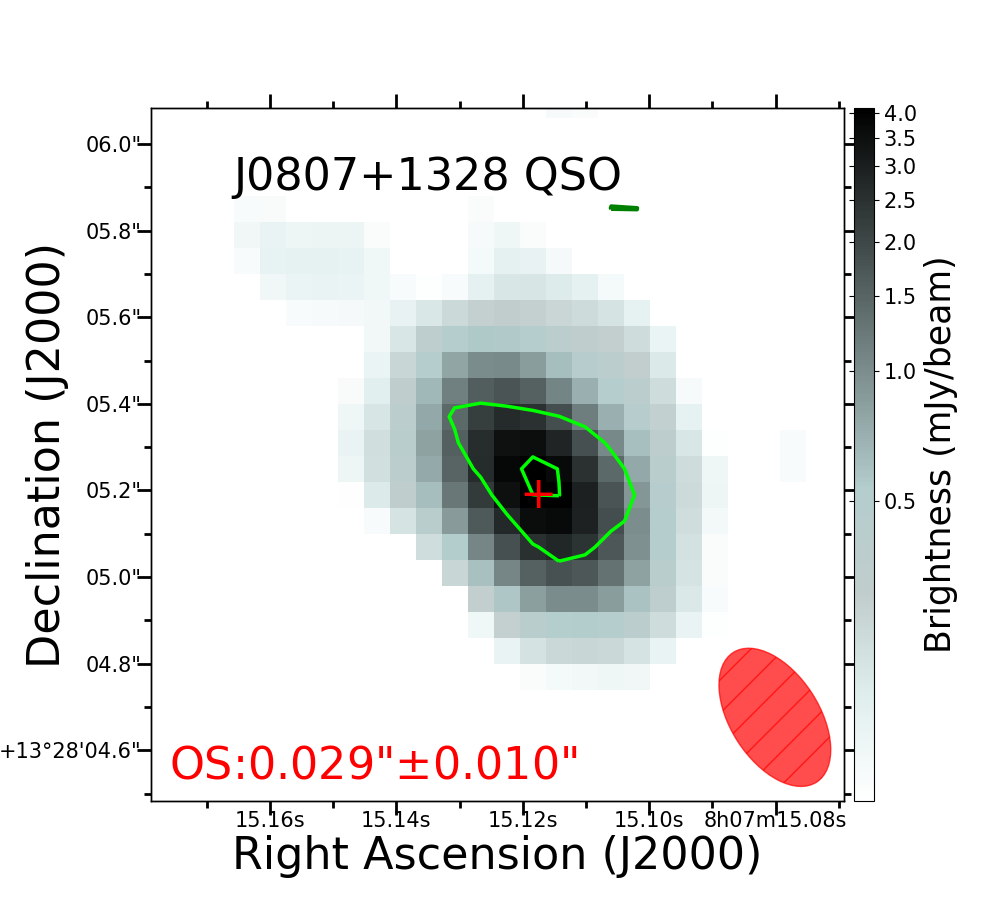} 
\includegraphics[scale=0.2]{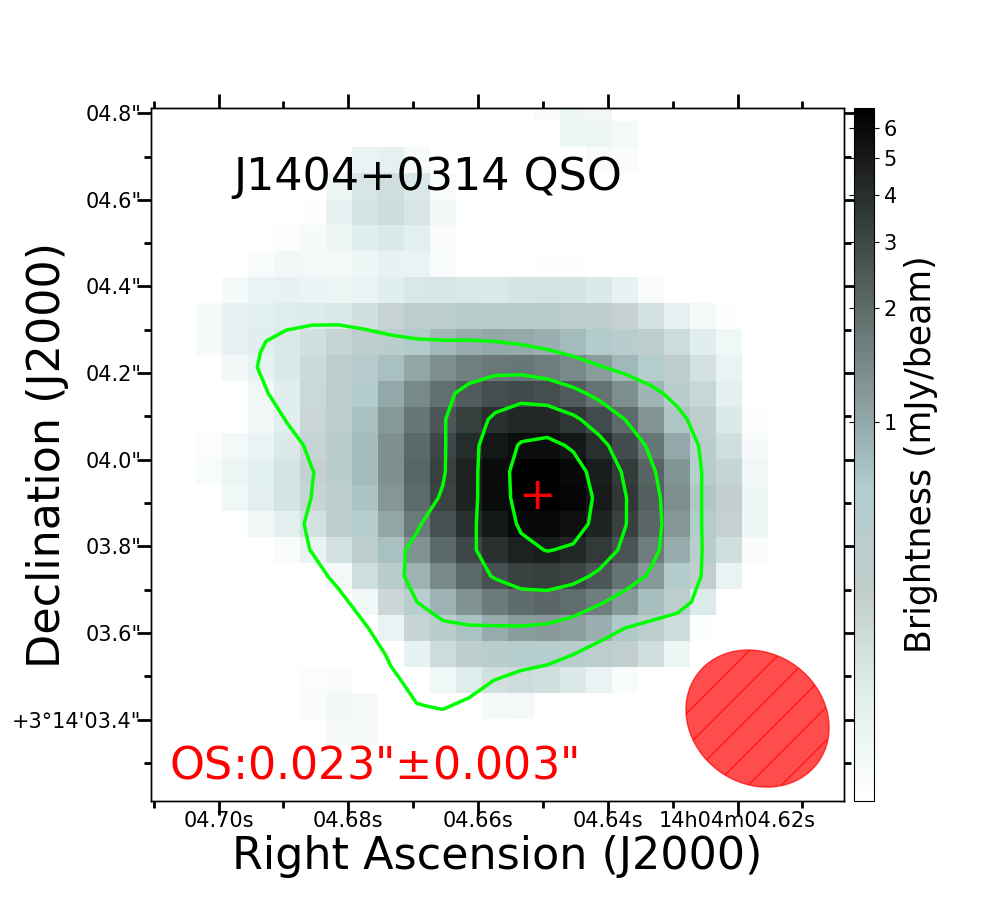}  
\includegraphics[scale=0.2]{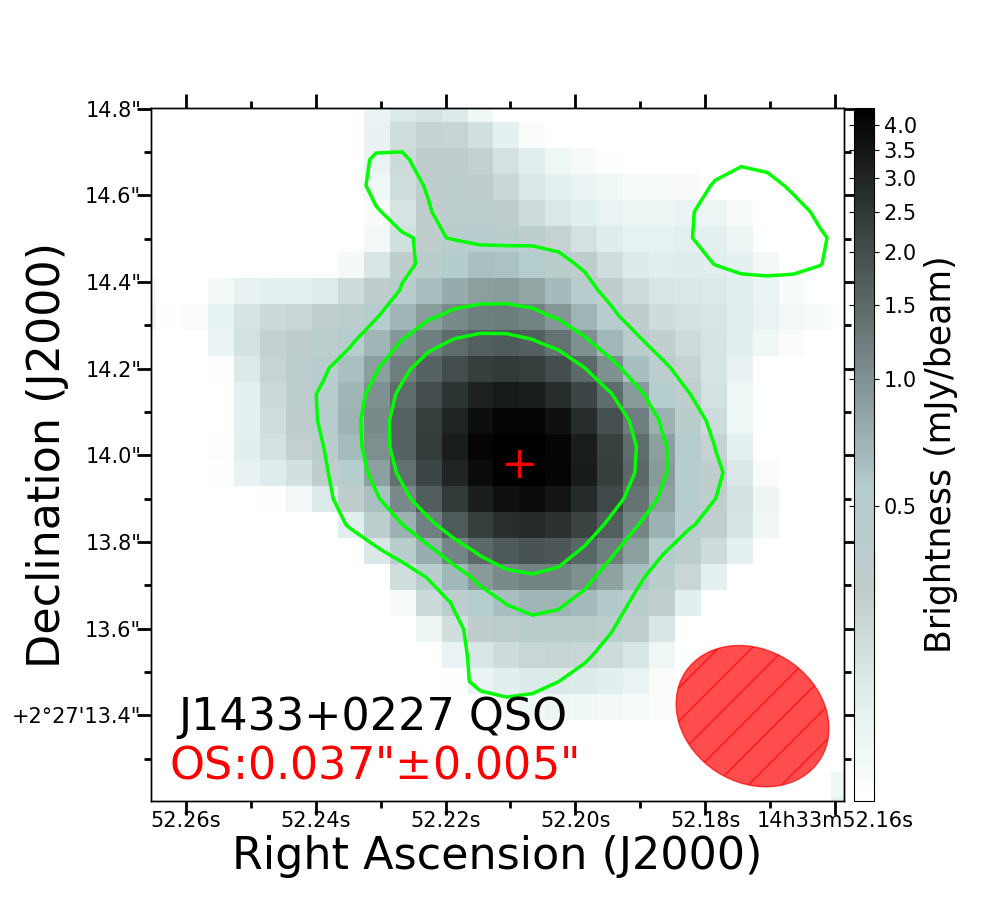} \\
\includegraphics[scale=0.2]{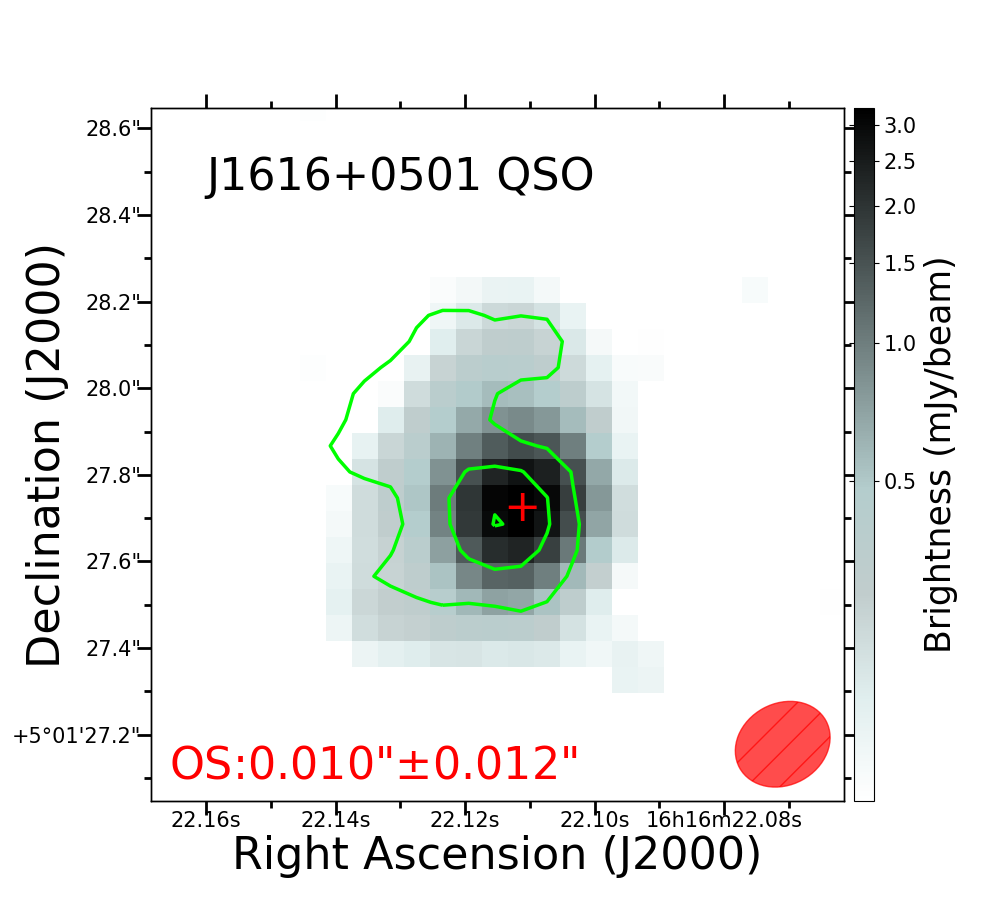}  
\includegraphics[scale=0.2]{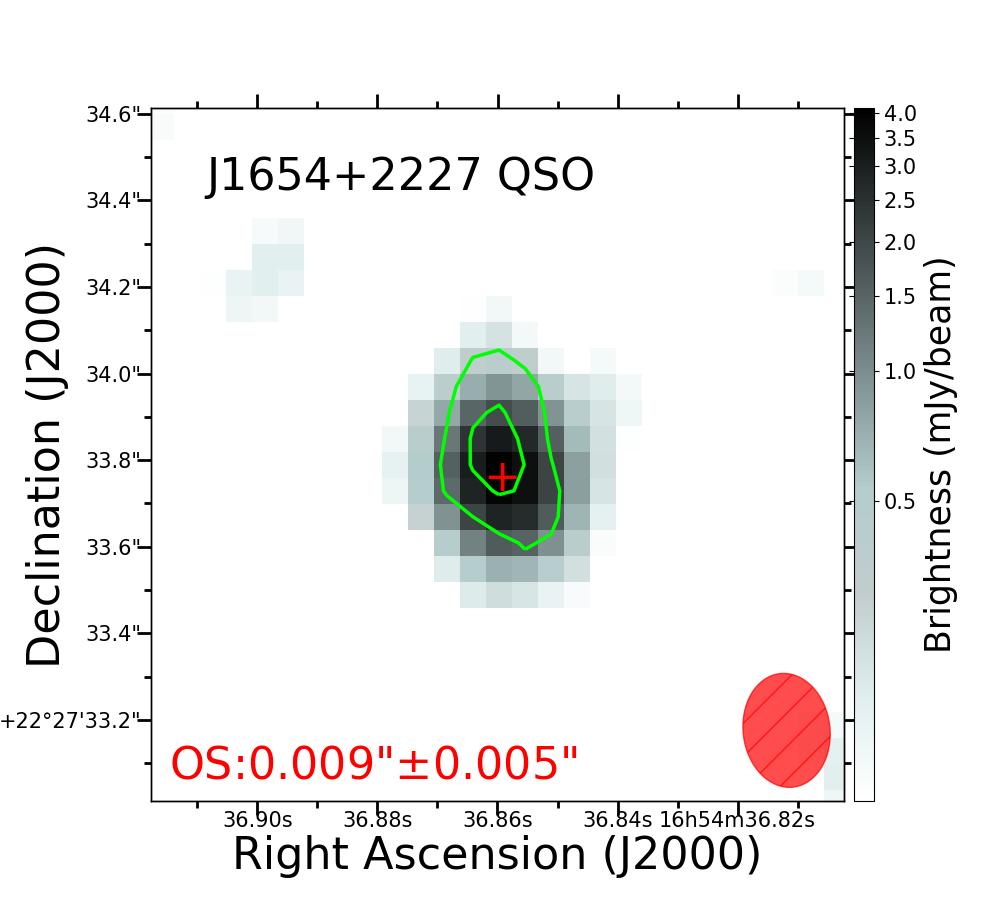} 
\includegraphics[scale=0.2]{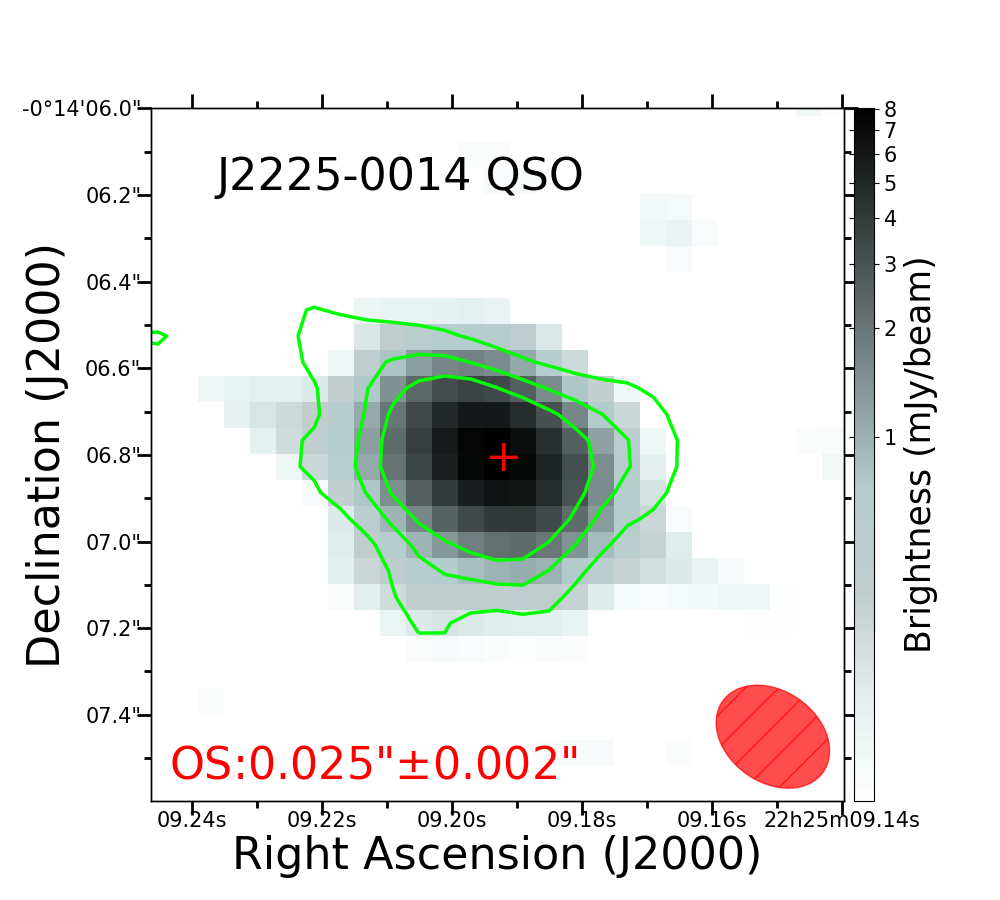} \\
FIR-Faint Objs.\\ 
\includegraphics[scale=0.2]{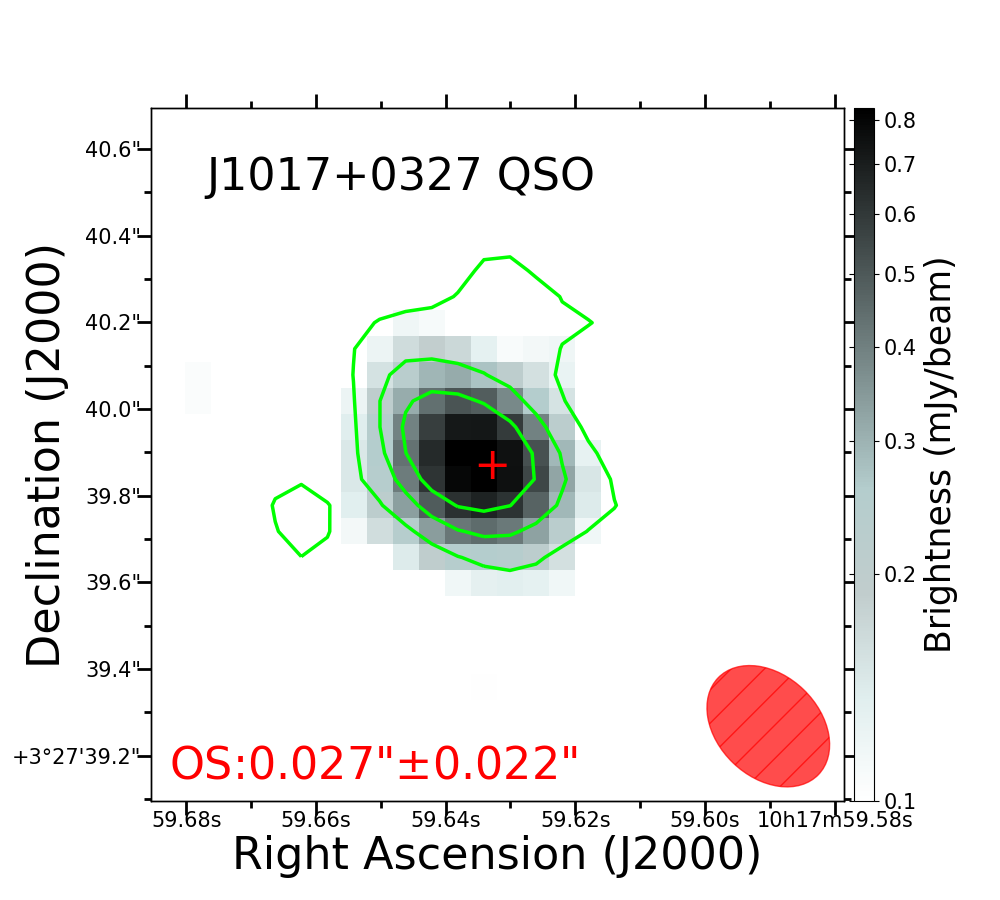} 
\includegraphics[scale=0.2]{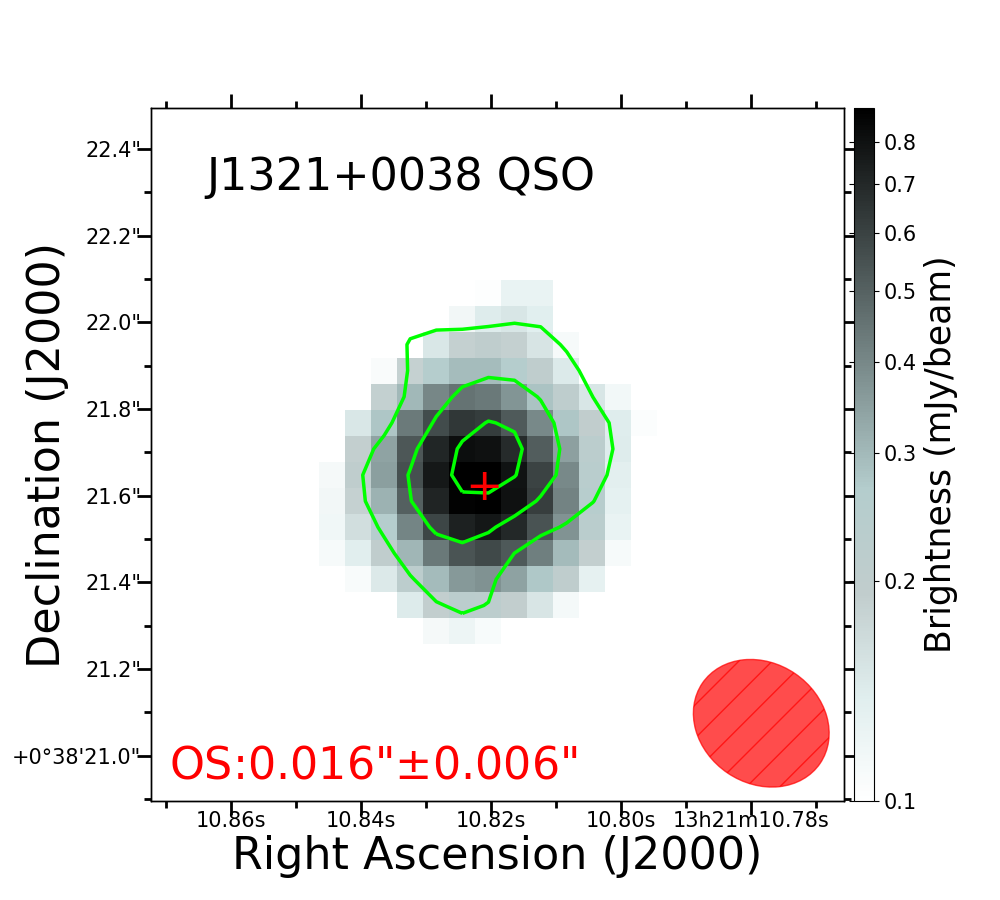} \\
\includegraphics[scale=0.2]{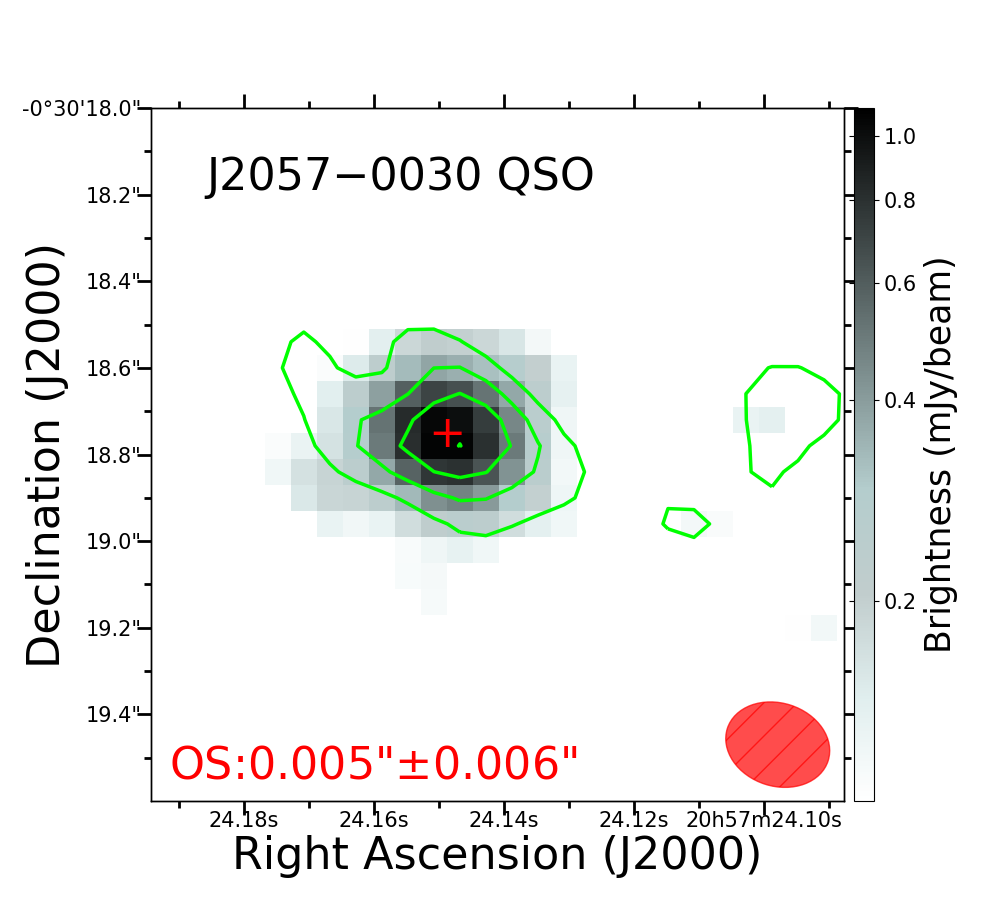} 
\includegraphics[scale=0.2]{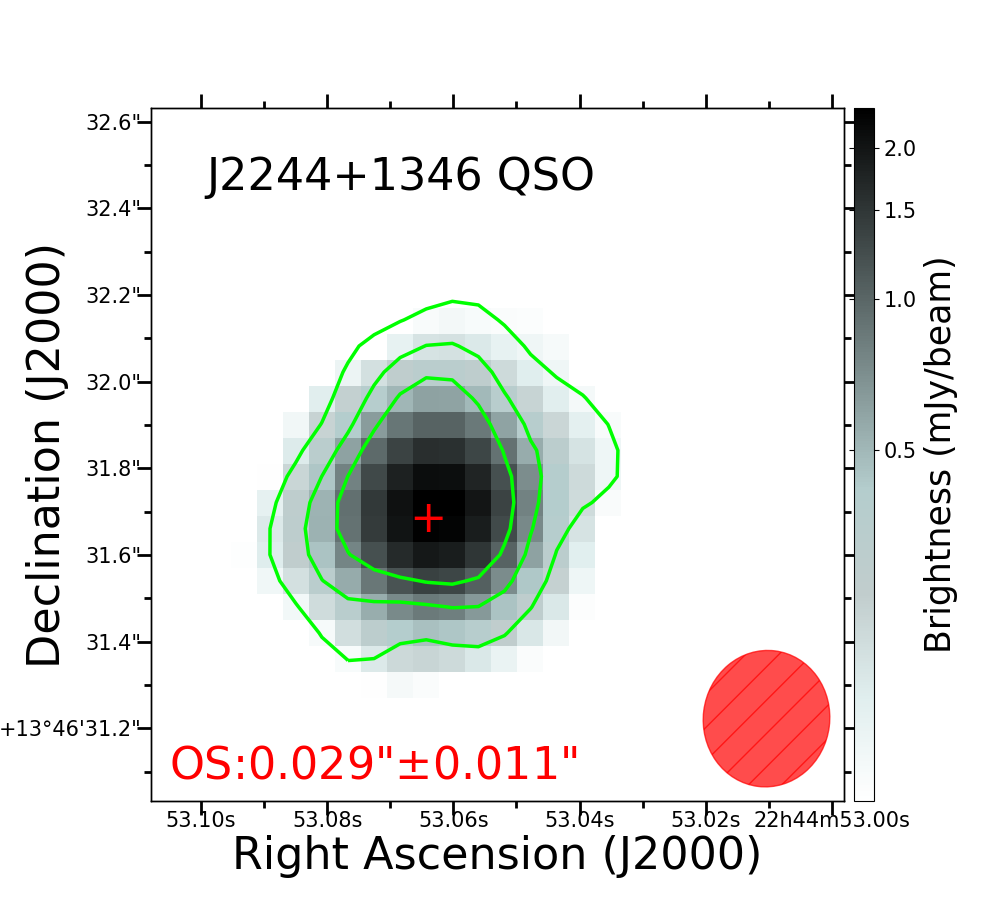}  \\

\caption{Small-scale continuum and \cii\ line emission maps derived from the Cycle-6 ALMA data, for all the sources with clear detection of \cii\ line emission. The FIR-bright sources in our sample are the top two rows, the FIR-faint sources are the bottom two rows.
For each source, the gray-scale map traces the continuum emission, while 
the contours trace the \cii\ line emission (i.e., surface brightness) at significance levels of 3, 6, 9, and 12$-\sigma$.
For each source, the line fluxes used for the contours were extracted from a spectral window spanning $\pm500$ \kms\ around the \cii\ line peak. The ALMA beams are shown as red ellipses near the bottom-right of each panel. The optical position from GAIA is marked with a red cross (+). In the bottom left of each image we list the Optical Separation (OS) along with the associated error.}
\label{fig:cont_os}
\end{figure*}

\begin{figure*}
\center
FIR-Bright Objs. \\ 
\includegraphics[scale=0.2]{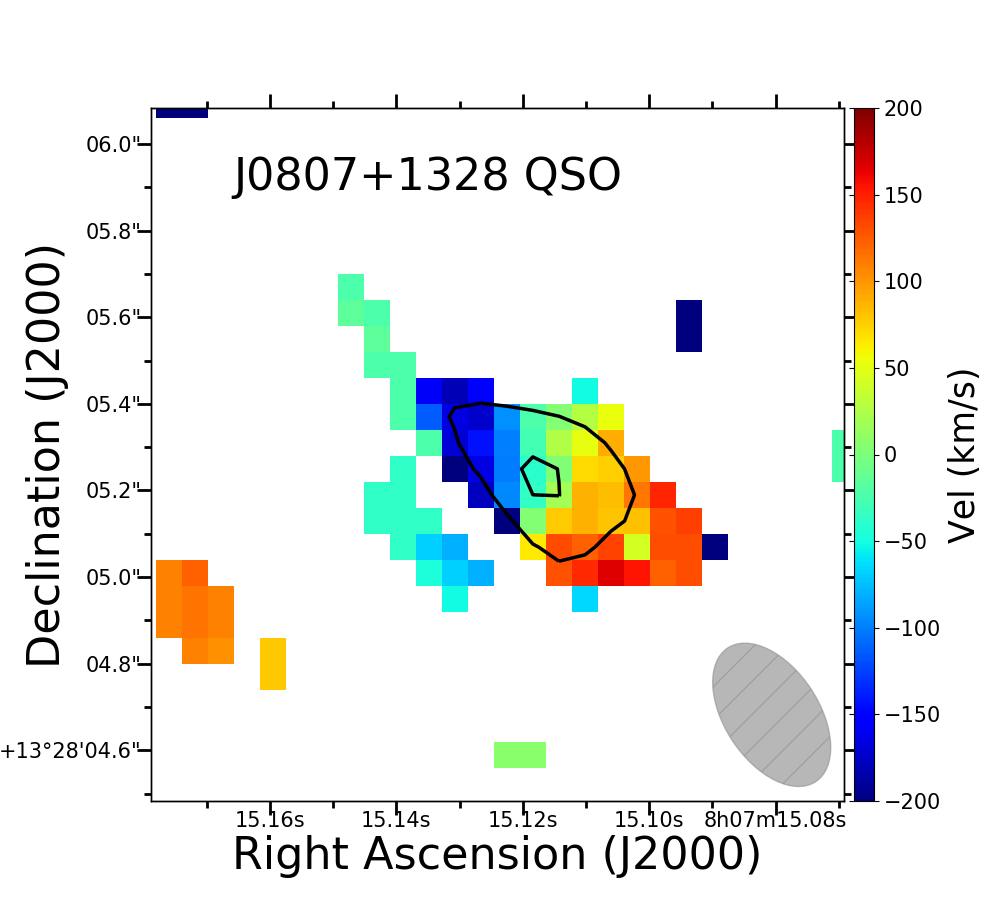} %
\includegraphics[scale=0.2]{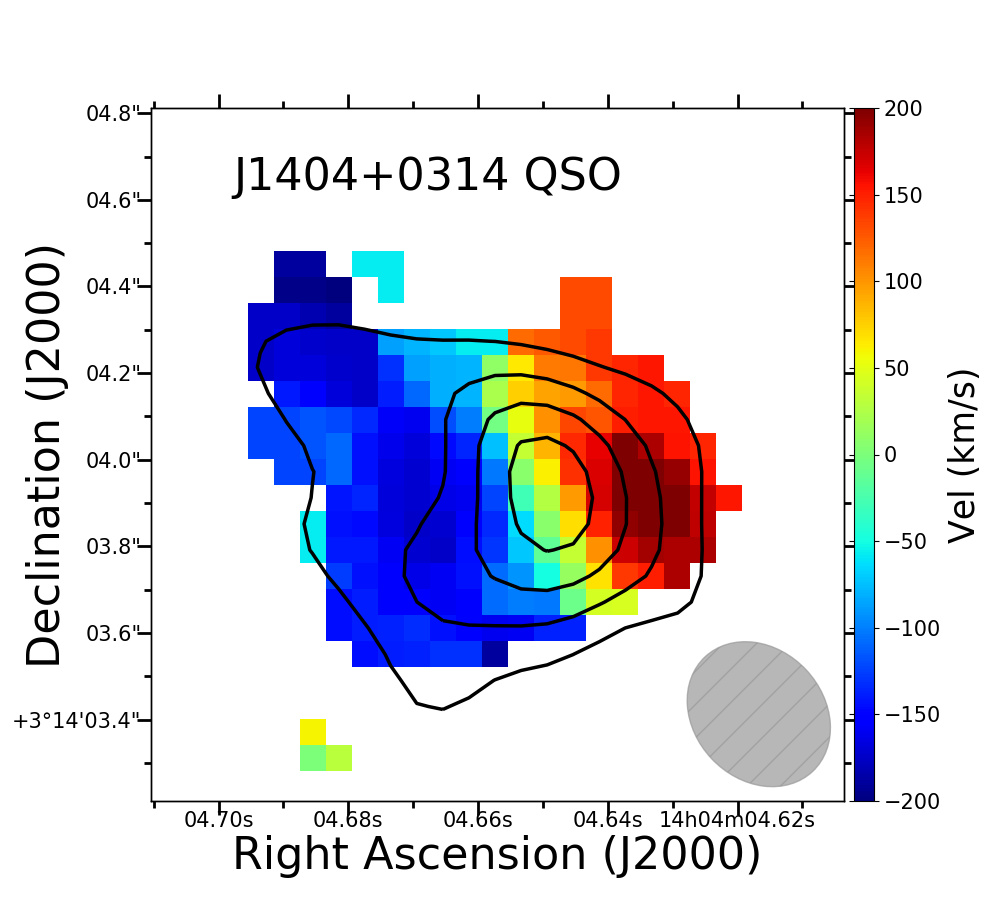} %
\includegraphics[scale=0.2]{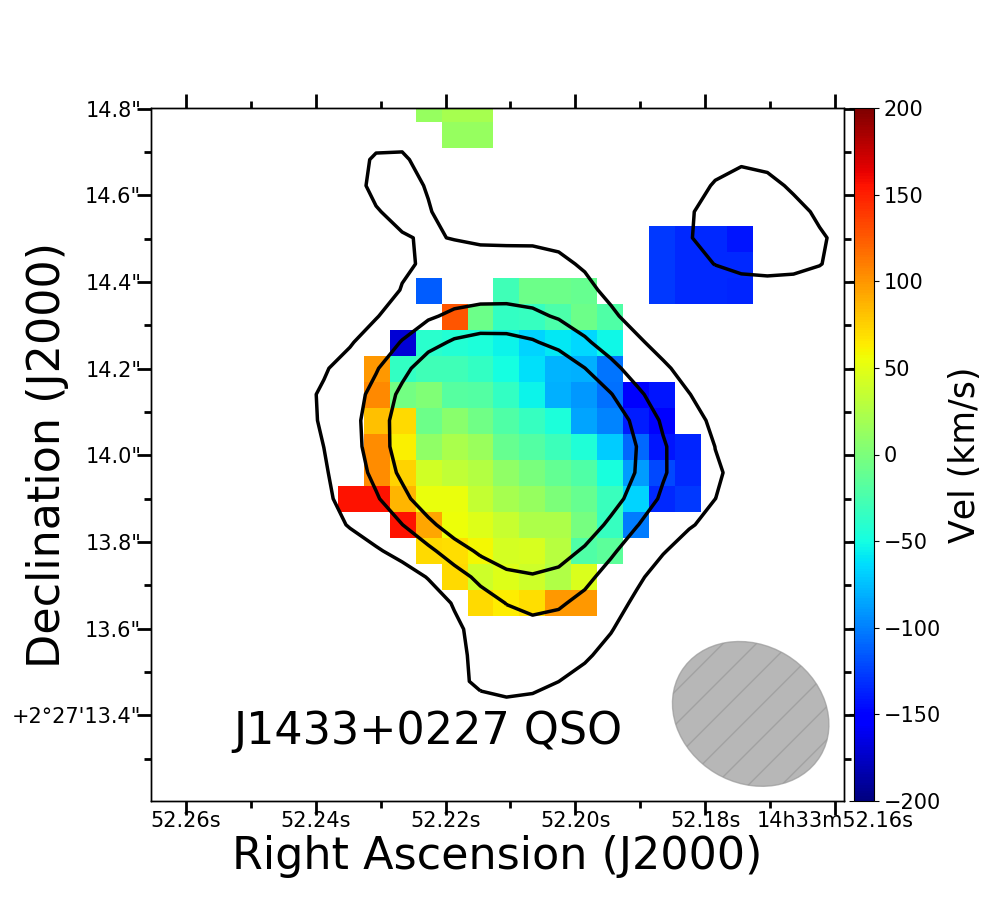} \\
\includegraphics[scale=0.2]{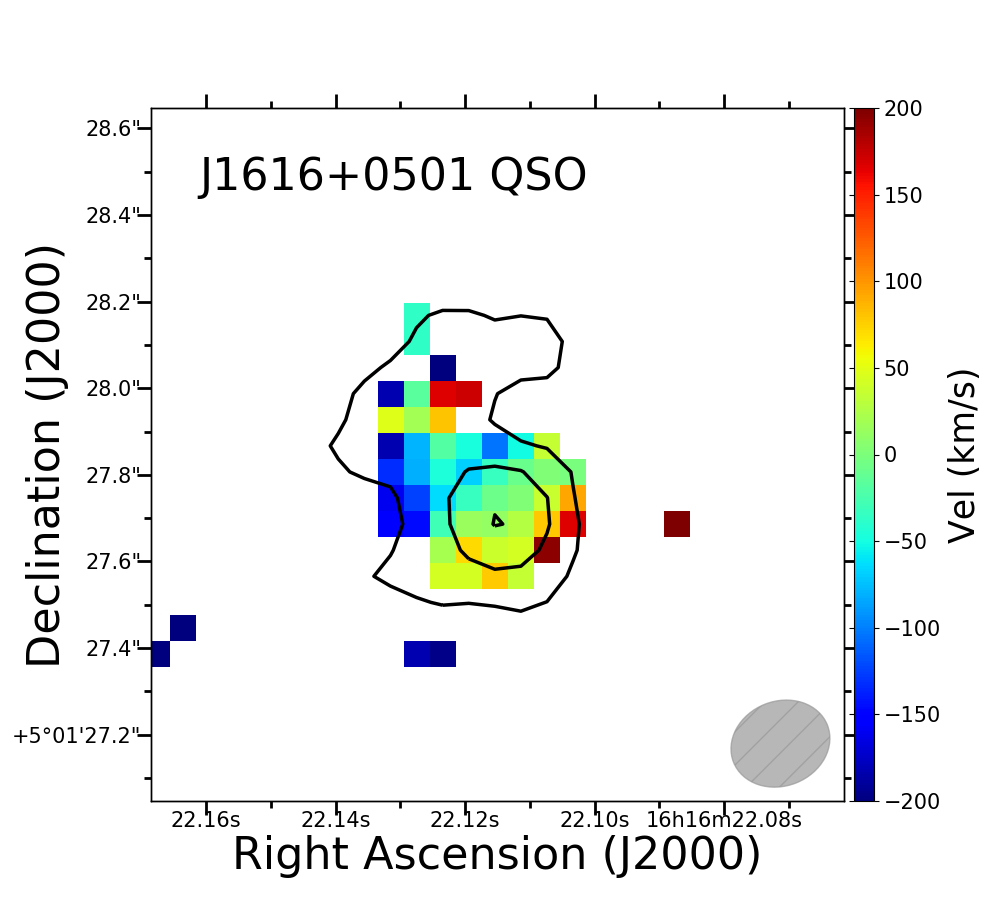} %
\includegraphics[scale=0.2]{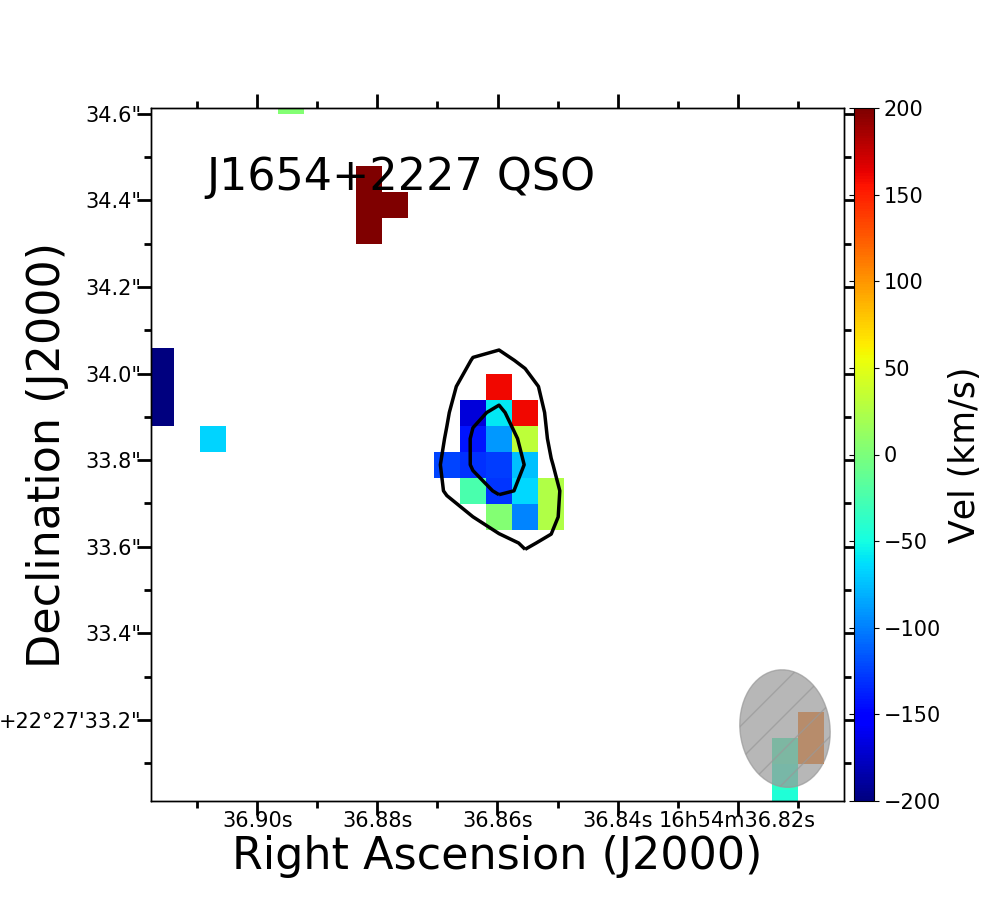} %
\includegraphics[scale=0.2]{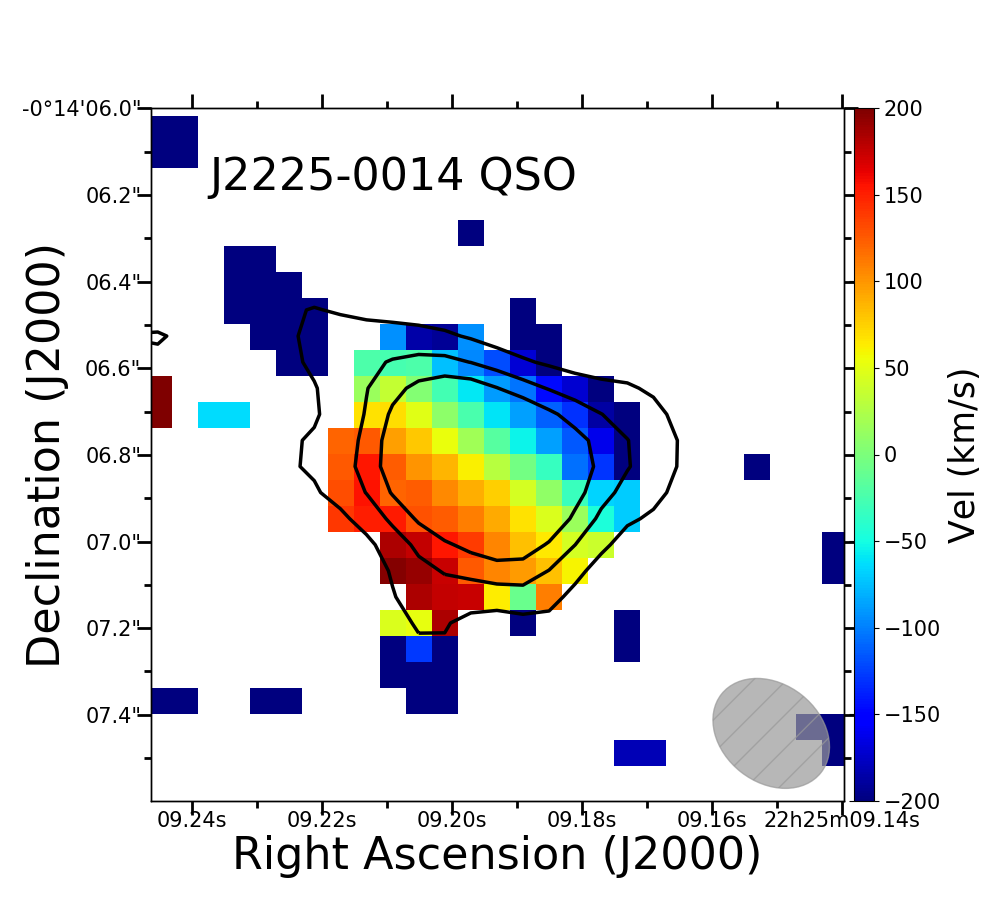} \\ 
FIR-Faint Objs. \\ 
\includegraphics[scale=0.2]{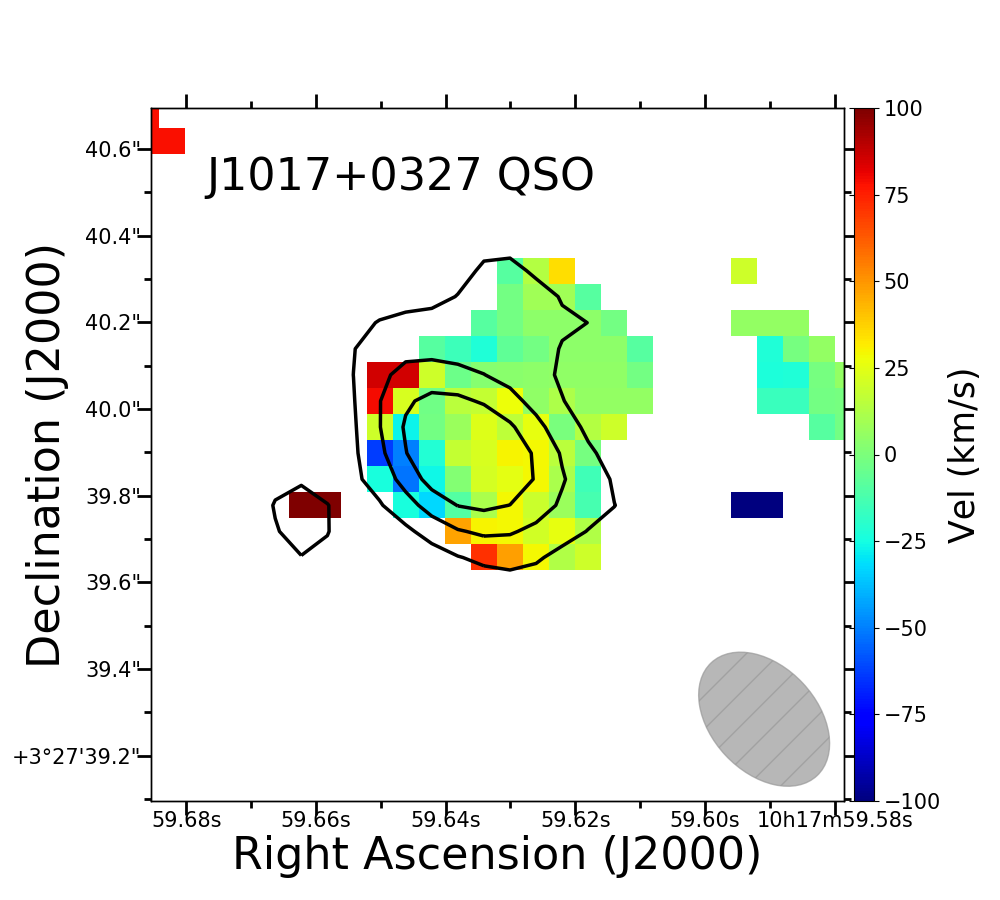} %
\includegraphics[scale=0.2]{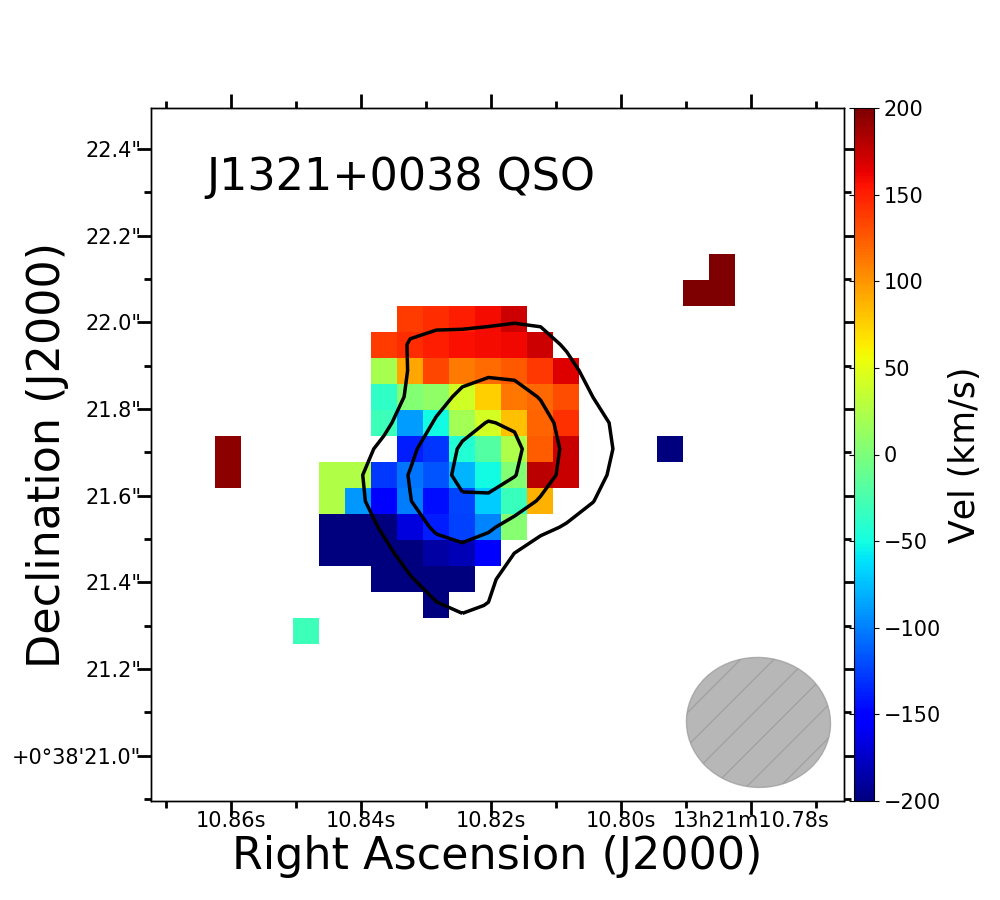} \\
\includegraphics[scale=0.2]{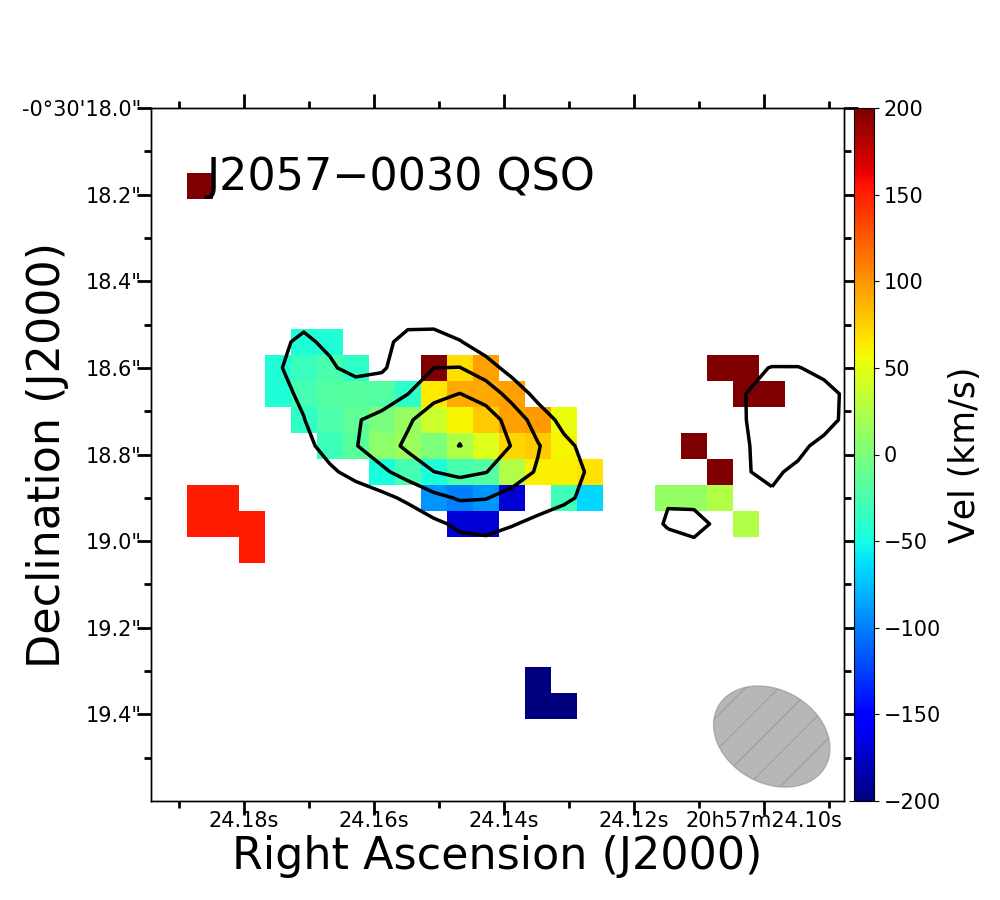} %
\includegraphics[scale=0.2]{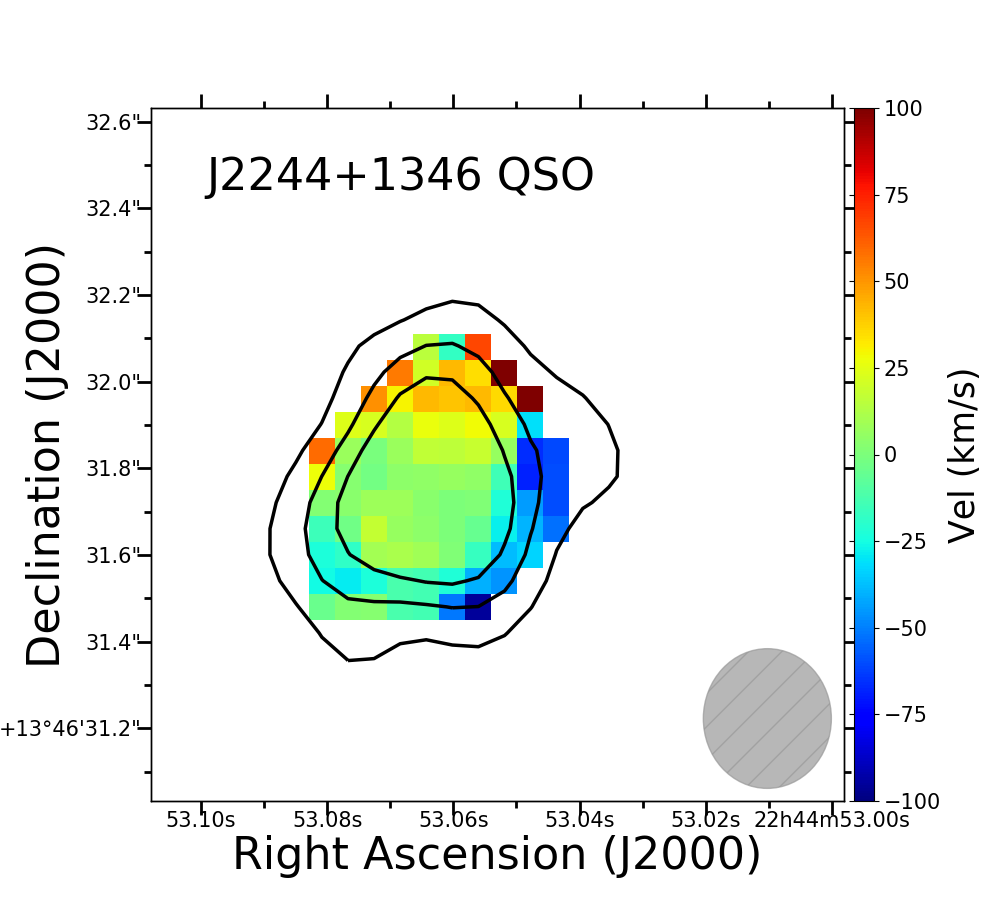} \\
SMGs \\
\includegraphics[scale=0.2]{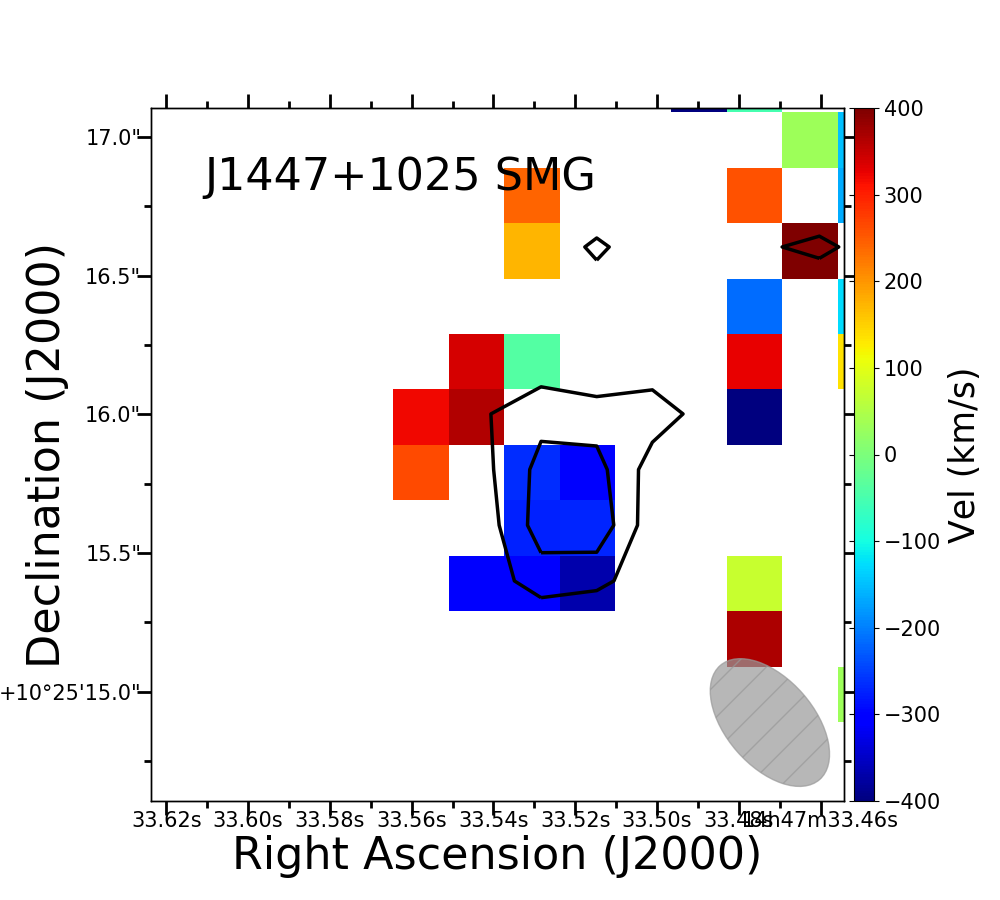} %
\includegraphics[scale=0.2]{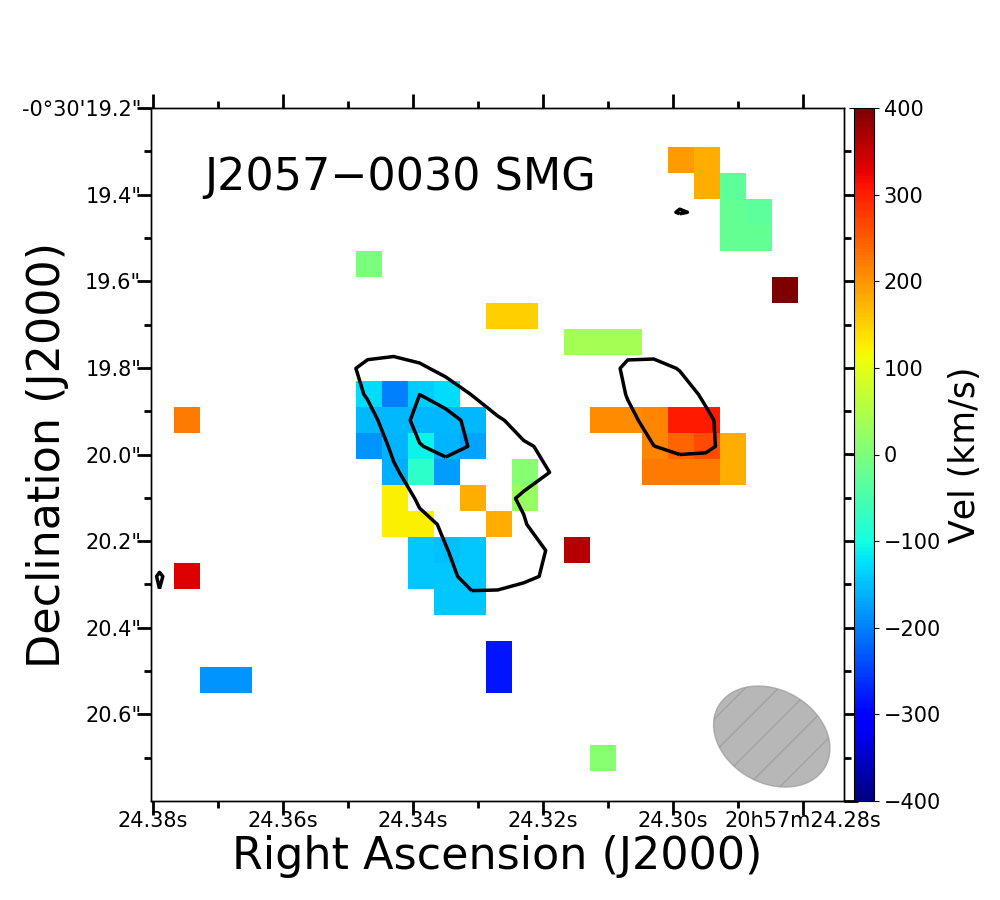} \\
\caption{\cii\ velocity maps for the FIR-bright sources in our
  sample (top two rows), FIR-faint sources (middle two rows), and the
  companion SMGs (bottom row). Black contours trace the
  \cii\ emission line surface brightness at significance levels of 3,
  6, 9, and 12$-\sigma$. The ALMA beams are shown as hatched gray
  ellipses near the bottom-right of each panel.}
\label{fig:vel}
\end{figure*}

\subsection{\cii\ Line Properties}
\label{sec:line_props}

In Figure \ref{fig:spectra}, we plot the continuum subtracted
\cii\ spectral region for all 12 quasar hosts presented in this work,
including J1151 which was undetected in both continuum and \cii, and
J1447 which had a 2$\sigma$ level detection in \cii. We also include
spectra for two SMGs accompanying J2057 and J1447. A best-fit line
model using a Gaussian profile is overlaid. The Root Mean Square (RMS)
spectra is plotted below each emission line spectra.

Figure \ref{fig:vel} shows velocity maps for the ten quasar hosts
significantly detected in \cii\ and the two SMGs accompanying J1447
and J2057. The weak \cii\ emission from J1447 was not sufficient to
determine moment maps. The morphologies of our targets are not as
uniform as in T17, possibly due to some of our sources being observed
through spectral sub-mm windows with worse transmission, as mentioned
in Section \ref{sec:alma_obs}. Well behaved velocity maps, with a
clear velocity gradient across the system, which suggests rotation of
a flat gaseous structure, is only seen in about half of systems. The
remaining sources show noisier, more irregular maps, although evidence
for a velocity gradient is still present.

As in T17, some of our quasar hosts show increased velocity dispersions
in the centers of the \cii\ - emitting regions, with $\sigma_{v} \sim
100\ \kms$, which can be an indication of beam smearing. This could
lead us to overestimate the rotation kinematics we see in Figure
\ref{fig:vel}. However we do not expand on correcting this smearing as
other studies of sub-mm sources have done, as our targets are only
partially resolved and modeling the rotation is not possible. In fact,
as many of our sources do not exhibit clear rotation dominated
kinematics (e.g., J1017 and J1654), other factors could be affecting
the kinematics of our hosts. Possible alternatives such as a turbulent
component have been demonstrated in several recent studies of resolved
ISM kinematics in high-redshift galaxies
\cite[e.g.,][]{Gnerucci2011,Williams2014}.

The majority of our objects have a single peak line profile except for
J1404 which exhibits double peak emission in the \cii\ line, and the
SMG companion to J1447. The double feature seen in J1404 has two peaks
separated $\~350$ \kms\ from each other, while the SMG of J1447 shows
two components to the \cii\ line separated by $\~600$ \kms.

The velocity map of J1404 in Figure \ref{fig:vel} shows a single
source with strong rotational signatures and a large total velocity
amplitude of $\sim 400\,\kms$, roughly the same separation we see in
the spectrum. This FIR-bright source does not show the presence of
companions, but the double peak could signal the late evolutionary
stage of a merger event.

On the other hand, the double feature seen in the SMG of J1447 most
likely corresponds to a double source. This is seen in the bottom
right panel of Figure \ref{fig:vel} where two spatially separated
kinematic components appear. The north-east peak is rather weak, as it
is below the 3 $\sigma$ threshold of the \cii\ contours, but it is
clearly recovered in the spectrum shown in Figure \ref{fig:spectra}
and coincides with strong emission seen in dust continuum (see Figure
\ref{fig:smgs}).

Finally, J2057 also presents some interesting dynamical
features. Besides the presence of two dust continuum peaks and a
complex velocity map shown by the companion SMG, the quasar itself
shows strong evidence for dynamical disruption: its \cii\ emission
appears as consistent with a $\sim 100\, \kms$ rotating disk plus
debris material and a $\sim 20$ kpc-long collimated ‘tadpole-like’
structure orientated roughly in the E-W direction, which is
constrained to a very narrow velocity range. This structure is not
apparent in Figure \ref{fig:vel} because of the velocity
binning. J2057 and its SMG companion will be the subject of a future
paper.
\subsection{Optical Center Separation}
\label{sec:opt}
In Figure \ref{fig:cont_os} we plot the continuum maps of our quasars, along \cii\ emission contours. From The 2nd data release of the Gaia mission \citep{Gaia2018}, crossed referenced with the Pan-STARRS 1 data base \citep{Flewelling2016}, we
obtain the optical centers of our objects. 

We then compute Optical Separation (OS) as the separation between the Gaia optical center and the peak of the dust continuum emission from our ALMA data, as determined by Gaussian fits in Section \ref{sec:reduction}. Each image in Figure \ref{fig:cont_os} lists the OS along with the associated error. OS values have a range of 0.005'' - 0.062'' for our entire sample. The median positional uncertainty of quasars in DR2 of Gaia is 0.4 milli-arcsecond \citep{Mignard2018}, while those objects which are cross referenced with the Pan-STARRS data base have median uncertainties of 3.1 and 4.8 milli-arcsecond for $\Delta ra$ and $\Delta dec$, respectively \citep{Chambers2016}. The associated error of our OS considers uncertainties associated with the optical position as well as our Gaussian fits to the continuum emission, giving an overall median error of $\~$ 11 mas.

Offsets of the optical center could be an indicator of dual-AGN or late stage major mergers \citep{Orosz2013,Makarov}. However these studies found OS values on scales of hundreds of milli-arcsecond scales, much larger than what we see. We also see that there is no correlation between host galaxy velocity gradients and OS. J1328-0224 is our object with the highest OS (62 mas) in our sample, but Figure \ref{fig:vel} shows that it has a very low gradient of velocity with rather uniform values. In contrast, J2057-0030 has the lowest OS but we believe it to be a perturbed system with a tidal tale. Thus we do not consider the OS to be an indicator of mergers or  perturbations for our sources.

\section{Results and Discussion}
\label{sec:results}

In what follows we divide the discussion into those results that are
robust and do not rely on unconstrained assumptions and those that are
more speculative and that need further observations in order to prove
their veracity. In particular, the determination of gas and dynamical
masses for our quasar hosts are highly uncertain, and therefore all
discussion based on these determinations should be taken withe extra
caution.

\subsection{Main Findings}

\subsubsection{Emission Line Velocity Offsets} 
\label{sec:blr_offsets}

\begin{figure}
\center
\includegraphics[scale=0.35,trim=80 0 0 0]{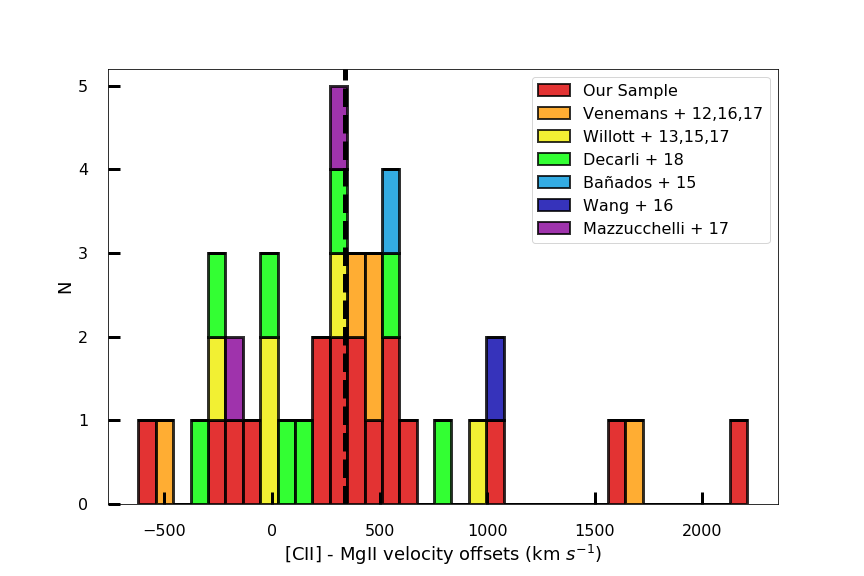}
\caption{Histogram presenting the distribution of the velocity shifts of
the \mgii\ with respect to the \cii\ emission lines of the quasars. Our
  twelve observations and those presented in T17 are at \zfpe, while
  those in \citet{Venemans2016} and
  \citet{Willot2013,Willot2015,Willott2017} are at $z \gtrsim 6$. The
  \cii\ line is clearly redshifted with respect to the
  \mgii\ measurements with a mean and standard deviation of $372 \pm
  582\,\kms$. The vertical line denotes the median of 337 \kms. }
\label{fig:redshift_hist}
\end{figure}

\begin{figure}
\center
\includegraphics[scale=0.35,trim=80 0 0 0]{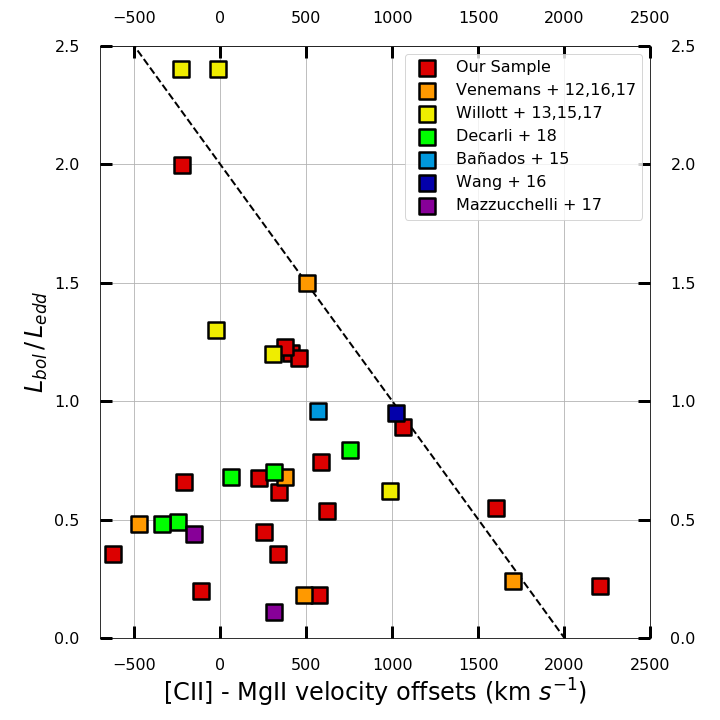}
\caption{Eddington ratios of our compiled quasars against the observed
  \mgii\ offsets. References for the \mgii\ measurements and the
  Eddington ratios can be found in Table 6. The average Eddington
  Ratio is 0.83. We plot a line in black to help illustrate that there
  are no objects with both high \mgii\ offsets and high Eddington
  ratios. 36 quasars are plotted in total. Only 5 of the 8 Decarli
  quasars have published Eddington ratios.}
\label{fig:edd_v_offset}
\end{figure}
Neutral carbon has a low ionization potential (11.3 eV) and can be excited by electron collisions. Therefore, \cii\ emission can be found in the ISM throughout a galaxy, particularly tracing photo-dissociation regions, that is, naturally diffuse and partially ionized gas. Although from observations in the local universe it is seen that the \cii\ line is broader than molecular gas (e.g., \citealt{Goicoechea2015}), because of its high brightness and narrow intrinsic width it is a good measure of the systemic redshift of the quasar host galaxy. While the \mgii\ line, produced in the vicinity of the SMBHs in the so called Broad Line Region (BLR), is dominated by the gravitational SMBH as well as other central bulk nuclear winds or turbulences.
In Table \ref{tab:z_cii} we compare the redshifts obtained from the
\cii\ and \mgii\ lines ($\Delta v_{\mgii}$) for our 17 quasar hosts
with detected \cii. For unobscured AGN at moderate redshifts ($z < 2$), the BLR
\mgii\ line is found within $\sim 200\ $\kms\ of the systemic
redshifts \citep{Richards2002,Shen2016,Mejia-Restrepo2016}, and
centered around $\sim 0$ \kms (see Figure 2 in \citealt{Shen2016}). The
large dispersions in the line shifts are clearly due to the broad
nature of the BLR lines and hence the difficulties in determining
precise line centers.

For comparison, we also list the SDSS-based redshift determinations
published in \cite{Hewett_2010}, along with the difference with
respect to the \cii\ line ($\Delta v_{SDSS}$). At $z \sim 5$
SDSS-based redshifts would be determined using the BLR UV \Lya,
\siv\ and \civ\ emission lines, which are usually considered
problematic because of the absorbed \Lya\ profile, the weakness of the
\siv\ line, and the well established blueshifts present in the
\civ\ line. In fact, we see no correlation between $\Delta v_{SDSS}$
and $\Delta v_{\mgii}$, most likely because of the uncertainties
associated to the $z_{SDSS}$ determinations \citep{Mason2017,dix2020}.

From table \ref{tab:z_cii} we can see that most objects in our total
sample of 17 quasar hosts have significant blueshifts of the
\mgii\ line with respect to the \cii\ line ($\Delta v_{\mgii} > 0$).
The average value for $\Delta v_{\mgii}$ is 464 \kms, with a standard
deviation of 657 \kms while the median is found to be at 379 \kms. As \cite{Venemans2016} already pointed out, since the distribution of offsets is not centered around 0 \kms, we can assume that they are not due to the uncertainty associated with fitting the broad emission line of \mgii. This is further supported by \cite{Shen2016}, where they state that the intrinsic uncertainty of using the \mgii\ broad-line for estimating redshifts is 200 \kms, smaller than our median offset. We find no noticeable correlation between \mgii\ offsets and the presence of companions. 

\citet{Venemans2016} compiled a list of $z > 6$ quasars and compared
the redshift measurements from the $\mgii$ line and those of the CO
molecular line or the \cii\ atomic line. The median of the $z_{\rm
  \cii/CO} - z_{ \mgii}$ distribution for their sample is 467
\kms\ with a standard deviation of 630 \kms, almost identical to our
findings. We created our own compilation, but used exclusively quasars
with a measured $\cii$ line for the sake of congruity. The compilation
is populated by our total sample of 17 quasars, eight quasars taken
from \citet{Decarli2018}, five quasars found in
\citet{Willot2013,Willot2015,Willott2017}, five quasars from
\citet{Venemans2012,Venemans2016,Venemans2017}, two from
\citet{Mazzucchelli2017}, and one each from \citet{Banados2015} and
\citet{Wang2016}. We present the \mgii\ offsets of this compilation as
a histogram in Figure \ref{fig:redshift_hist} and in Table
\ref{tab:comp_offset}. For this compilation we found a mean $z_{\cii}
- z_{\mgii}$ of 372 \kms, a median of 337 \kms, and a standard
deviation of 582 \kms. It should be noted that only our sample is at $
z \simeq 5$, while the quasars from the literature are all at $z
\gtrsim 6$. The mean and median of the $z \gtrsim 6$ only quasars are
300 and 309 \kms, respectively, very close to the results from our
full compilation. This result strongly suggests a velocity difference
between the BLR and quasar host galaxies of several hundred \kms.

Blueshifts are usually associated to outflowing gas which is
approaching the observer. Blueshifts seen in the \civ\ line, for
example, are usually interpreted as evidence for nuclear outflows and
they seem to correlate well with accretion rate
\citep{Coatman2016,Sulentic2017,Vietri2018,Sun2018,Ge2019}. We have
looked for such correlation for the objects in our compilation and
found none. Figure \ref{fig:edd_v_offset} presents the accretion rate
in units of Eddington (as reported in the literature) versus the
measured \cii-\mgii\ shifts. A rather low correlation coefficient is
determined, with $r = -0.23$. In fact, Figure \ref{fig:edd_v_offset}
suggests that low accretion sources can show a wide range of possible
shifts, while high-Eddington sources tend to show small offsets, if
any. We also tested a correlation of the offsets with infrared
luminosities (compiled values can also be found in Table
\ref{tab:comp_offset}) but no significant result was found ($r =
0.26$).
\par
Like in Section \ref{sec:opt} we search for correlations with the presence of companions. On average the mean offset of objects with companions is lower than the entire sample (92.4 \kms). Interestingly, of the four quasars with companions presented in \cite{Decarli2018}, two have tabulated \cii-\mgii\ offsets in Table 6, giving a mean offset of $\~0$ \kms. Since the number of sources with companions is very small, these are by no means conclusive findings, but suggest a possible link between merger activity and smaller \cii-\mgii\ shifts.

\begin{centering}
\begin{table}
\caption{Redshifts and $\cii$ Line Shifts}
\label{tab:z_cii}
\begin{tabular}{l l c c c c c}
\hline \hline 
Sub- & Target & $z_{\cii}$ & $z_{SDSS}$ & $\Delta v_{SDSS}$ & $z_{\mgii}\tablenotemark{a}$ & $\Delta v_{\mgii}$ \\
sample & & &  & \kms\ & & \kms \\ 
\hline 
Bright 	
       
         & J0807     & 4.879 & 4.871 & $+$378  & 4.874 & $+$256\\
         & J1404     & 4.923 & 4.871 & $+$2208 & 4.880 & $+$2208\\
         & J1433     & 4.728 & 4.685 & $+$2281 & 4.721 & $+$379\\ 
         & J1616     & 4.884 & 4.863 & $+$1061 & 4.872 & $+$620\\
         & J1654     & 4.728 & 4.707 & $+$1081 & 4.730 & $-$112\\
         & J2225     & 4.716 & 4.883 & $+$508  & 4.886 & $+$340\\
         & J0331\tln & 4.737 & 4.732 & $+$257  & 4.729 & $+$412\\ 
         & J1341\tln & 4.700 & 4.682 & $+$981  & 4.689 & $+$573\\
         & J1511\tln\tablenotemark{*} & 4.679 & 4.677 & $+$88   & 4.670 & $+$456\\
\hline 
Faint  
         & J1017     & 4.949 & 4.918 & $+$1559 & 4.917 & $+$1605\\
         & J1151     & ---   & 4.699 & ---     & 4.698 & --- \\
         & J1321     & 4.722 & 4.739 & $-$882  & 4.716 & $+$337\\
         & J1447\tablenotemark{*}     & 4.682 & 4.688 & $-$329  & 4.686 & $-$224\\	
         & J2057\tablenotemark{*}     & 4.683 & 4.685 & $-$97   & 4.663 & $+$1064\\
         & J2244     & 4.661 & 4.621 & $+$2153 & 4.657 & $+$225\\
         & J0923\tln\tablenotemark{*} & 4.655 & 4.650 & $+$257  & 4.659 & $-$213\\
         & J1328\tln\tablenotemark{*} & 4.646 & 4.650 & $-$188  & 4.658 & $-$621\\
         & J0935\tln & 4.682 & 4.699 & $-$911  & 4.671 & $+$588\\
\hline
\end{tabular}
\tablenotetext{a}{\MgII-based redshifts taken from T11.}
\tablenotetext{T17}{\ \ \ \ Sources from T17.}
\tablenotetext{*}{Sources with the presence of companions.}

\end{table}
\end{centering}

\subsubsection{SEDs and SFRs}
\label{sec:seds_sfrs}

\begin{figure*}
\center
\begin{tabular}{c c c}
\includegraphics[width=5cm,height=4cm]{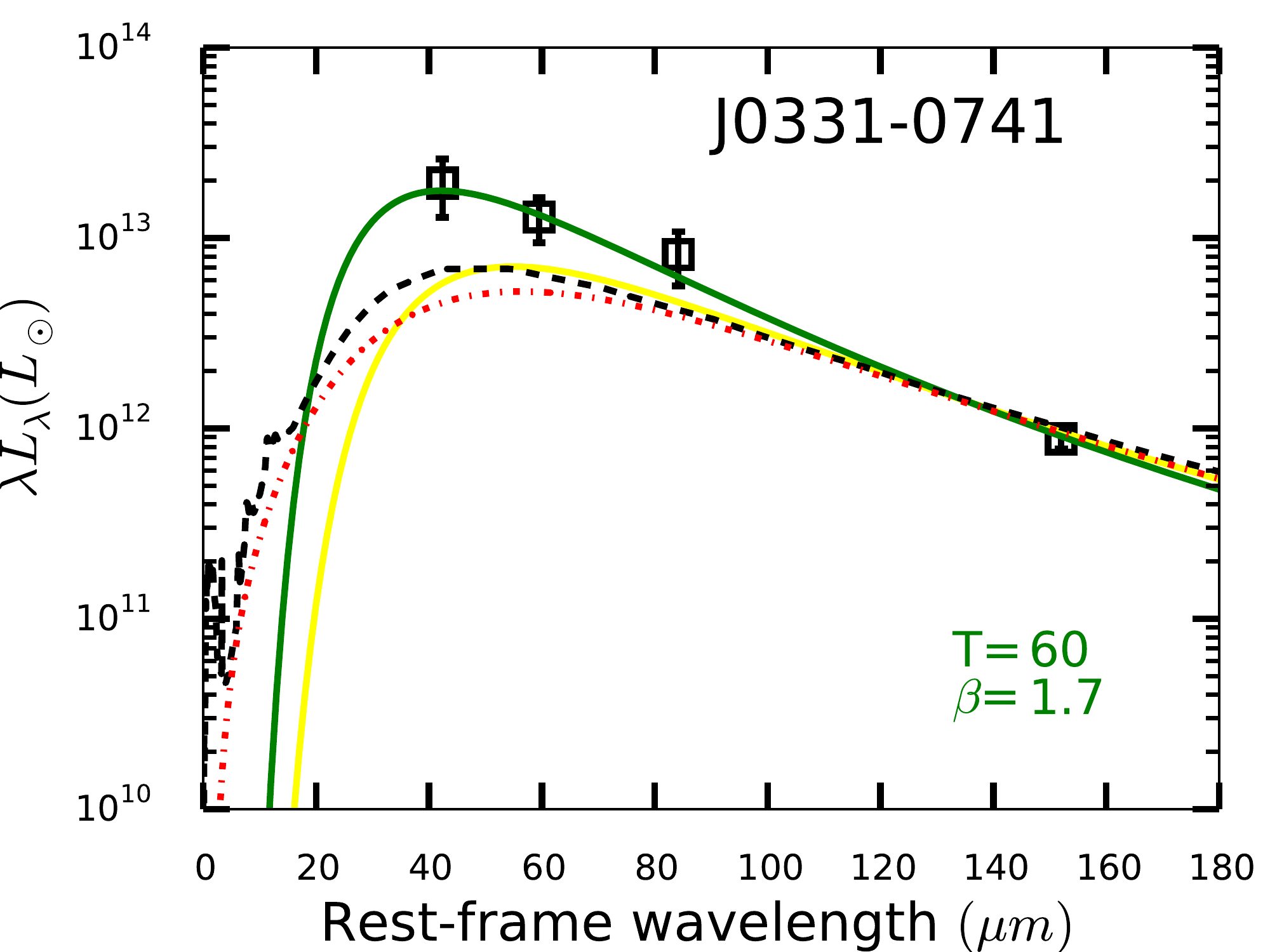} & \includegraphics[width=5cm,height=4cm]{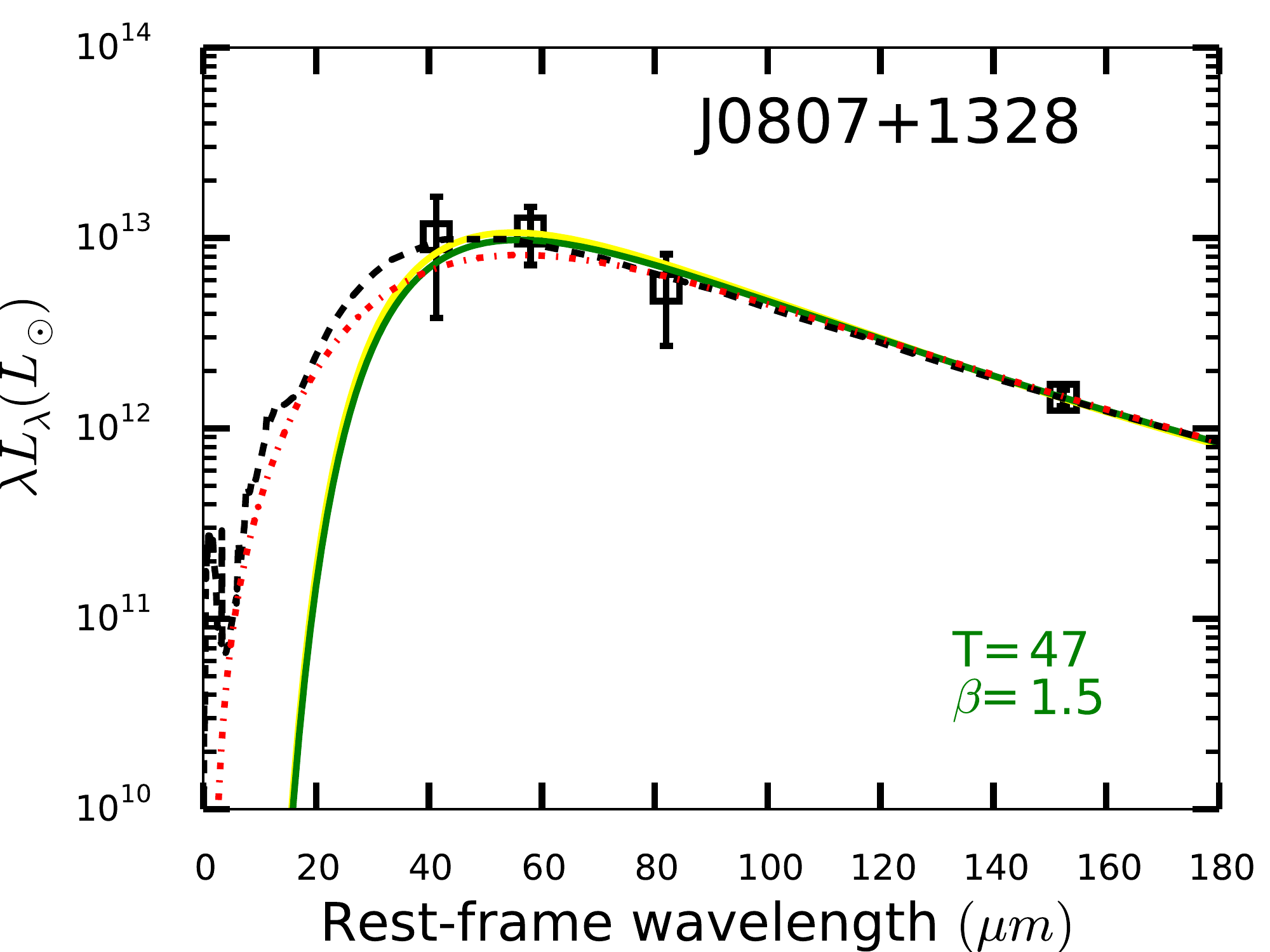} & \includegraphics[width=5cm,height=4cm]{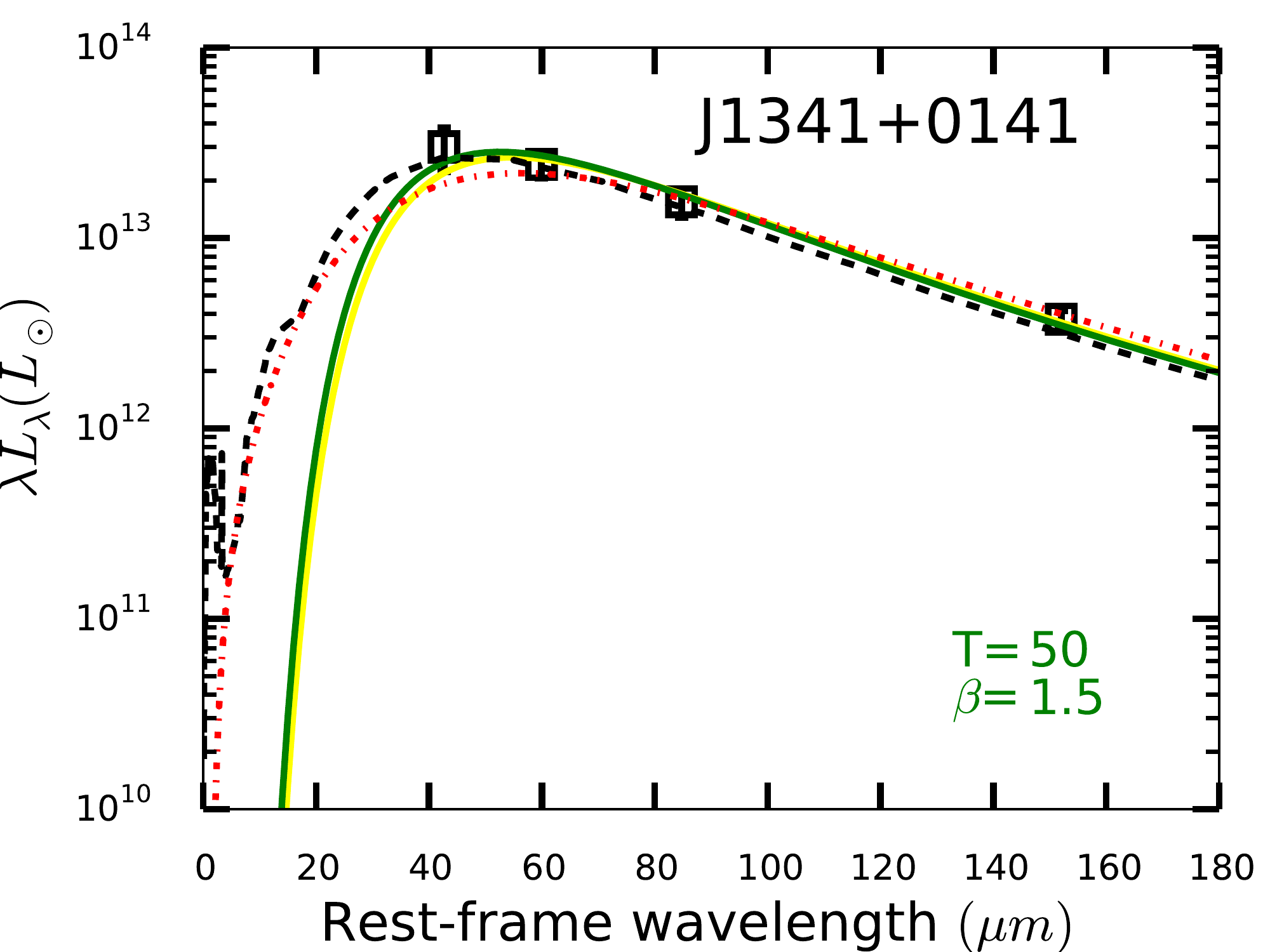} \\
\includegraphics[width=5cm,height=4cm]{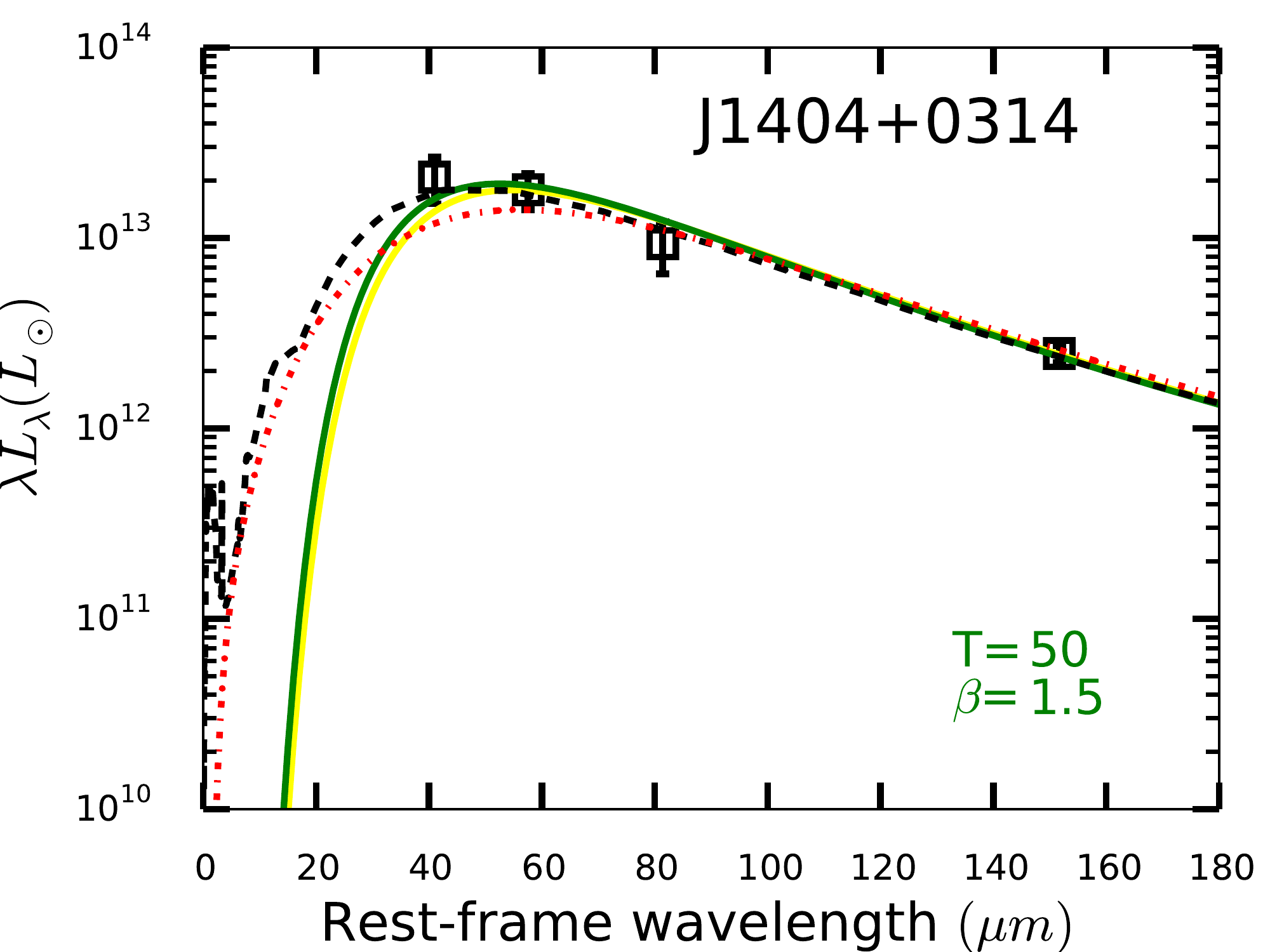} & \includegraphics[width=5cm,height=4cm]{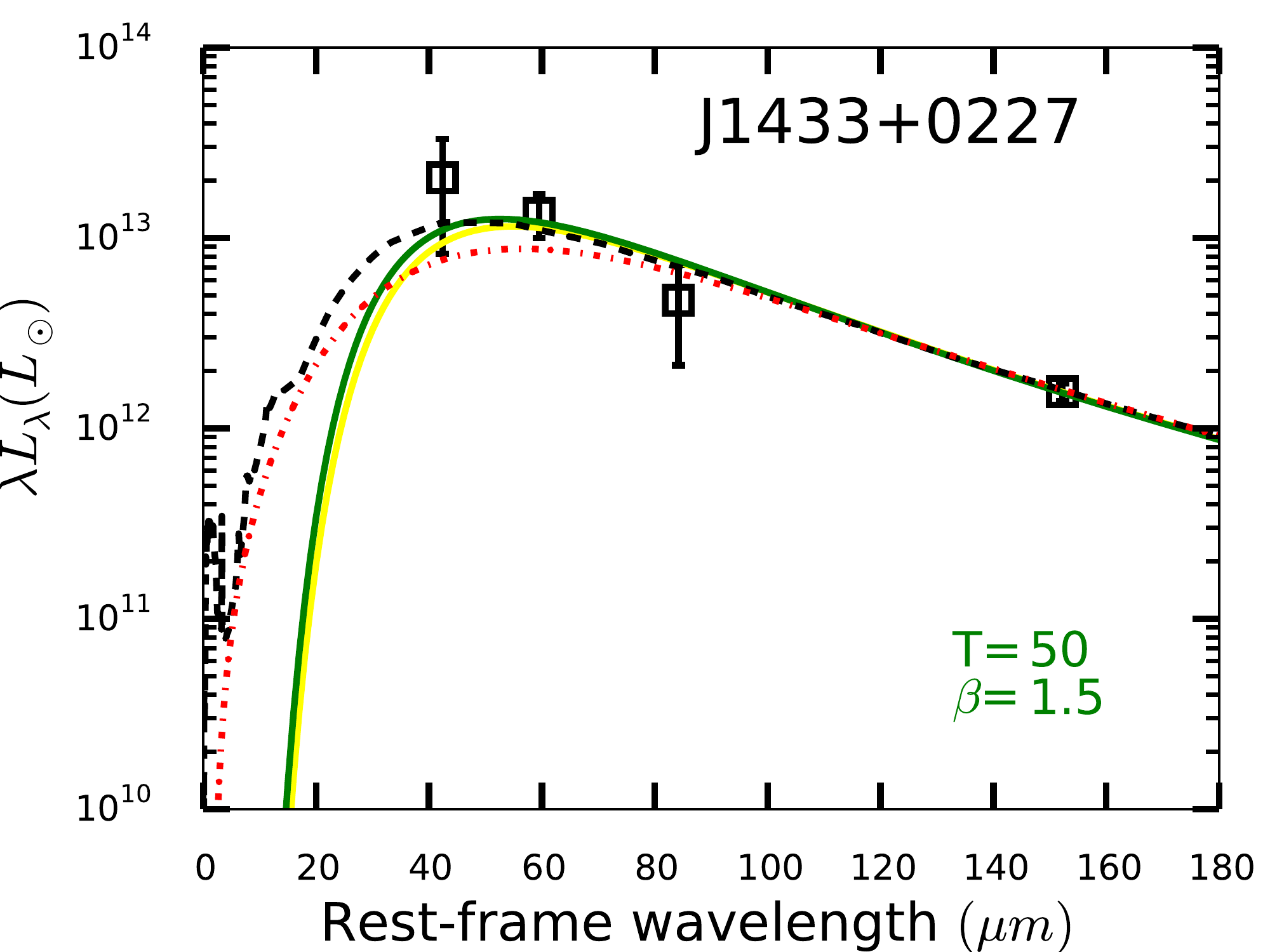} & \includegraphics[width=5cm,height=4cm]{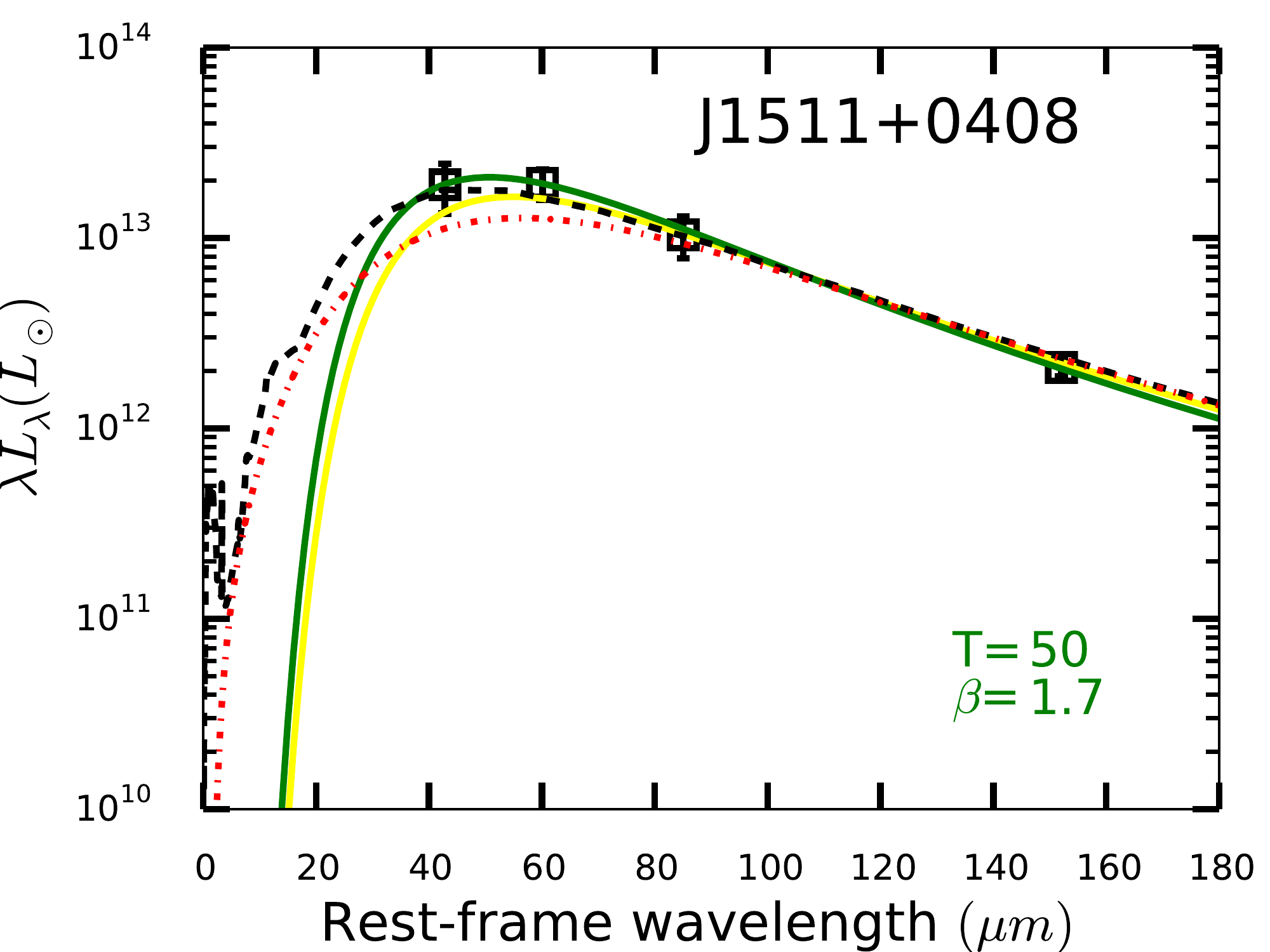} \\
\includegraphics[width=5cm,height=4cm]{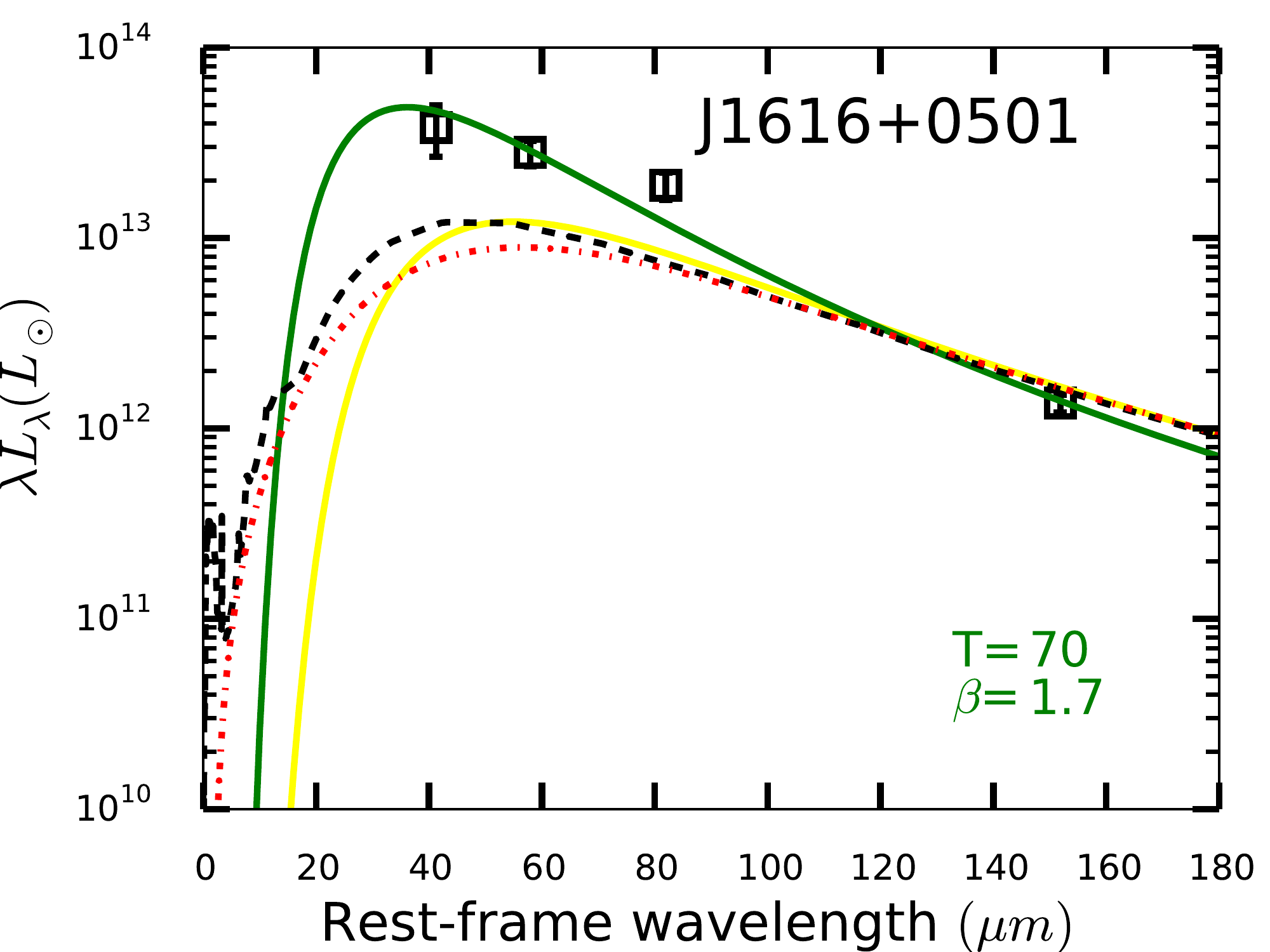} & \includegraphics[width=5cm,height=4cm]{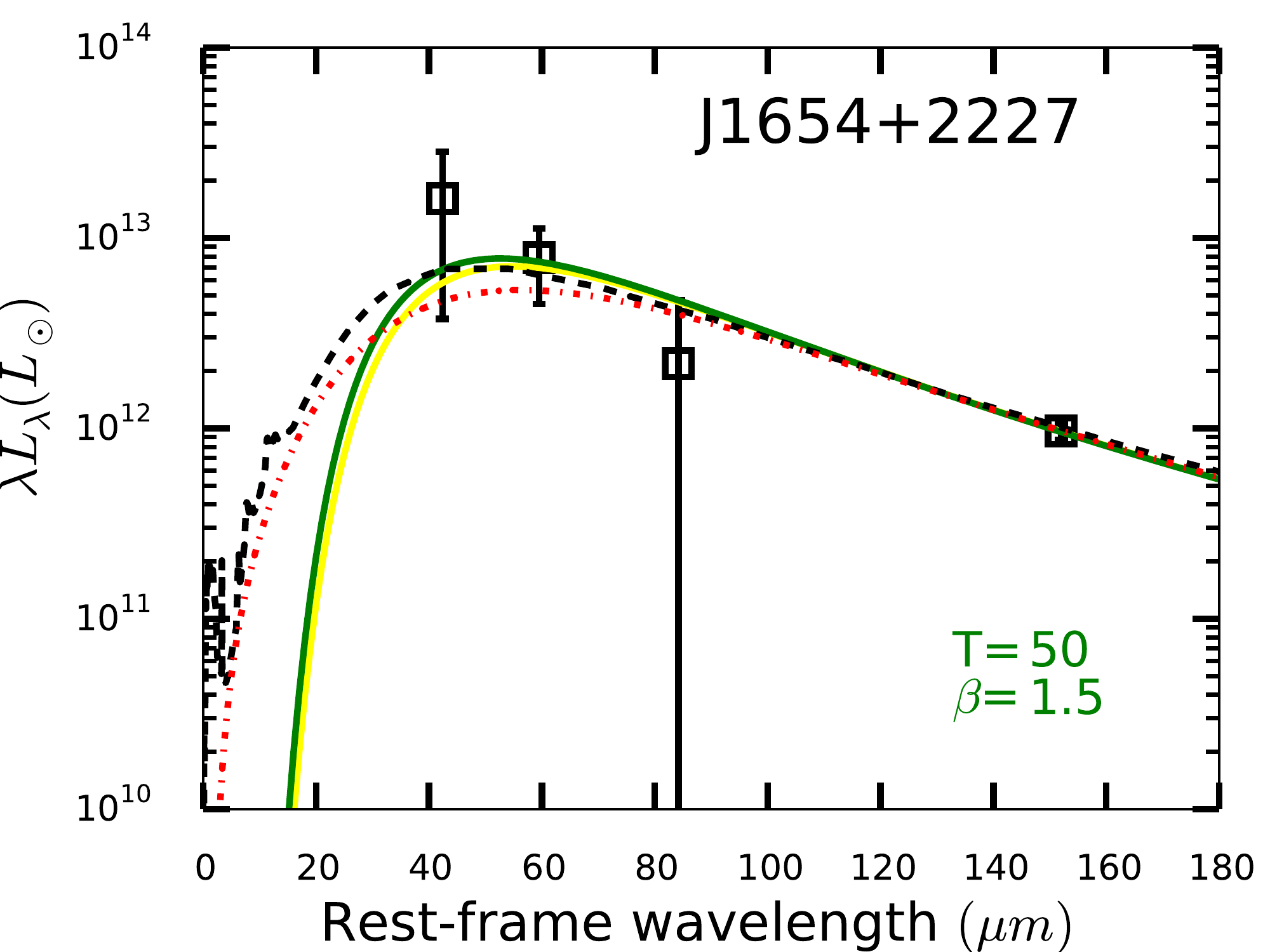} & \includegraphics[width=5cm,height=4cm]{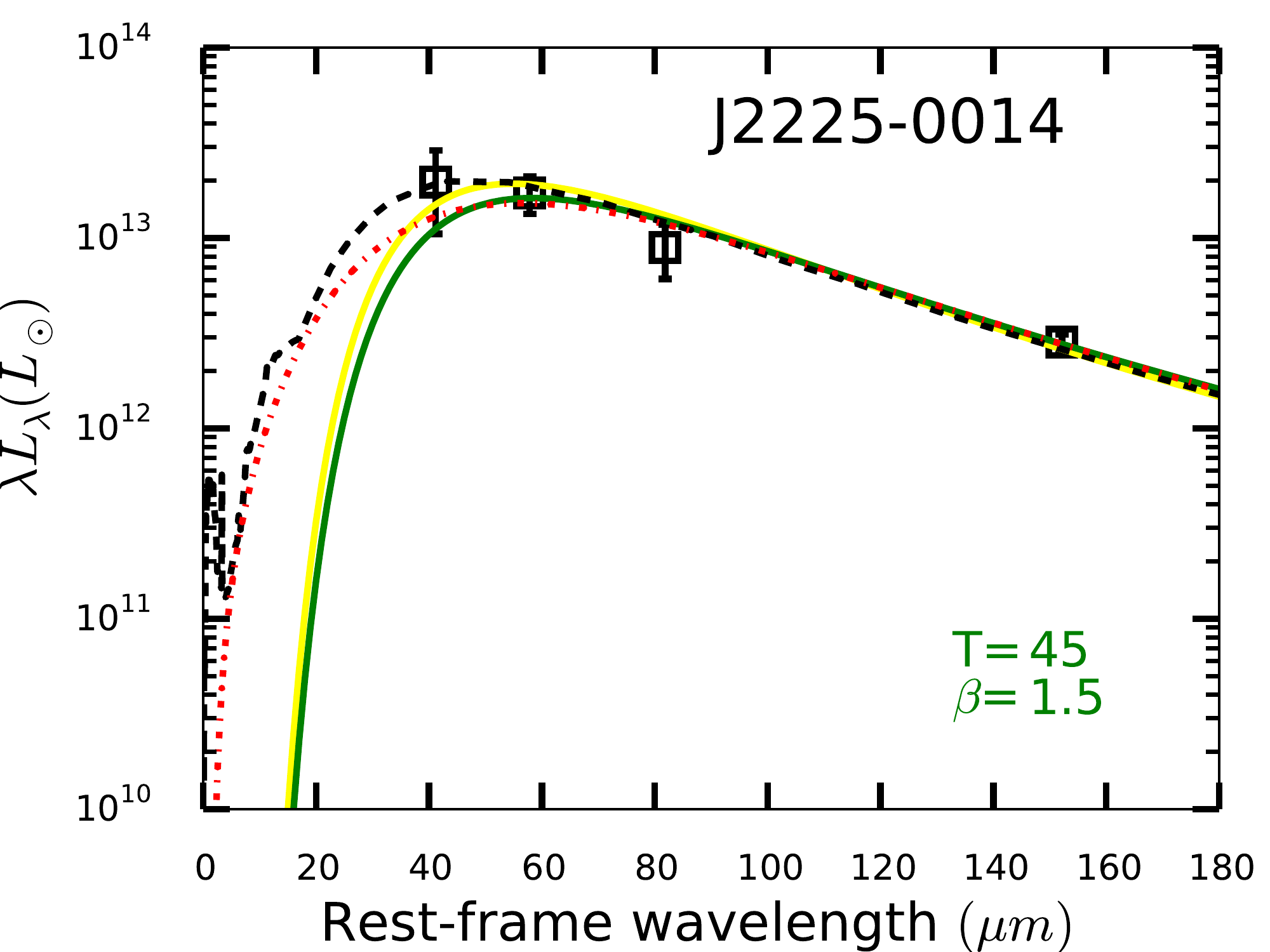} \\
\end{tabular}
\caption{FIR SEDs for the nine FIR-bright quasars in our sample,
  including those already presented in T17. Data points correspond to
  \herschel/SPIRE measurements at 250, 350 and 500 \mic\ and ALMA
  detections at 895 \mic\, (in the observed frame). For each source
  four model SEDs are presented: black-dashed lines represent the
  best-fitting FIR template from \cite{CharyElbaz2001} while
  red-dotted lines represent the scaled SED from
  \cite{Magnelli2012}. A scaled gray-body SED with $T_{\rm d}=47\,{\rm
    K}$ and $\beta=1.6$ is shown with a solid-yellow line, while a
  best fit model gray-body SED is shown with solid-green lines. The
  gray-body best fit parameters are included for each source.}
\label{fig:sfr_figs}
\end{figure*}

We will rely on the rest frame FIR continuum emission to estimate the
total FIR emission of our objects. This will allow us to determine the
SFRs of the host galaxies and nearby SMGs using the well established
relation between the FIR luminosity and the SFR
\citep{Kennicutt1989}. We include in this analysis the objects already
presented in T17.

For our FIR-faint objects this determination will be based only on the
ALMA detection. For the FIR-bright objects, we will also use the
\herschel\ measurements. We do not aim at performing a full modeling
of the FIR Spectral Energy Distribution (SED), as the number of
photometric points available do not allow for a determination of the
several physical parameters necessary for that, but rather determine
which set of SEDs better represent the observations.

The contribution to the FIR SED from the AGN should be small, commonly
given as $\sim$10 $\% $ (e.g.,
\citealt{Schweitzer2006,Mor_Netzer2012,Rosario2012,Lutz2016}). In this
paper we assume that the FIR emission of the sources in our sample is
dominated by dust heated by SF activity (see full discussion and many
references in \citealt{Netzer2016} and \citealt{Lani2017}). The
alternative view which involves AGN heated dust contributing
significantly to the FIR SED has been discussed in several
publications
\citep[e.g.,][]{Duras2017,Leipski2014,Siebenmorgen2015,Schneider2015}
but will not be addressed in this work. However, we do account for an
additional error of the 250 \mic\ \herschel/SPIRE band due to
contribution from AGN-heated dust (as explained in N14, we add in
quadrature an uncertainty estimated as 0.32 times the AGN luminosity
at 1450\AA). Taking this effect into consideration increases the error
of the 250 micron measurement on average by a factor of 1.67. ALMA
absolute flux calibration in band-7 is claimed to be of the order of
10\%. We add this uncertainty in quadrature to the errors quoted in
Table \ref{tab:spec_measurements}.

We use three different methods to produce model FIR SEDs for our
sources. Because of the lower uncertainties in the ALMA measurements,
this value usually dominates the fits for all three methods
discussed.

For the first method we use the grid of FIR SEDs provided by
\citet[][CE01]{CharyElbaz2001}.  These templates are unique in shape
and scaling. The best fit model is determined using the ALMA
monochromatic luminosity and its associated uncertainty, while for the
FIR-bright objects we also include the \herschel\ measurements (values
and errors from N14, with the 250 \mic\ flux error corrected as
explained above). For FIR-faint objects the fit relies only on the
ALMA measurement.

For the second method we scale the SED determined by
\cite{Magnelli2012}, which corresponds to an average from the most
luminous SMGs in their work. As before, the \herschel\ measurements are
included for those quasars with detections at 250, 350 and 500 \mic.

For the third method we use a gray-body SED. Following T17 and other
works, we use a temperature of $T_d = 47$ K and dust emissivity
coefficient $\beta = 1.6$. However, since some of our sources are not
well fitted using this set of parameters, we also try gray-body SEDs
with a wider range of temperatures (40, 45, 50, 55, 60 and 70 K) and
$\beta$ values (1.5 and 1.7). The determination of a best-fit
temperature and $\beta$ is only possible for the FIR-bright,
\herschel\ detected sources. The mean $\chi^2$ for our nine FIR-bright
quasars is $\sim 2$ (for one degree of freedom). We show the SED fits
in Figure \ref{fig:sfr_figs} together with the best fit values, which
are also reported in Table \ref{tab:gal_props}. We find that out of
nine objects, seven are well fit by temperatures in the 40 to 50 K
range. 

\begin{longrotatetable}
\begin{deluxetable*}{lllcccccccccc}
\tablecolumns{13}
\tablecaption{Galaxy Properties I \label{tab:gal_props}}
\tablehead{
\colhead{Sub-sample} &
\multicolumn{2}{c}{Target} &
\colhead{$\log L_{\rm CE}$} &
\colhead{$\log L_{\rm Mag}$} &
\colhead{$\log L_{\rm 47K \beta 1.6}$} &
\colhead{$\log L_{\rm best GB}$} &
\colhead{$T_{\rm best GB}$} &
\colhead{$\beta_{\rm best GB}$} &
\colhead{SFR$_{\rm CE}$} &
\colhead{SFR$_{\rm Mag}$} &
\colhead{SFR$_{\rm 47K \beta 1.6}$} &
\colhead{SFR$_{\rm best GB}$} \\
\cline{2-3}
        & ID        & Object    &(\Lsol)&(\Lsol)&(\Lsol)&(\Lsol) &(K)&        &(\mpyr)&(\mpyr)&(\mpyr)&(\mpyr) }
\startdata
Bright & J0807 & QSO     & 13.15 & 13.07 & 13.07 & 13.0 & 47 & 1.50 &	 1405 &	 1175 &	 1170 &	 1082 \\
       & J1404 & QSO     & 13.40 & 13.31 & 13.29 & 13.3 & 50 & 1.50 &	 2496 &	 2033 &	 1959 &	 2135 \\
       & J1433 & QSO     & 13.23 & 13.10 & 13.10 & 13.1 & 50 & 1.50 &	 1688 &	 1268 &	 1262 &	 1394 \\
       & J1616 & QSO     & 13.23 & 13.11 & 13.13 & 13.7 & 70 & 1.70 &	 1688 &	 1289 &	 1336 &	 5275 \\
       & J1654 & QSO     & 12.99 & 12.89 & 12.89 & 12.9 & 50 & 1.50 &	  985 &	  770 &	  778 &	  865 \\
       & J2225 & QSO     & 13.44 & 13.34 & 13.32 & 13.3 & 45 & 1.50 &	 2766 &	 2201 &	 2113 &	 1796 \\
       & J0331 & QSO\tln & 12.99 & 12.88 & 12.89 & 13.3 & 60 & 1.70 &	  985 &	  756 &	  776 &	 1922 \\
       & J1341 & QSO\tln & 13.56 & 13.50 & 13.46 & 13.5 & 50 & 1.50 &	 3613 &	 3164 &	 2911 &	 3137 \\
       & J1511 & QSO\tln & 13.40 & 13.26 & 13.26 & 13.4 & 50 & 1.70 &	 2496 &	 1838 &	 1805 &	 2262 \\
       & J1511 & SMG\tln & 12.25 & 12.40 & 12.41 & ---  & ---  & ---  &   176 &   250 &   256 &   --- \\
\hline\\ [-2.0ex]
Faint  & J1017 & QSO     & 12.25 & 12.37 & 12.38 & ---  & ---  & ---  & 176 &	 237 &	 242 &	 --- \\
       & J1151 & QSO     & 11.93 & 12.12 & 12.13 & ---  & ---  & ---  &  86 &	 131 &	 134 &	 --- \\
       & J1321 & QSO     & 12.28 & 12.41 & 12.42 & ---  & ---  & ---  & 192 &	 254 &	 260 &	 --- \\
       & J1447\dag&QSO&$<$11.06&$<$11.28&$<$11.29& ---  & ---  & ---  &$<$12&	 $<$19&	 $<$19&	 --- \\
       & J1447 & SMG     & 12.68 & 12.79 & 12.80 & ---  & ---  & ---  & 482 &	 620 &	 634 &	 --- \\
       & J2057 & QSO     & 12.39 & 12.51 & 12.52 & ---  & ---  & ---  & 246 &	 326 &	 333 &	 --- \\
       & J2057 & SMG     & 11.83 & 11.99 & 12.00 & ---  & ---  & ---  &  67 &	  98 &	 100 &	 --- \\
       & J2244 & QSO     & 12.65 & 12.73 & 12.74 & ---  & ---  & ---  & 444 &	 536 &	 548 &	 --- \\
       & J0923 & QSO\tln & 12.56 & 12.68 & 12.69 & ---  & ---  & ---  & 362 &	 477 &	 487 &	 --- \\
       & J0923 & SMG\tln & 12.16 & 12.27 & 12.28 & ---  & ---  & ---  & 144 &    187 &   191 &   --- \\
       & J1328 & QSO\tln & 12.32 & 12.43 & 12.44 & ---  & ---  & ---  & 207 &	 270 &	 276 &	 --- \\
       & J1328 & SMG\tln & 11.86 & 12.04 & 12.05 & ---  & ---  & ---  &  72 &    109 &   112 &   --- \\ 
       & J0935 & QSO\tln & 12.28 & 12.41 & 12.42 & ---  & ---  & ---  & 192 &	 255 &	 261 &	 --- \\
\enddata
\tablenotemark{}
\tablenotetext{T17}{\ \ \ \ Sources from \cite{Trakhtenbrot2017}.}
\tablenotetext{\dag}{Upper limit available for ALMA continuum flux.}
\end{deluxetable*}
\end{longrotatetable}

%\begin{longrotatetable}
\begin{deluxetable*}{lllccccccccc}[h]
\tablecolumns{9}
\tablecaption{Galaxy Properties II \label{tab:gal_dyns}}
\tablehead{
\colhead{Subsample} &
\multicolumn{2}{c}{Target}&
\colhead{$\log\mdyn^{uncorr}$} &
\colhead{$\log\mdyn$ \tablenotemark{a}} & 
\colhead{$\log M_{Disp.}$} &
\colhead{$\log M_{dust}$ \tablenotemark{b}} &
\colhead{$\log M_{BF}$\tablenotemark{c}}  &
\colhead{$\log\mbh$ \tablenotemark{d}} &
\colhead{$\mdyn/\mbh$} &  
\colhead{$\dot{M}_*/\Mdotbh$ \tablenotemark{e}} \\ [-1ex]
\cline{2-3}
        & ID &Object&($\Msol$)&($\Msol$)&($\Msol$)&($\Msol$)&($\Msol$)&($\Msol$)&($\Msol$)&($\Msol$)}
\startdata
Bright  & J0807  &  QSO        & 10.7 &  10.8 &10.5& 9.0 &9.0 & 9.2  &  33 &  65 \\
        & J1404  &  QSO        & 10.9 &  11.4 &10.7& 9.2 &9.2& 9.5  &   81 & 130 \\
        & J1433  &  QSO        & 10.6 &  11.0 &10.4& 9.0 &9.0& 9.1  &  131 &  38 \\
        & J1616  &  QSO        & 10.9 &  11.1 &10.7& 8.9 &8.7& 9.4  &   49 &  71 \\
        & J1654  &  QSO        & 10.8 &  10.8 &10.6& 8.8 &8.7& 9.6  &   18 &  51 \\
        & J2225  &  QSO        & 10.7 &  11.0 &10.5& 9.2 &9.2& 9.3  &   53 & 82 \\
        & J0331  &QSO\tln      & 10.6 &  10.8 &10.4& 8.8 &8.6& 8.8  &   88 &  57 \\
        & J1341  &QSO\tln      & 10.7 &  10.9 &10.5& 9.4 &9.4& 9.8  &   11 & 111 \\
        & J1511  &QSO\tln      & 10.8 &  10.9 &10.6& 9.1 &9.0& 8.4  &  264 & 183 \\
        & J1511  &SMG\tln      & 10.8 &  10.8 &10.6& ...  & ... & ...  & ...  & ... \\
\hline \\ [-3ex]
Faint   & J1017  &  QSO        & 10.0 & 11.0 &9.8 & 8.3 &   & 8.7  & 178  &  32 \\
        & J1151$\sharp$  &  QSO  &  ... & ...  & ... & $<$8.0& ... & 8.8  & ...  &  27 \\
        & J1321  &  QSO        & 10.8 & 11.0 & 10.6& 8.3 & ...  & 9.0  & 110  &  30 \\
        & J1447$\dag\sharp$&QSO& 10.2 & 11.1 & 10.0& $<$7.2& ... & 8.0  & 1214  &   3 \\
        & J1447  &  SMG        & 10.0 & 10.2 & 9.8 & 8.7& ...  & ...  & ..   & ... \\
        & J2057  &  QSO        & 10.5 & 10.6 & 10.3& 8.4& ...  & 9.2  &  21  &   8 \\
        & J2057  &  SMG        & 11.0 & 11.0 & 10.8& 7.9& ...  & ...  & ..   & ... \\
        & J2244  &  QSO        & 10.3 & 10.7 & 10.1& 8.6& ...  & 8.8  & 126  &  84 \\
        & J0923  &QSO\tln      & 10.5 & 10.9 & 10.4& 8.6& ...  & 8.7  & 158  &  60 \\
        & J0923  &SMG\tln      & 10.2 & 10.3 & 10.1& ...& ...  & ...  & ...  &  ...\\
        & J1328  &QSO\tln      & 10.1 & 10.8 & 9.8 & 8.3& ...  & 9.1  &  50  &  24 \\
        & J1328  &SMG\tln      & 10.8 & 11.0 & 10.7& ...& ...  & ...  & ...  & ... \\
        & J0935  &QSO\tln      & 10.4 & 10.6 & 10.3& 8.3& ...  & 8.8  &  56  &  20 \\
\enddata

\tablenotetext{T17}{\ \ \ \ Sources from \protect \cite{Trakhtenbrot2017}.}
\tablenotetext{\dag}{Dynamical masses based on estimate of the size of the J1447 host.}
\tablenotetext{\sharp}{No dust continuum detections.}
\tablenotetext{a}{Calculated using the inclination-angle corrections derived from the sizes of the \cii-emitting regions.}
\tablenotetext{b}{Calculated assuming the CE01-based SFRs.}
\tablenotetext{c}{Best fit values are $T_{bestGB}$ and $\beta_{bestGB}$ from Table Galaxy Properties I}
\tablenotetext{d}{Black hole masses taken from T11.}
\tablenotetext{e}{Calculated assuming $\Mdotbh= \left(1-\eta\right) \Lbol/\eta\, c^2$, with $\eta=0.1$.}
\end{deluxetable*}
%\end{longrotatetable}
Two require higher temperatures: J0331–0741 is best fit by a
$\beta = 1.7$ and $T_d = 60$ K gray-body SED while J1616+0501 needs
$\beta = 1.7$ and $T_d = 70$ K. We briefly discuss these two cases
next.

For J1616 all the \herschel\ photometric points are found more than
3$\sigma$ above the CE01 best-fit template, which is dominated by the
scaling to the ALMA measurement. The corresponding SFR from the
gray-body best fit is 5275 \mpyr\, even higher than the $\sim 4200$
\mpyr\ found by N14 based on \herschel\ data only. A similar, although
not as extreme case is J0331, whose data were already presented in
T17. For J0331 N14 determined a SFR of $\sim 2100$ \mpyr, in good
agreement with the value of 1922 \mpyr\ we determine from the
gray-body best fit. Clearly, the high SFRs determined for these
sources are driven by their very high \herschel\ luminosities. The
high gray-body temperatures, on the other hand, are the result of the
correspondingly steep SEDs, which are found once the ALMA data are
also taken into account.

Gray-body temperatures as high as $T_{d} = 60-70$ are not expected for
star-forming sources. However, temperatures as high as 70 or 80 K have
been recently determined for a very small fraction of SMGs at
high-redshift \citep{Miettinen2017}, so these rather high ISM
temperatures might not be totally unusual in the most luminous sources,
although more observations are necessary in order to confirm this.

Once the total IR luminosity is determined by integrating the SED over
the 8-1000 \mic\ range, the SFR is obtained using $\rm SFR / \mpyr =
L_{\rm FIR}\ / 10^{10} \Lsol$, which assumes a Chabrier Initial Mass
Function (IMF). The results are presented in Table \ref{tab:gal_props}
as $L_{\rm CE}$ and SFR$_{\rm CE}$ for the CE01 fits, $L_{\rm Mag}$
and SFR$_{\rm Mag}$ for the \cite{Magnelli2012} fits, $L_{\rm 47K
  \beta 1.6}$ and SFR$_{\rm 47K \beta 1.6}$ for the gray-body fit with
fixed parameters $T_d = 47$ K and $\beta = 1.6$, and as $L_{\rm best
  GB}$ and SFR$_{\rm best GB}$ for the gray-body fit with $T_d$ and
$\beta$ left as free parameters. 

For our total sample of 18 quasars,
we see that the FIR-bright targets have a SFR range of $\sim 900 -
3200$ \mpyr, the FIR-faint objects have a range of $\sim 200 - 500$
\mpyr, and the SFRs of the SMGs cover $\sim 90 - 600$ \mpyr. The
difference in the determined SFRs using the different methods
illustrates the systematic uncertainties of these calculations.

Besides the FIR-bright sources presented in Figure \ref{fig:sfr_figs},
Table \ref{tab:gal_props} also lists the SFRs obtained for the
FIR-faint sources. The SED best-fit values for J1447 are based on the
ALMA continuum upper limit previously determined. We found this object
to have an extremely low SFR of $< 20$ \mpyr, maybe indicating that
effective starformation quenching has already occurred. The detection
of \cii\ in this host showcases how this line can be detected in the
ISM of galaxies with very little on-going star formation.
In the following sections we will take the average SFR obtained from
these methods as the representative SFR for each object. Errors will
be computed as the maximum and minimum derived SFR.

\subsubsection{The \Lagn\ versus \LSF\ plane}
\label{sec:lagn_lsf}

\begin{figure}
\center
\includegraphics[scale=0.35,trim=80 0 0 0]{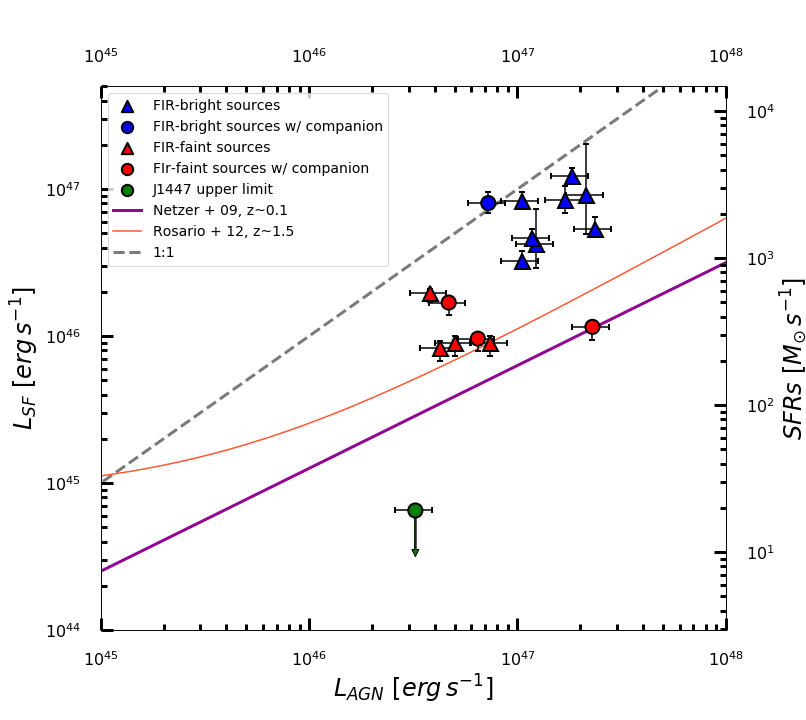}
\includegraphics[scale=0.60,trim=40 0 0 0]{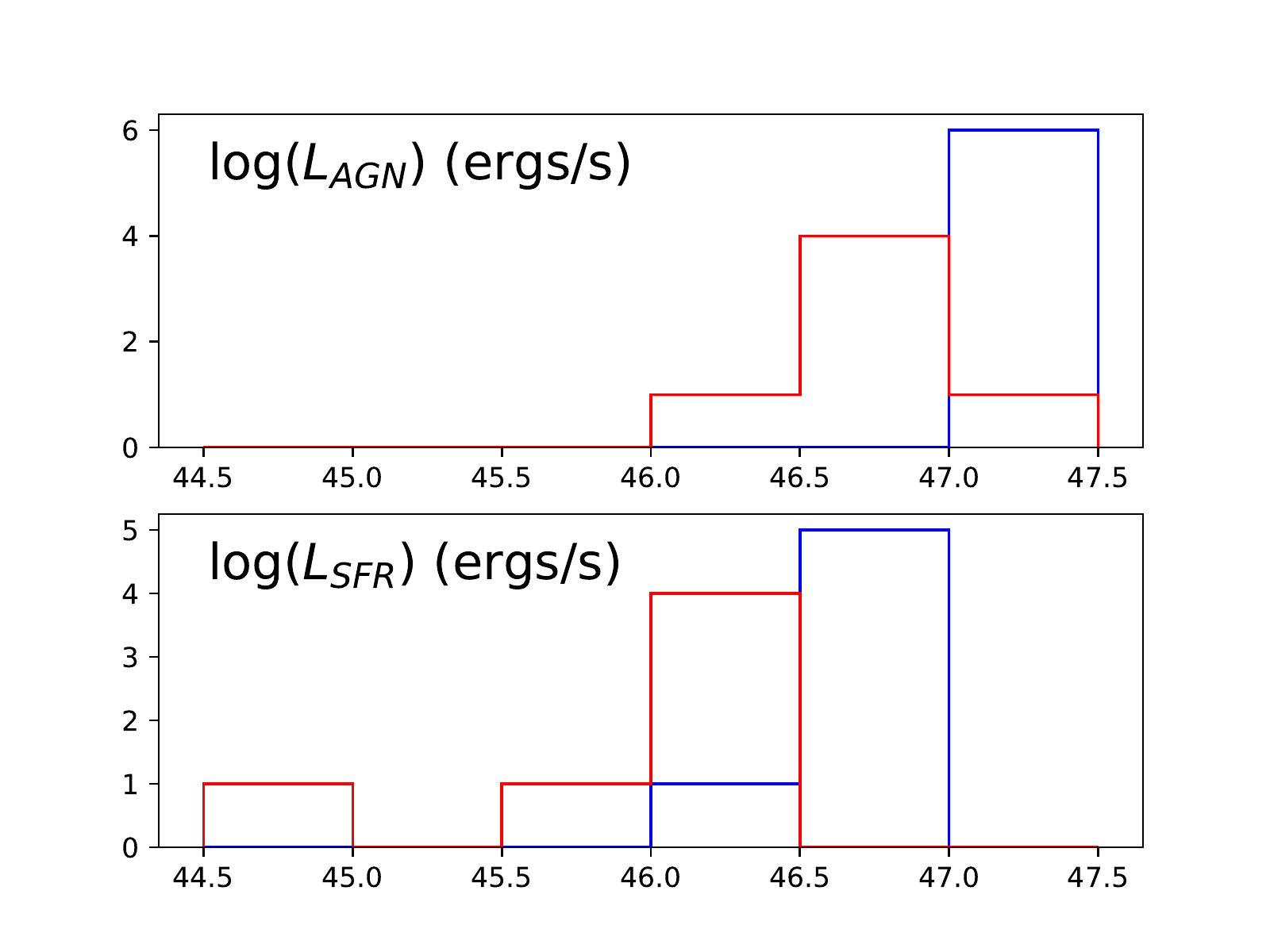}
\caption{Top: \LSF\ versus \lagn\ for our FIR-bright (blue markers)
  and FIR-faint sources (red markers), as well as the upper limit of
  J1447 (green marker) with an arrow to indicate it is an upper limit.
  Those sources with companions are marked as a circle. 
  The  orange curve for redshift $0.8-1.5$ from \cite{Rosario2012} and
  scaled up by a factor of two to allow for the difference between
  L(60 $\mu$m) used in that paper and the \LSF\ used in our work. The
  correlation for AGN dominated sources is shown as a solid purple and
  is taken from \cite{Netzer2009} as $\LSF \simeq 10^{43} (\lagn /(10^{43} \ergs))^{0.7} $. The dashed straight line
  corresponds \lagn\ = \LSF, shown for reference. Bottom: Logarithmic
  distributions of \LSF\ and \lagn\ in units of ergs s$^{-1}$.}
\label{fig:l_agn_sf}
\end{figure}
% Hollow black markers are for the 23 detected sources from \cite{Decarli2018}.   
Figure \ref{fig:l_agn_sf} presents the \lagn\ versus \LSF\ plane.
FIR-bright and FIR-faint objects are shown with different colors,
while the presence of companions is shown using different symbols.

Since our luminosity ranges in \lagn\ and \LSF\ are rather narrow, it
is not possible to draw conclusions about how our sources compare with
those trends found by previous works for \lagn\ and \LSF-dominated
sources. In fact, it has been a matter of great debate why the
\Lagn\ versus \LSF\ plane shows significantly different trends
depending on the way samples are defined (e.g., see discussion in
\citealt{Netzer2016}). The answer to the apparent contradictory results
seems to reside in the stochastic nature of AGN activity, with duty
cycles much shorter than those that characterize star formation
\citep[e.g.,][]{Hickox2014,Volonteri2015,Stanley2015}. In short,
selecting samples based on SFR and binning them in AGN power, will
give a representative \lagn-\LSF\ relation since the rapid ($\sim
10^{5-6}$ yr) changes in AGN power will be smoothed out, while
selecting them in \lagn\ and binning in \LSF\ will mix-and-match
objects selected from their `unrepresentative' AGN luminosity and with
very different SF power. These observed differences, however, seem to
saturate at the highest luminosities.

In Figure \ref{fig:l_agn_sf} we include a 1:1 \lagn-\LSF\ line as well
as the trends determined by \cite{Netzer2009} from observations at a
wide redshift range and \cite{Rosario2012} at $z \la 2$, both of which
defined for bright AGN but dominated by local samples in the case of
\cite{Netzer2009} and from samples drawn from deep field surveys in
the case of \cite{Rosario2012}, hence not including the most powerful
AGN. It is therefore not surprising that our optically flux-limited
selected sample of quasars on average, sits above both the
\cite{Netzer2009} and \cite{Rosario2012} relations. This is
particularly true for the bright-FIR subsample.
It would be of great interest to compare the \Lagn\ and \LSF\ distributions of our sources with those of higher redshift, like that of \cite{Decarli2018}. However, quasars found at $z > 5$ are very hard to find because of a strong contamination of late brown dwarfs, which introduces severe and complex selection biases to those systems \citep{Banados2016}.

Inspection of Figure \ref{fig:l_agn_sf} shows that the our FIR-bright and FIR-faint populations
occupy distinctively different regions of the diagram, even though the
individual distributions of these two properties do not show evidence
for two separated populations, as can be seen in the bottom panel of
Figure \ref{fig:l_agn_sf}. This can also be seen in Figure 2 of
\cite{Netzer2014} (as $\lambda L_{\lambda}$(3000\AA)) who analyzed the
\herschel\ observations of $z \sim 4.8$ quasars, including all our
FIR-bright and FIR-faint sources, and 20 further FIR-faint sources.
From this work it becomes clear that FIR-bright and FIR-faint sources
dominate at the high and low-end of the BH mass distribution,
respectively, but with no indication of a bimodality. 
In fact, the distributions of the source properties indicate a clear relation between BH mass, \Lagn, and SFRs with the median \mbh, \Lagn, and \LSF\ of the FIR-bright sources (9.28 \Msol, $10^{47.13}$ \ergs, and $10^{46.91}$ \ergs, respectively) being higher than those of the FIR-faint sources(8.85 \Msol, $10^{46.78}$ \ergs, $10^{46.23}$ \ergs). For a more in depth discussion see \cite{Netzer2014}.

We find a weak correlation coefficient between \lagn\ and \LSF\ for our entire sample ($r = 0.55$), however more extensive studies (e.g., \citealt{Stanley2015,Stanley2017,Lanzuisi2017}) indicate that much larger samples are required to draw conclusions. We will return
to the issue of the possible segregation observed in the \Lagn\ versus
\LSF\ plane in Section \ref{sec:major_mergers}.

\subsubsection{Dust masses}
\label{sec:dust_masses}

The continuum emission at rest wavelength $\~$ 152 \mic\ can also be
used to calculate dust masses for our objects assuming that the FIR
continuum flux originates from optically thin dust at these
wavelengths. Using the same methods as in \citet{Dunne2000} and
\citet{Beelen2006} (see also, \citealt{Scoville2016}), the dust mass can
be calculated as:

\begin{equation}
M_d = \frac{S_{\lambda rest} D_{L}^2}{\kappa_{d}(\lambda_{rest})B(\lambda_{rest}, T_d)}
\label{eq:mdust}
\end{equation}

\noindent where $\kappa_{d}(\lambda) \propto \lambda^{- \beta}$ is the
wavelength dependent dust mass opacity, $S_{\lambda rest}$ is the
continuum flux density at $\lambda rest$, $B(\lambda_{rest}, T_d)$ is
the monochromatic value of the Planck function at $\lambda_{rest}$ for
temperature $T_d$, and $D_{L}$ is the luminosity
distance. $\kappa_{d}$ is found to be 0.077 m$^{2}$ kg$^{-1}$ at 850
\mic\ \citep{Dunne2000}, and hence, $\kappa_{d}(\lambda_{rest})=
0.077\ (850 / \lambda_{rest})^{\beta}$ m$^{2}$ kg$^{-1}$. To calculate
the dust mass we assume $T_d = 47\,$ K and $\beta = 1.6$.

We note from equation (1) that the only formal error comes from the
measurement of the continuum flux, while systematic errors will arise
from our assumption of the adopted SED and the opacity coefficient,
which will dominate. However, as we are using very similar parameters
to those adopted in the literature a direct comparison of results is
possible.

We derive dust masses for our full sample of 16 continuum detected
quasars and find a range of $M_{dust} \sim 2-15\times 10^8 \Msol$ (see
Table \ref{tab:gal_dyns}). Upper limits of $\sim 10^8 \Msol$ and
$\sim 10^7 \Msol$ are found for J1151 and J1447 hosts,
respectively. The average value is larger for FIR-bright objects than
for FIR-faint objects, with dust masses of $10^{9.0}$ and $10^{8.4}
\Msol$, respectively. In Table \ref{tab:gal_props} we also determine dust masses for the FIR-bright objects using the best fit values of $T_{bestGB}$ and $\beta_{bestGB}$ discussed in Section \ref{sec:seds_sfrs}. However, we  note that due to the small range of $T$ and $\beta$ and the dominance of the continuum flux density and luminosity distance, the differences in these calculations from assuming $T_d = 47\,$ K and $\beta = 1.6$ are minor.

We derive dust masses for our full sample of 16 continuum detected
quasars and find a range of $M_{dust} \sim 2-15\times 10^8 \Msol$ (see
Table \ref{tab:gal_props}). Upper limits of $\sim 10^8 \Msol$ and
$\sim 10^7 \Msol$ are found for J1151 and J1447 hosts,
respectively. The average value is larger for FIR-bright objects than
for FIR-faint objects, with dust masses of $10^{9.0}$ and $10^{8.4}
\Msol$, respectively. In Table \ref{tab:gal_props} we also determine dust masses for the FIR-bright objects using the 
best fit values of $T_{bestGB}$ and $\beta_{bestGB}$ discussed in Section \ref{sec:seds_sfrs}. However, we  note that due to the small range of $T$ and $\beta$ and the dominance of the continuum flux density and luminosity distance, the differences in these calculations from assuming $T_d = 47\,$ K and $\beta = 1.6$ are minor. 

\subsubsection{Companion detections}

Current cosmological models recognize high-z quasars as sign-posts of
high-density environments (see \citealt{Costa2014} and references
therein). It is therefore not unexpected that our sample shows a
larger number of companions when compared to ALMA observations of
blank fields.

Recent blank deep field surveys conducted with ALMA
\citep{Carniani2015,Aravena2016a,Fujimoto2016} imply that each ALMA
pointing of 18\arcsec\ should have of the order of $\sim$ 0.1 SMGs at
a flux limit of 15 $\mu$Jy at 1.2mm. Other measurements of the HST
Legacy Fields \citep{Bouwens2015} and the Great Observatories Origins
Deep Survey (GOODS) Fields \citep{Stark2009} give surface densities on
the order of 0.01 galaxies per single ALMA band-7 pointing (for SMGS
with SFR $\sim 100\,\mpyr$). Though they have not been confirmed with
higher S/N, \citet{Aravena2016b} cites a number count of roughly 0.06
\cii-emitting $z \sim 5-8$ galaxies per ALMA pointing of the Hubble
Ultra Deep Field.

As quasars, SMGs are also highly clustered and seem to be hosted by
massive dark matter halos \citep{Wilkinson2017}. Besides, several
works have found that a substantial fraction of sub-mm sources with
multiple components, varying from 35 to 80 percent, depending on
resolution and flux limit
\citep{Hodge2013,Bussmann2015,Scudder2016,Hayward2018}.

In a recent study of multiplicity of far-infrared bright quasars,
\citet{Hatziminaoglou2018} assembled a random sample of 28
infrared-bright SDSS quasars with detections in \herschel/SPIRE. This
sample of detected quasars would correspond to our FIR-bright objects
in terms of $L_{AGN}$, $M_{BH}$, and Eddington ratios, but with $z
\sim 2-4$. Using the ALMA Atacama Compact Array (ACA)
\citet{Hatziminaoglou2018} found that 30 percent of their targets were
found to be multiple. However, their observations do not provide the same depth or
resolution as our own, and the redshifts of their sub-mm sources were not confirmed.

\citet{Decarli2017,Decarli2018} present a similar study of \cii\ and
dust continuum at a similar redshift to our study, where the ALMA
observations provide enough information to indicate whether the nearby
sources are real companions. They found that 4/25 rapidly star-forming
galaxies have a companion, i.e., 16 percent. Based on the
IR-luminosities reported by \citet{Decarli2018}, 20 quasars hosts
would be classified as FIR-faint for a threshold FIR luminosity of
$10^{12.9}$ \Lsol, and 3/4 of the companions would be associated to
FIR-faint quasar hosts.

With the two newly observed companions we present here, our total
observed sample of 18 quasars has 5 sources with companions, 1
FIR-bright (J1511) and 4 FIR-faint (J0923, J1328, J2057, and J1447),
i.e., 28 percent. J0923 and J1328 have no nearby sources in
\textit{Spitzer}/IRAC, while J2057 and J1447 were not observed by
\textit{Spitzer}. J1511 (T17) has two further nearby
\textit{Spitzer}/IRAC sources. It is interesting that we only find
that 1 FIR-bright target is multiple in ALMA observations, a rate much
lower than that found in the randomly selected FIR-bright sample of
\citet{Hatziminaoglou2018}, and that we find a percentage of
companions slightly higher than that reported by \citet{Decarli2017}.

\subsubsection{Major mergers among hosts}
\label{sec:major_mergers}

Different lines of evidence suggest that mergers among gas-rich
galaxies should drive the most luminous AGN and the most powerful
starformation of their hosts. This is proposed by numerical
simulations \citep{Hopkins2005,Hopkins2008} and also backed by
observations at low and high-$z$
\citep{Treister2012,Glikman2015,Koss2018}. Thus our initial
expectations were to find that our ALMA observations would show that
the FIR-bright sources are powered by major mergers of gas-rich
galaxies, and that the FIR-faint sources, found closer to the main
sequence of galaxies, could be evolving through a secular process or
also involved in mergers. The evidence would emerge from the presence
of close companions to our quasars.

We find that of the $\sim 28$ \% host galaxies with companions the
majority are FIR-faint sources (1 FIR-bright and 4 FIR-faint). One
FIR-bright source, J1404, presents an unusual \cii\ double peak that
could signal a late stage merger. \cite{Bischetti2018} found three
companions around their targeted $z=4.4$ quasar, two of which have
double-peaked line emission, while in \cite{Willott2017}, the high
spectral and spatial resolution allows them to attribute different
peaks in the \cii\ line to the quasar source, a 5 \kpc\ separated
companion, and a "central excess" component between the two.

The lack of companions to FIR-bright quasars is in fact problematic,
as it is usually assumed major mergers between gas-rich galaxies to be
the triggering mechanism for starbursting galaxies. Note, however,
that recent ALMA observations at $z \sim 4.5$, suggest that
minor-mergers might also locate systems above the main sequence
\citep{Gomez-Guijarro2018}.

The preference for companions in FIR-faint sources could then be
explained if these correspond to very early stages in the merger
process, while the FIR-bright systems correspond to much later stages,
when the progenitor galaxies are no longer resolved by our ALMA
observations. The lack of disturbances in the velocity fields of our
systems does not oppose this argument, as observations of the ISM in
low-$z$ mergers demonstrate that the central core of mergers rapidly
settles into a rotating-dominated system \citep{Ueda2014}.

The lack of clear, `on-going' mergers among our systems could be
explained as a sample bias since our quasars were optically selected.
\cite{Glikman2015} has shown that for a sample of 2MASS selected
dust-reddened quasars at $z \sim 2$, 8/10 hosts show clear evidence
for very close, interacting companions. Similar results were found by
\cite{Urrutia2008} for dust-reddened quasars at $z \sim 0.4 -
1.0$. The nuclei are so heavily dust-enshrouded that HST follow up
clearly revealed the perturbed hosts. These type of quasars would not
be found in our parent sample. It is then possible that the distinct
populations observed in the \Lagn-\LSF\ plane (Figure
\ref{fig:l_agn_sf}) reflect the properties of the very early and very
late mergers just mentioned.

\subsection{Other Determinations}

\subsubsection{Dynamical Masses}
\label{sec:dyn_masses}

The $\cii$ line can be used to estimate the dynamical masses (\mdyn)
of the quasar host galaxies and the companion SMGs. We use the same
method as in T17 and several other studies of \cii\ and CO emission in
high-redshift sources, which assumes the \cii-traced ISM is arranged
in an inclined, rotating disk
\citep{Wang2013,Willot2015,Venemans2016}, and determine \mdyn\ as:

\begin{equation}
\mdyn = 9.8 \times 10^{8} \left(\frac{D_{\cii}}{\kpc}\right) \left[\frac{\fwcii}{100\,\,\kms}\right]^2\,\frac{1}{\sin^2\left(i\right)} \,\,\Msol \,\, .
\label{eq:mdyn_fwcii}
\end{equation}

\noindent In this relation $D_{\cii}$ is the size of the \cii-emitting
region measured by the deconvolved major axis of the Gaussian fit of
said region (see Table \ref{tab:spec_measurements}). The
$\sin\left(i\right)$ term reflects the inclination angle between the
line of sight and the polar axis of the host gas disks, with the
circular velocity given as $v_{\rm
  circ}=0.75\times\fwhm/\sin\left(i\right)$.  \textit{i} is determined
from the ratio $\cos\left(i\right)=\left(a_{\rm min}/a_{\rm
  maj}\right)$, where $a_{\rm min}$ and $a_{\rm maj}$ are the
semi-minor and semi-major axes of the \cii\ emitting regions,
respectively. These masses can be found in Table \ref{tab:gal_dyns},
where we list \mdyn\ as well as its inclination uncorrected value
(i.e., $\mdyn^{uncorr} = \mdyn \times \sin^2(i)$). We also include the
values determined in T17.

We find that the FIR-bright and FIR-faint systems have comparable
$M_{dyn}$ values. The mean is $9 \times 10^{10}$ \Msol. We also note
that among the interacting SMGs reported in this work the companion to
J1447 is of particular interest. Its two \cii\ spectral components
taken individually, each with unresolved sizes, would correspond to
systems with comparable dynamical masses found at the lower end of the
observed range presented in Table \ref{tab:gal_dyns}. Therefore they
would represent a major merger between these two components, but a
likely minor merger with the quasar host. Note, however, that the
dynamical mass of the J1447 host is also particularly uncertain, due
to the weakness of the \cii\ detection.

This method of deriving the dynamical mass carries significant
uncertainties, due to the several assumptions required to derive them,
and to the limited spatial resolution data available for our
systems. A large contributor to the error is our measurement of the
major and minor axis of the \cii\ emitting region, from which we
derive $i$ and $D_{\cii}$. We estimate a mean error of 0.44 dex by
propagating systematic uncertainties and the uncertainties of our
measured values.

However, the most significant assumption is that we are observing
inclined rotating disks. Only 4/6 of our FIR-bright and possibly 2/4
of our FIR-faint objects show clear indications of a smooth and
coherent velocity gradient, as can be seen in Figure
\ref{fig:vel}. Furthermore, note that even a smooth and coherent
velocity gradient does not guarantee a rotation dominated host galaxy.
We can compute the dynamical masses assuming the case of pure dispersion-dominated gas \citep{Decarli2018}:
\begin{equation}
    M_{\rm Disp} = \frac{3}{2} \frac{a_{\rm maj}\sigma_{line}^{2}}{G}
\end{equation}
\noindent where $\sigma_{line}$ is the line width of the Gaussian fit of the \cii\ spectrum, G is the gravitational constant, and $a_{\rm maj}$ again is the major axis of the \cii-region. We find that the dynamical masses we derive from assuming dispersion-dominated gas is lower than both the inclination corrected and non-corrected dynamical masses derived from assuming a rotating disk, with a mean of $2.4\times 10^{10}$ \Msol\ for $M_{\rm Disp}$ (see Table \ref{tab:gal_props}). Dynamical masses derived from assuming dispersion-dominated gas can be regarded as a lower limit to the true dynamical mass. We will use the dynamical masses obtained assuming an inclined, rotating disk throughout this rest of the work in order to be comparable to similar studies in the literature.

\subsubsection{Gas Masses}
\label{sec:gas_masses}

We can determine gas masses, $M_{gas}$, making use of a gas-to-dust
ratio (GDR) of 100, as determined at low-$z$ \citep{Draine2007}.
Recent studies comparing gas mass estimates obtained from CO line
measurements and dust masses obtained from FIR emission have given a
wide range of GDRs for high-redshift systems ($\sim 30-100$)
\citep{Ivison2010,Aravena2016c,Banerji2017}. This is an unexpected
result, since it is well established that high-$z$ galaxies are
characterized by lower metallicities at all galaxy masses
\citep{Lian2018} and that the GDR is inversely proportional with
metallicity \citep{Remy-Ruyer2014,De_Vis2019}. However, as discussed
in \cite{Aravena2016c} and \cite{Banerji2017}, another interpretation
for these results is to assume a `normal' GDR and revisit the
determination of the CO luminosity to total gas conversion
factor. Both, the GDR and CO luminosity to total gas fraction are
highly dependent on galaxy properties, such as surface density,
compactness, and particularly, metallicity. In summary, and for a more
straight forward comparison with other works, we adopt a GDR  of 100.

Gas masses are derived from dust masses in Table \ref{tab:gal_dyns} and are found
to be large, in the $10^{10-11} \Msol$ range. For four of our
FIR-bright systems $M_{gas}$ are larger than the dynamical masses by factors of up to three,
while for only one FIR-faint system $M_{gas} \sim 0.9 \times \mdyn$,
the remaining showing factors ranging from 0.7 to 0.2.

In general the estimated ISM masses for our quasars are comparable to
their dynamical masses. For our FIR-bright sources, 6/9 show $M_{gas}
\ge M_{dyn}$, by factors 1-3 (the unphysical finding that $M_{gas} >
M_{dyn}$ would be alleviated had we adopted a GDR as low as 30, as
discussed above). This is not seen for the FIR-faint sources,
suggesting that FIR-bright objects are more gas rich than FIR-faint
systems. Defining $f_{gas} \equiv M_{gas}/M_{dyn}$, we find for those
objects where $M_{gas} < M_{dyn}$ that $f_{gas} = 0.2-1.0$.

\begin{figure}
\center
\includegraphics[scale=0.6,trim=20 0 0 0]{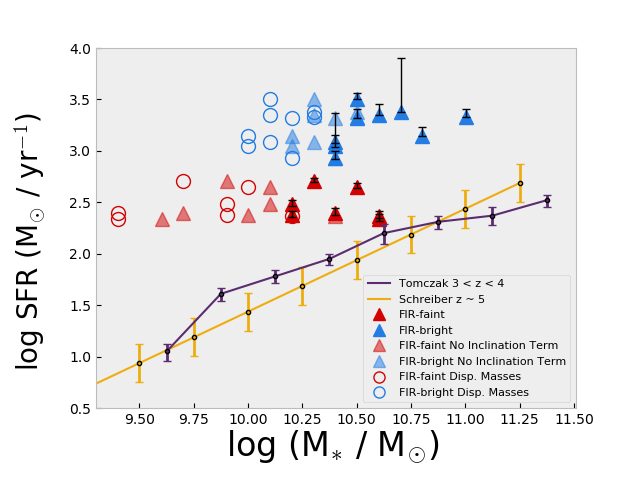}
\caption{Stellar mass versus Star Formation Rate -- the main sequence
  of starforming galaxies -- for our quasars. Bright and faint FIR
  sources are shown with different colors. Only seven FIR-faint
  objects are plotted as J1447 is not detected significantly in continuum or \cii, 
 while J1151 is a complete non-detection. The dynamical mass and SFR values are taken
  from Table \ref{tab:gal_props}. We assume $M_{\star} = 0.4\ \mdyn$,
  as explained in Section \ref{sec:dyn_mass}. We include the MS curves
  given in Equation (9) of \cite{Schreiber2015} (shown in yellow) and
  that of \cite{Tomczak2016} (in purple). The opaque red and blue
  triangles are our inclination corrected dynamical masses, while the
  transparent triangles are the inclination \textit{uncorrected}
  dynamical masses. The dynamical masses calculated assuming dispersion dominated gas are plotted as circles.}
\label{fig:ms}
\end{figure}

\subsubsection{The Main Sequence at $z \sim 5$}
\label{sec:dyn_mass}

We want to compare our full quasar sample with galaxies found on the
stellar mass -- SFR sequence for starforming systems, the `main
sequence' (MS), at similar redshifts. However, we only have estimates
for the total dynamical and gas masses of our quasar hosts, not of
their stellar masses. In principle, these could be obtained
calculating $M_{\star} = \mdyn-M_{gas}$. From the measured values
there is a strong indication that most of the quasars hosts are very
gas rich, with $M_{gas} = 0.3-4.0\ M_{\star}$, for those objects where
$M_{gas} < M_{dyn}$, and possibly higher for those objects where
$M_{gas} > M_{dyn}$. As already explained, the uncertainties on these
values are significant.

An alternative approach is to adopt a gas fraction measured in
non-active high-$z$ galaxies where the stellar mass can be determined
directly, which is not possible for our sample because of the
dominance of the AGN continuum at rest-frame near-IR and optical
bands. These determinations have been done out to $z \sim 4$
\citep{Schinnerer2016,Dessauges2017,Darvish2018,Gowardhan2019}, and
found $f_{gas} \sim 0.5-0.8$ (considering no dark matter), where a
strong dependence with redshift and no correlation with environment
are also seen \citep{Darvish2018}. We can then conservatively assume
that for our systems $f_{gas} = M_{gas}/M_{dyn} = 0.6$ and therefore
$M_{\star} = 0.4\ \mdyn$.

In Figure \ref{fig:ms}, we plot two MS curves. One is the
parameterization given in Equation (9) of \cite{Schreiber2015} for
redshift ranges $4 < z < 5$, after correcting for the different
adopted IMF (\citealt{Schreiber2015} uses a conversion factor of SFR
to $L_{SF}$ 1.7 times larger than our own). The second curve is from
\cite{Tomczak2016}, for galaxies at redshifts $0.5 < z < 4$. Both MS
curves agree well with each other.

We find that the majority of our sources lie above the MS
curves. If we used dynamical mass values derived from assuming dispersion-dominated gas,
our objects would shift to the lower stellar mass regime and sit even higher above the MS, as seen in Figure \ref{fig:ms}. Clearly, all of our FIR-bright quasars are found in the
starbursting domain and at least 1 dex from the MS. Their SFRs are
only comparable to the brightest known SMGs. Some of the FIR-faint sources sit within 1$\sigma$ of the MS of starforming galaxies at those early epochs, but again, the majority of our faint sources sit above the MS. Note that our
division into FIR-bright and FIR-faint sources is completely
arbitrary, and the determined SFRs for our full sample is indeed a
continuous distribution, as shown in the bottom panel of Figure \ref{fig:l_agn_sf}.

\subsubsection{SMBH--Host Galaxy Mass Relation}
\label{sec:smbh_galaxy_relation}

\begin{figure}
\center
\includegraphics[scale=0.35,trim=0 0 0 0]{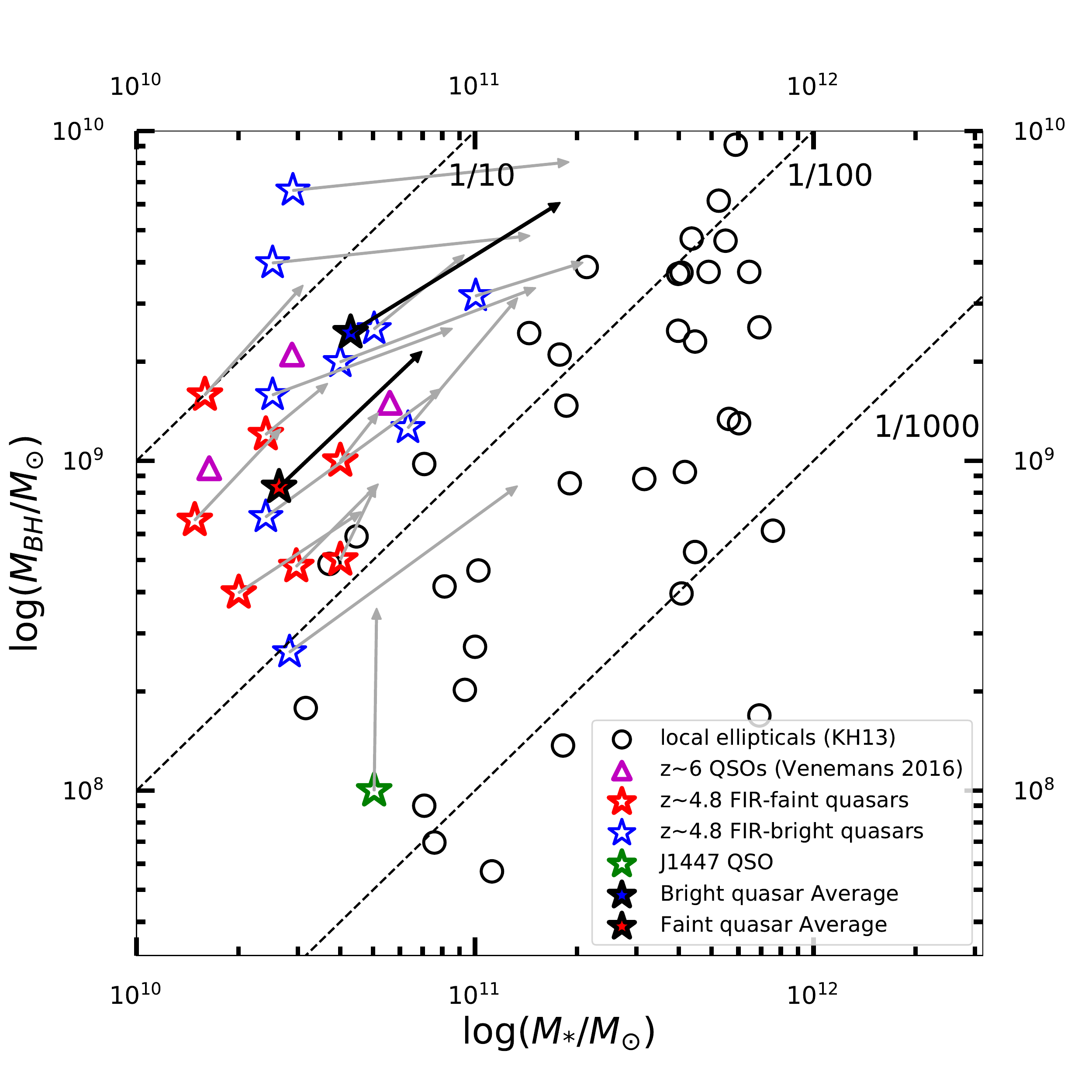}
\caption{ Black hole masses, \mbh, vs. host galaxy stellar masses,
  $M_{\star}$, for our sample of $z \sim 4.8$ quasars. FIR-bright
  objects are marked with blue stars, FIR-faint objects are marked
  with red. For comparison we also plot a sample of $z \simeq 0$
  elliptical galaxies taken from \cite{Kormendy2013} shown as black circles. The dotted diagonal lines trace different
  constant BH-to-host mass ratios. Grey arrows indicate the
  possible evolution in both the BH and stellar components, assuming
  constant mass growth rates over a period of 50 Myr. Filled stars
  with black arrows indicate average values and growth for
  both FIR-faint and bright objects. Our plotted sources have a typical error of  0.44 dex for 
  $M_{\star}$ (from our own estimates) and 0.4 dex for \mbh\ (derived in T11).}
\label{fig:bh_ratio}
\end{figure}

In Figure \ref{fig:bh_ratio}, we plot the stellar masses of our quasar
hosts against their black hole masses for the full sample of nine
FIR-bright and eight FIR-faint quasars detected in \cii. As before, we
have adopted $M_{\star} = 0.4\ \mdyn$. However, it is likely that the
real values of $M_{\star}$ would broaden the observed distribution,
which now corresponds to a net shift of the observed
\mdyn\ distribution. Black Hole masses were taken from T11 and
are based on \mgii\ measurements. We find that the average black hole mass of our sample is $10^{9.2}$\Msol, with a slight difference between the black hole properties of FIR-bright and -faint objects. FIR-bright objects having an average \mbh\ of $10^{9.4}\Msol$ and Eddington ratio of $\lledd \~ 0.65$, while FIR-faint objects have an average \mbh\ and Eddington ratio of $10^{8.9}\Msol$ and 0.78 respectively. We find a mean $M_{BH}/M_{\star}$
ratio of 1/19, with FIR-bright sources having $M_{BH}/M_{\star} =
1/15$ and FIR-faint systems 1/28.

We compare our sample with the local massive elliptical galaxies 
\citep{Kormendy2013} , Figure \ref{fig:bh_ratio} shows the positions of local galaxies with a  $M_{BH}/M_{\star}$ ratio ranging from $\sim 1/100$
to $\sim 1/1000$, the ratios being strongly correlated with
mass. While the black hole masses of our sample of high-$z$ luminous
quasars are found at similar values as seen in the local universe, the
stellar masses are on average one order of magnitude lower. This is
similar to the results found by other groups and is in good agreement
with the direct detection of two quasars hosts at $z \sim 4$
\citep{Targett2012}. A higher redshfit sample of three quasars at redshift $z\~6.8$ with available \mbh, $M_*$, \lledd, and SFRs is found in \cite{Venemans2016}. \mbh\ values were taken from \cite{derosa2014}.
When compared to our own sample in Figure \ref{fig:bh_ratio} we see that they sit between the FIR-bright and FIR-faint objects with an average $M_{BH}/M_{\star}$ ratio of 1/19, the same as our sample. In terms of their AGN properties they are found at the top end of the mass distribution of $z\simeq6$ sources presented in T11, but at the low end in terms of \lledd. Their SFRs are somewhat in between our FIR-bright and FIR-faint objects. We include these 3 sources in our Figure \ref{fig:bh_ratio}. As we already pointed out, it is not possible to derive any conclusions from a direct comparison between sources at $z\~5$ and $z\~6$ because of the very different way these samples have been defined.

Assuming that the stellar mass of the quasar host galaxies grows only
due to the formation of new stars (i.e., neglecting possible mergers),
we can use our SFR estimates from Section \ref{sec:seds_sfrs} to
calculate the growth rate of $M_{\star}$, i.e., $\dot{M}_{\star}$. The
instantaneous growth rate of the black holes can be computed as the
mass accreted onto the black hole which does not convert into energy:
$\dot{M}_{BH} = \frac{1 - \eta}{\eta} \frac{L_{bol}}{c^2}$, where
$L_{bol}$ is the bolometric luminosity from T11 using the rest-frame
UV continuum emission. We assume the radiative efficiency to be $\eta
= 0.1$.

As in T17 we find that all systems have $\dot{M}_{BH}/\dot{M}_{\star}
> 1/200$ and typical values are found to be $\sim 1/54$ (see Table
\ref{tab:gal_props}), with the FIR-bright and FIR-faint systems having
medians of $\dot{M}_{BH}/\dot{M}_{\star} \simeq 1/71$ and $1/27$
respectively. 

Assuming that the calculated instantaneous growth rates continue for a
period of time, we can determine the migration that our sources would
undergo on the $M_{BH}$ vs $M_{\star}$ plot. The time span needs to be
determined under reasonable assumptions. Typical starformation time
scales derived at lower redshifts might not be applicable to our
sample. Using the determined $M_g$ and SFRs we can find the depletion
time for the observed reservoir of gas. This is found to be between
20 to 100 Myr. Hence, we will adopt a general time span of 50 Myr,
which is also what was used in T17.

As already discussed, because of the stochastic nature of AGN
activity, with duty cycles shorter than those of star-formation by one
or perhaps up to two orders of magnitude
\citep[e.g.,][]{Hickox2014,Volonteri2015,Stanley2015}, the
instantaneous $\dot{M}_{BH}$ values measured for single objects might
not be the best proxy to characterize black hole growth over the time
required for the build up a sizable stellar mass due to star
formation. Instead, the value averaged over our entire sample will
result in better determination of the `typical' $\dot{M}_{BH}$. The
resulting `growth tracks' are shown in Figure \ref{fig:bh_ratio}.  We
also obtained the means of $\dot{M}_{BH}$ and $\dot{M}_{\star}$
separately for the FIR-bright and FIR-faint subsamples and have
plotted them in Figure \ref{fig:bh_ratio}. For most objects these
tracks suggest a larger future growth of stellar mass over BH mass,
which is necessary to bring them closer to the local population of
elliptical galaxies.

\section{Summary and Conclusion}
\label{sec:summary}

We have presented new band-7 ALMA observations for twelve new luminous
quasars at \zfpe, to reach a total sample size of 18 sources, which
are divided into \herschel/SPIRE detected (FIR-bright) and
\herschel/SPIRE undetected (FIR-faint) systems. The data probes the
rest-frame far-IR continuum emission that arises from dust heated by
SF in the host galaxies of the quasars, and the \CII\ emission line
from the host ISM. The ALMA observations resolve the continuum- and
line-emitting regions on scales of $\sim2$ \kpc.

Our main findings for our total sample of 18 targets is as follow.

\begin{enumerate}

 \item\ 5/18 of our quasars have companions, four of the quasars are
   FIR-faint and one is FIR-bright. The companions are separated by 15
   - 60 kpc. The quasar hosts with companions have a SFR rate of $\sim$ 220 - 3200
   \mpyr. The companions are forming $90 - 580\,\mpyr$  which is generally lower than the SFR measured for the quasar hosts.

 \item\
 The dynamical masses of the quasar hosts, estimated from the
 \cii\ lines, are within a factor of $\sim 3$ of the masses of the
 interacting companions, supporting an interpretation of these
 interactions as major mergers.

 \item\
 For all our sources, we find that the gas mass is comparable to the
 dynamical mass, suggesting that some of them could be kinematically
 dominated by the ISM component.

\item\
The \cii-based dynamical masses show that our systems are
above the ``main sequence'' of star-forming galaxies. When comparing \Lagn\ vs
\LSF\ we find evidence that the FIR-bright and FIR-faint subsamples are separated. We tentatively interpret this result as an evolutionary sequence within merger evolution, but great caution must be exercised as this is based on small number statistics.

 \item\
 Compared with the BH masses, the \cii-based dynamical host masses are
 generally lower than what is expected from the locally observed
 BH-to-host mass ratio. 
 
 \item\ 
 We have found a clear blueshift of \mgii\ with respect to our
 \cii\ measurements which is not observed at lower redshifts. No
 correlation is found between the shift and the presence of companions
 or the accretion rate of the supermassive black holes.
 
\item\ The lack of companions to most of our quasar hosts may suggest that
processes other or besides major mergers are driving the significant
SF activity and fast SMBH growth in these systems. Alternatively, the
systems could be observed at very different stages of the merger
process, with most FIR-faint sources found at the early stages, while
FIR-bright are found at very late phases.
 
\end{enumerate}

\acknowledgments This paper makes use of the following ALMA data:
ADS/JAO.ALMA\#2016.1.01515.S. ALMA is a partnership of ESO
(representing its member states), NSF (USA) and NINS (Japan), together
with NRC (Canada) and NSC and ASIAA (Taiwan) and KASI (Republic of
Korea), in cooperation with the Republic of Chile. The Joint ALMA
Observatory is operated by ESO, AUI/NRAO and NAOJ. N.N. and P.L. acknowledge funding from Fondecyt Project \#1161184.
B.T. is a Zwicky Fellow. R.M. acknowledges support from the ERC Advanced Grant 695671
`QUENCH’ and support by the Science and Technology Facilities Council
(STFC).

\bibliography{z48_ALMA_c2c4_refs}

\begin{table}
\caption{Compiled Offset List}
\label{tab:comp_offset}
\begin{tabular}{l l c l l c}
\hline
Source & Target & $\cii\ - \mgii\ $ & $L/L_{\rm Edd}$ & log $L_{\rm IR}$ & \mgii\ Paper \\
       &        & \kms\             &               &                & \\
\hline 
\citealt{Trakhtenbrot2017} & SDSS~J033119.67–074143.1     & $+$412  & 1.202  & 13.05 & \citealt{Trakhtenbrot2011}\\     
                           & SDSS~J134134.20+014157.7     & $+$573  & 0.1819 & 13.51 &...\\                           
                           & SDSS~J151155.98+040803.0     & $+$456  & 1.1819 & 13.31 &...\\                           
                           & SDSS~J092303.53+024739.5     & $-$213  & 0.6606 & 12.65 &...\\                           
                           & SDSS~J132853.66-022441.6     & $-$621  & 0.3548 & 12.40 &...\\                           
                           & SDSS~J093508.49+080114.5     & $+$588  & 0.741  & 12.37 &...\\                            
This Work   	           & SDSS~J080715.11$+$132805.1   & $+$256  & 0.447  & 13.08 &...\\
		           & SDSS~J140404.63$+$031403.9   & $+$2208 & 0.219  & 13.33 &...\\
		           & SDSS~J143352.21$+$022713.9   & $+$379  & 1.230  & 13.15 &...\\ 
		           & SDSS~J161622.10$+$050127.7   & $+$620  & 0.537  & 13.38 &...\\
		           & SDSS~J165436.85$+$222733.7   & $-$112  & 0.199  & 12.93 &...\\
		           & SDSS~J222509.19$-$001406.9   & $+$340  & 0.617  & 13.35 &...\\
      	                   & SDSS~J205724.14$-$003018.7   & $+$1064 & 0.891  & 12.48 &...\\
		           & SDSS~J132110.81$+$003821.7   & $+$337  & 0.355  & 12.37 &...\\
		           & SDSS~J224453.06$+$134631.6   & $+$225  & 0.676  & 12.71 &...\\
		           & SDSS~J101759.63$+$032739.9   & $+$1605 & 0.549  & 12.34 &...\\
		           & SDSS~J144734.09$+$102513.1   & $-$224  & 1.995  & 11.23 &...\\ \hline
\citealt{Decarli2018}      & SDSS~J084229.43$+$121850.4   & $+$310  & 0.7    & 12.20 & \citealt{Derosa2011}\\      
                           & SDSS~J130608.26+035626.3     & $+$757  & 0.792  & 12.50 & \citealt{Kurk2007}\\              
                           & CFHQS~J1509-1749             & $+$63   & 0.68   & 12.59 & \citealt{Willott2010} \\
                           & CFHQS~J2100-1715             & $-$245  & 0.49   & 11.77 &...\\                            
                           & PSO~J231.6576–20.8335        & $-$340  & 0.48   & 13.04 & \citealt{Mazzucchelli2017}\\     
                           & VIKING~J1048-0109            & $+$583  & ...    & 12.92 & Venemans in Prep.\\
                           & VIKING~J2211-3206            & $+$139  & ...    & 12.24 &...\\
                           & VIKING~J2318-3113            & $-$20   & ...    & 12.92 &...\\ \hline
\citealt{Willot2013}       & CFHQS~J0210$-$0456           & $-$230  & 2.4    & 11.41 & \citealt{Willott2010}\\
\citealt{Willot2015}       & CFHQS~J0055$+$0146           & $+$988  & 0.62   & 11.69 &...\\
                           & CFHQS~J2229$+$1457           & $-$13   & 2.4    & 11.09 &...\\
\citealt{Willott2017}      & CFHQS~J2329$-$0301           & $-$24   & 1.3    & 10.95 &...\\
                           & PSO~J167.6415$-$13.4960      & $+$308  & 1.2    & 12.43 & \citealt{Venemans2015} \\ \hline
\citealt{Venemans2012}     & ULAS~J112001.48$+$064124.3   & $-$474  & 0.48    & 12.30 & \citealt{derosa2014} \\
\citealt{Venemans2016}     & VIKING~J234833.34$-$305410.0 & $+$486  & 0.18   & 12.73 &...\\
                           & VIKING~J010953.13$-$304726.3 & $+1690$ & 0.24   & 12.19 &...\\ 
                           & VIKING~J030516.92$-$315056.0 & $+$374  & 0.68   & 12.93 &...\\
\citealt{Venemans2017}     & ULAS~J134208.10$+$092838.6   & $+$503  & 1.5    & 11.98 & \citealt{Banados2018} \\ \hline
\citealt{Mazzucchelli2017} & PSO~J338.2298$+$29.5089      & $+$313  & 0.11   & 12.45 & \citealt{Mazzucchelli2017}\\
                           & PSO~J323.1382$+$12.2986      & $-$154  & 0.44   & 12.11 &...\\ \hline
\citealt{Banados2015}      & PSO~J036.5078$+$03.0498      & $+$567  & 0.96   & 12.88 & \citealt{Venemans2015} \\ \hline
\citealt{Wang2016}         & SDSS~J010013.02$+$280225.8   & $+$1019 & 0.95   & 12.54 & \citealt{Wu2015} \\ \hline 
\end{tabular}
\end{table}

\end{document}